\documentclass[useAMS,usenatbib]{mn2e}

\usepackage[toc, page]{appendix} 
 
\usepackage{epsfig} 
 
\usepackage{graphicx} 
 
\usepackage{color} 
 
\usepackage{times} 
 
\usepackage{amsmath} 
 
\usepackage{amssymb} 
 
\usepackage{xspace} 
 
\usepackage{txfonts} 
 
\usepackage{enumitem} 
 
\usepackage{natbib} 
 
\usepackage{algorithm} 
 
\usepackage{algorithmic} 
 
\usepackage{subfloat} 
 
\usepackage{subfig}

\def\Mpch{h^{-1} {\rm Mpc}} 
\def\d3k{{\displaystyle {\rm d}{\bf k} \over \displaystyle (2\pi)^3}}

\newcommand{\Manifold}   {\mm{{\mathbb M}}} 
\newcommand{\Rspace}     {\mm{{\mathbb R}}}

\newcommand{\diff}       {\mm{\rm \,d}} 
 
\newcommand{\Probability}{\mm{\rm Prob}} 
\newcommand{\ProbDensity}{\mm{\cal P}}

\newcommand{\ff}         {\mm{\bf f}}

\newcommand{\xx}         {\mm{\bf x}} 
 
\newcommand{\XX}         {\mm{\bf X}}

\newcommand {\mm}[1] {\ifmmode{#1}\else{\mbox{\(#1\)}}\fi} 
 
\newcommand{\Mspace}         {\mm{{\mathbb M}}} 
 
\newcommand{\Volume}[1]      {\mm{\rm vol\,}{#1}} 
 
\newcommand{\persistence}[1] {\mm{\rm pers}{({#1})}} 
 
\newcommand{\low}[1]         {\mm{\rm low}{({#1})}} 
 
\newcommand{\Ddgm}[2]        {\mm{\rm Dgm}_{#1}{({#2})}}

\begin{document}

\title{The Topology of the Cosmic Web \\ in Terms of Persistent Betti Numbers} 
\author[Pratyush Pranav, Herbert Edelsbrunner, Rien van de Weygaert, Gert Vegter] 
	{Pratyush Pranav$^{1,2}$\thanks{pranav@astro.rug.nl}, 
	Herbert Edelsbrunner$^{3}$, 
        Rien van de Weygaert$^{1}$, 
        Gert Vegter$^{4}$, 
        \and	 
        Michael Kerber$^{6}$, 
        Bernard J.T. Jones$^{1}$, 
	and Mathijs Wintraecken$^{4,5}$\\ 
$^1$   Kapteyn Astronomical Institute, University of Groningen, P.O. Box 800, 9700AV Groningen, the Netherlands \\ 
$^2$   Technion - Israel Institute of Technology, Haifa, Israel 32000\\ 
$^3$   IST Austria (Institute of Science and Technology Austria), Klosterneuburg, Austria\\ 
$^4$   J.B. Institute for Mathematics and Computer Science, University of 
Groningen, the Netherlands\\ 
$^5$   INRIA  Sophia Antipolis-M\'editerran\'ee, 2004 route des Lucioles - BP 93, 06902 Sophia Antipolis Cedex, France\\ 
$^6$   Institute of Geometry, Graz University of Technology, Kopernikusgasse 24, 8010 Graz, Austria
}  
\maketitle

\begin{abstract} 
We introduce a multiscale topological description of the Megaparsec weblike cosmic matter  
distribution. Betti numbers and topological persistence offer a powerful means of  
describing the rich connectivity structure of the cosmic web and of its multiscale  
arrangement of matter and galaxies. Emanating from algebraic topology and Morse theory,  
Betti numbers and persistence diagrams represent an extension and deepening of the  
cosmologically familiar topological genus measure, and the related geometric Minkowski functionals.  
In addition to a description of the mathematical background, this study presents the  
computational procedure for computing Betti numbers and persistence diagrams for  
density field filtrations. The field may be computed starting from a discrete spatial distribution of galaxies or  
simulation particles. The main emphasis of this study concerns an extensive and systematic exploration of the  
imprint of different weblike morphologies and different levels of multiscale clustering in  
the corresponding computed Betti numbers and persistence diagrams. To this end, we use Voronoi  
clustering models as templates for a rich variety of weblike configurations, and the fractal-like  
Soneira-Peebles models exemplify a range of multiscale configurations. We have identified the clear  
imprint of cluster nodes, filaments, walls, and voids in persistence diagrams, along with that  
of the nested hierarchy of structures in multiscale point distributions. We conclude by outlining  
the potential of persistent topology for understanding the connectivity structure of the cosmic  
web, in large simulations of cosmic structure formation and in the challenging context of the  
observed galaxy distribution in large galaxy surveys.  
\end{abstract} 
 
 
 
 
\section{Introduction} 
 
\label{sec1} 
 
 
 
This study presents a substantial extension of the topological description of the galaxy and cosmic matter  
distribution. It involves a fundamental topological description of the cosmic mass distribution  
oriented towards quantifying the complex connectivity properties of the cosmic web \citep{BKP96,WeygaertLectureNotesI,CWJF14}.  
By means of Betti numbers, this study quantifies the various classes of topological features that result from  
the spatial organization of the various morphological components -   nodes, filaments, walls, and voids -  
in the cosmic web. The complex multiscale topology that is manifestation of the hierarchical  
buildup of cosmic structures is quantified by the powerful language of persistent topology  
\citep{EdHa10}. The present work follows up on earlier preliminary work \citep{eldering05,WPVE10,ISVD10}.  
The persistent analysis of the cosmic web is closely related to other studies applying aspects of  
Morse theory, in particular via the watershed transform, to describe the cosmic web \citep{colombi1,PWJ07,colombi2,colombi3,aragon2010,Sousbie1,Sousbie2}.  
 
\subsection{The Cosmic Web} 
The Megaparsec scale distribution of matter revealed by galaxy surveys features a complex network of interconnected filamentary galaxy  
associations. This network, which has become known as the {\it Cosmic Web} \citep{BKP96}, contains structures from a few megaparsecs up  
to tens and even hundreds of megaparsecs of size. Galaxies and mass exist in a wispy web-like spatial arrangement consisting of dense  
compact clusters, elongated filaments, and sheet-like walls, amidst large near-empty voids, with similar patterns existing at earlier epochs,  
albeit over smaller scales. The multiscale nature of this mass distribution, marked by substructure over a wide range of scales and  
densities, has been clearly demonstrated by the maps of the  
nearby cosmos produced by large galaxy redshift surveys such as the 2dFGRS, the SDSS, and the 2MASS redshift  
surveys \citep{Col03,TSB04,HMM12}, as well as by recently produced maps of the galaxy distribution at larger  
cosmic depths such as VIPERS \citep{vipers}. 
 
The Cosmic Web is one of the most striking examples of complex geometric patterns found in nature, and certainly the  
largest in terms of size. According to the {\it gravitational instability scenario} \citep{Pee80},  
cosmic structure grows from tiny primordial density and velocity perturbations. Once the gravitational clustering process  
has gone beyond the initial linear growth phase, we see the emergence of complex patterns and structures in the density field.  
 
Highly illustrative of the intricacies of the structure formation process are the results of the state-of-the-art N-body computer simulations of cosmic  
structure formation \citep[e.g.][]{springel2005,IRMP13,Illustris2014}. These simulations suggest that the observed cellular patterns are a prominent and natural aspect of  
cosmic structure formation. The simulations also reveal the distinct characteristics of the structure   
formation process: the anisotropic nature of the structures, as well as their hierarchical aggregation. 
 
The existence of the the Cosmic Web is the manifestation of the generic anisotropic nature of gravitational collapse, resulting from the  
intrinsic anisotropy of gravitational forces induced by the inhomogeneities in the cosmic mass distribution. For a full understanding of  
the intricacies of the cosmic web, the relationship between these gravitational tidal forces, and the resulting deformation of the matter  
distribution is of key importance \citep{BKP96,WeygaertLectureNotesI}.

Perhaps the most significant and characteristic property of the cosmic mass distribution is its hierarchical nature. As it develops out  
of a primordial density field of supposedly Gaussian fluctuations, structure builds up in a hierarchical fashion. The first objects to  
emerge are small. Their formation is followed by a gradual buildup of ever larger structures through the assembly of these smaller  
constituent features. In this way, the large massive galaxy or cluster halos have formed \citep[see e.g.][]{kauffmann1993,LC94}. 
The filaments  that dominate the observed cosmic web have been formed in a similar fashion, through the gradual merging of 
smaller tendrils. Even the population of the vast near-empty regions, the underndese voids that dominate and mark the topology of the universe  
on Megaparsec scales, have been recognized to follow the same hierarchical process \citep[][]{shethwey2004,aragon2013}. 

It culminates in a scenario in which voids grow, merge and shrink, much as bubbles do in soapsuds. The hierarchical buildup  
of the cosmic web thus produces a multiscale pattern of structures and objects, comprising a wide range of spatial and mass scales.  
 
It has remained a major challenge to characterize the structure, geometry and connectivity of the Cosmic Web. The complex spatial pattern -- marked  
by a rich geometry with multiple morphologies and shapes, an intricate connectivity, a lack of structural symmetries, an  
intrinsic multiscale nature and a wide range of densities -- eludes a sufficiently relevant and descriptive analysis by conventional  
instruments to quantify the arrangement of mass and galaxies. Many attempts to analyze the clustering of mass and galaxies  
at Megaparsec scales have been rather limited in their ability to describe and quantify, let alone identify, the features and components  
of the cosmic web. Measures like the two-point correlation function, which has been the mainstay of many cosmological studies over the past  
forty years \citep{Pee80}, are not sensitive to the spatial complexity of patterns in the mass and galaxy distribution.  
 
Only over the past few years have we seen the development and formulation of more sophisticated techniques that address the spatially complex Megaparsec 
scale patterns. Some of these involve the statistical evaluation of stochastic geometric concepts, such as the filament detection via  a generalization 
of the classical Candy model or Bisous model \citep{Stoica05,Stoica10,Tempel12}, others involve geometric inference formalisms \citep{Chazal09,GPVW10,Chazal14} 
while we also see the proliferation of tessellation-based algorithms \citep{Gonzalez09,weyschaap2009}. A large class of formalisms is based on local geometric properties, 
expressed via the signature of the Hessian of the density field, of the tidal field, or of the shear of the velocity field \citep[e.g.][]{colombi1,aragon07b,hahn07,colombi2,colombi3,bond10, 
forero09,libeskind12,Nexus}. While most of these existing methods have the downside of being defined on only one particular - and sometimes  
arbitrary - scale, the more elaborate MMF/Nexus framework explicitly takes into account the multiscale character of the cosmic mass  
distribution \citep{aragon07b,Nexus}. Most closely connected to the dynamics of the cosmic web formation process  
are several recently proposed formalisms that look at the phase-space structure of the evolving mass distribution  
\citep[][]{shandarin11,neyrinck12,abel11}. Noting that the emergence of nonlinear structures occurs at locations where different  
streams of the corresponding flow field cross each other, the phase-space sheet methods provide a dynamically based  
identification of their morphological nature. For example, walls correspond to three-stream regions while most filament regions involve  
five stream regions. A few other formalisms use the topological structure of the cosmic density field. The first examples  
are the Watershed Void Finder \citep{PWJ07} and ZOBOV \citep{Neyrinck2008}. They use the watershed transform to delineate the underdense void  
basins in the large scale universe \citep[also see][]{SLWW14}. \cite{aragon2010} expanded this to Spineweb, an elaborate framework for  
identifying all different morphological entities in the cosmic web. Spineweb shares its topological foundation with the Disperse formalism  
\citep{Sousbie1,Sousbie2}, which has proven to be particularly succesful in outlining the filamentary spine of the cosmic web  
\citep[for a further development also see][]{Shivashankar2015}.   
 
\subsection{Topology: Connectivity of the Cosmic Web} 
In this study we specifically address a central aspect of the Cosmic Web, the connectivity of its various structural  
components. The way in which matter has distributed itself over the various structural components - such as walls, filaments, 
cluster nodes and voids - and the manner in which they  connect up in the complex network of the cosmic web is a key aspect 
of the spatial structure of the cosmic mass distribution.  
 
The branch of mathematics that addresses issues of shape and connectivity is topology. The cosmic mass distribution  
emerging in different cosmological scenarios will entail different spatial patterns, and we should expect to find  
its expression in subtle yet highly significant differences in topological characteristics. Existing topological  
descriptions have not yet addressed these in any substantial detail.  
 
The first cosmological studies that focused on topological aspects of the cosmic mass distribution evaluated and analyzed the  
genus and Euler characteristic of the corresponding iso-density surfaces. Gott and collaborators \citep{GDM86,HGW86} studied the  
genus as a function of density threshold. Later, more discriminative topological information became available with the introduction  
of Minkowski functionals \citep{Mecke94,schmalzing1997}. However, nearly without exception these studies had a largely global character,  
often focussing on issues such as the statistical nature of the cosmic mass distribution. Following up on our earlier preliminary  
work \citep{eldering05,WPVE10,ISVD10}, the present study represents  
a substantial extension of the topological arsenal used for description of the galaxy and cosmic matter distribution. Most significantly,  
it takes into account the intricate hierarchical and multiscale weblike spatial patterns into which mass has organized itself  
on Megaparsec scales.    
 
Of particular interest and relevance for the present study is the way in which the different morphological features are spatially connected in the global  
weblike network. A few characteristic examples illustrate this. A configuration of interconnected walls that enclose low-density  
void cavities represents an entirely different topological pattern than a percolating network of mutually connected elongated filaments.  The  
latter would facilitate the connection of all underdense regions into a percolating valley with a sponge-like topology. The former  
is more reminiscent of a cheese-like configuration of cavities enclosed by high-density filaments and walls.  
 
For a more detailed assessment, we would therefore want to understand the role of individual walls, filaments and other mass concentrations in  
outlining the topological structure. A key aspect of this quest is the topological imprint of the multiscale nature of the weblike mass distribution.  
It concerns the way in which the smaller scale features of the structural hierarchy are embedded in or emanate from the prominent large  
scale features of the cosmic web, and in particular how this is reflected in its topological character. It involves questions such as how  
topology may help us to probe the nature and scale of the dominant filamentary network that defines the spine of the cosmic web and to  
quantify the extent to which it branches off in a multiscale tapestry of ever smaller tendrils \citep[see e.g.][]{aragon07a,CWJF14}.  
Equally interesting is the prospect of having a profound and well-defined quantitative characterization of the multiscale void population,  
the product of the hierarchically evolving soapsuds of voids outlining the segmentation of the Megaparsec scale Universe.  
 
\subsection{Homology} 
As indicated above, there is ample motivation to extend topological  
analysis beyond global characterizations such as genus, and to orient the description towards the identification of the underlying  
connections and details of the topological structure. Following this motivation and rationale, the prime purpose of our study is the introduction of a  
fundamental topological formalism that addresses the issues outlined above. These well-known mathematical concepts will equip  
cosmologists with new and potent methods for a more profound analysis of spatial patterns encountered in the Megaparsec scale universe.  
 
The formalism that we introduce here finds its roots in algebraic topology and in Morse theory \citep{Mil63,EdHa10}. Algebraic topology is the branch  
of mathematics that uses tools from abstract algebra to study topological spaces. It accomplishes this by establishing the correspondence between  
topological spaces and objects on the one hand and algebraic \emph{groups} on the other hand. This  
allows one to formulate statements about topological spaces into the language of group theory, offering substantial flexibility and  
a deeper understanding of spatial structure and connectivities. It provides us with a global characterization of structural  
topology in terms of \emph{Betti numbers} \citep[e.g.][]{Bet71,EdHa10}. It also forms the foundation for the subsequent investigation  
of the hierarchical aspects of the topological structure of the cosmic mass distribution. This leads us to the introduction of the formalism  
of \emph{persistent homology} \citep{ELZ02,Zom05,Carl05,CarlZom09,Carl09,EdHa10}.  
 
The specific formalism from algebraic topology that we use to describe the topological structure of the space defined by the cosmic  
density distribution is known as \emph{homology}. This is the mathematical formalism for the quantitative characterization of the connectivity of space  
by assessing the presence and identity of the holes, usually via the description of the boundaries of these holes \citep{munkres1984elements}. The original  
motivation for homology was the observation that two topological spaces may be distinguished by examining their holes. In homology, \emph{holes}  
are a key concept. In general,  
for a manifold or a more general topological space 
embedded in $d$-dimensional Euclidean space, there are 
$d$ different types of holes, of dimension $0$ to $d-1$. 
A three-dimensional topological space may contain  
3 different species of holes. Restricting to 3D space, these holes have an intuitive interpretation. A 0-dimensional hole is the  
\emph{gap} between two separate objects or \emph{components}. A 1-dimensional hole is a  
\emph{tunnel} through which one may pass in either direction without encountering a boundary.  
A \emph{cavity} or \emph{void} is a 2-dimensional hole, fully enclosed within a 2-dimensional surface or \emph{shell}.  
 
A central consideration of homology is that the identification of holes may be conveniently and unequivocally achieved on the basis of the  
boundary that surrounds them. For instance, while a disk is a 2-dimensional surface, a circle is only the 1-dimensional boundary of a disk.  The circle 
has a 1-dimensional hole formed by puncturing the disk; the disk has no such hole. Along the same vein, a sphere is not a circle because it encloses a two-dimensional hole while the circle encloses a one-dimensional hole. These considerations  
lead homology to describe and classify topological spaces according to their boundary. Homology characterizes the boundaries in terms of \emph{cycles}.   
Loosely speaking, cycles are closed loops, or submanifolds, that can be drawn on a given topological space. They are classified by dimension: a 0-cyle is a  
connected object or point, a 1-cycle is a closed \emph{loop}, and a 2-cycle is a \emph{shell}. Cutting  
along a 0-cycle corresponds to puncturing the topological space, while cutting along a 1-cycle yields either a disconnected piece or a simpler shape.  
 
The  concept of cycles can be translated into the language of group theory. Two $p$-cycles are called \emph{homologous} when together they bound 
a $(p+1)$-dimensional part of the space.  This is the technical sense in which the two cycles are considered to be the same. Extrapolating these observations, 
we find that cycles can be  arranged into homology groups. The collection of all $p$-dimensional cycles in the topological space forms the $p$-th \emph{homology group} $H_p$. 
In this paper, all homology groups will be vector spaces, and in this case the \emph{rank} of $H_p$ is its dimension, 
namely the number of \emph{independent} $p$-dimensional cycles in a topological space. 
This is the formal definition for the \emph{Betti numbers} $\beta_p$ \citep{Bet71,EdHa10}, where $p=0,1,\ldots,d$. Like the Euler characteristic,  
the Betti numbers are topological invariants of a space, meaning that they do not change under systematic transformations like  
rotation, translation and deformation. The first three Betti numbers have intuitive meanings: $\beta_0$ counts the number of  
isolated components, $\beta_1$ counts the numbers of loops enclosing independent tunnels and $\beta_2$ counts the number of shells  
enclosing separate voids. Betti numbers contain more topological information than the Euler characteristic $\chi$,   
as may be directly appreciated and inferred from the fundamental \emph{Euler-Poincar\'e Formula} \citep{EdHa10,Adl10}. This states that  
$\chi$ is the alternating sum of all $d$ Betti numbers. In other words, any one given value of the Euler characteristic lies on a $(d-1)$-dimensional  
hyperplane of corresponding possible combinations of Betti numbers 
$(\beta_0,\beta_1,\ldots,\beta_{d-1})$. 
This has important repercussions for the  
topological description of the cosmic mass distribution: even when having the same Euler characteristic or genus, a space - such as defined by 
the level set of a density field - may differ topologically in terms of their Betti numbers.  
 
\subsection{Persistence} 
The details of the spatial connections between the various topological spaces, holes or boundaries underlying the global homology properties,   
leads to the concept of \emph{persistence} \citep{ELZ02,EdHa10}. Persistence formalizes topology as a hierarchical concept,  
and represents a substantially richer characterization of the topological structure of the cosmic mass distribution than that specified by  
conventional descriptions in terms of genus and even Betti numbers. It is based on the realization that there is a wealth of topological information to be gained from  
a systematic analysis of the singularity structure of a field.  
 
A central role is played by Morse theory, the branch of mathematics that studies the singularity structure of a field, ie. the position of minima, maxima and saddle  
points and their mutual connections. Of fundamental importance in this is the mathematical tenet that there is a close relationship between 
the topology of the space \footnote{The \emph{space} here refers to a \emph{topological space}, and not the space in the sense of \emph{space-time}, that cosmologists are more familiar with.} and the critical points of any 
smooth function on the topological space \citep{Mil63,EdHa10}. Following this observation, Morse theory describes the topology of the  
space by studying the critical points of a corresponding \emph{Morse function}, i.e a smooth scalar function defined on the topological space. Submanifolds defined  
as the regions where the Morse function is in excess of a particular functional threshold value (superlevel sets) are topologically equivalent or, 
more precisely, \emph{diffeomorphic} when the interval between the two defining threshold values does not contain any critical point. The important implication of this  
is that all changes in topology of a space occur only at critical points.  
 
Armed with this knowledge, one may identify the connection of individual topological features to the overall cosmic mass distribution. To this end, we use  
the fact that the critical points of the density field, or other fields related to the mass distribution, are not only responsible for the \emph{formation} of  
a feature, but also for their \emph{destruction}. By varying the density threshold, a topological feature --- e.g. a component, tunnel, or a cavity --- may emerge,  
disappear, or connect up with other features, as the topology of the space changes while passing through a critical value \footnote{a \emph{critical value} is  
the value of a function at the \emph{critical point}}. In the language of persistence,  this marks the \emph{birth} or \emph{death} of a feature. In the case of 
a merger of features, the \emph{elder rule} specifies that the elder feature survives. It is the nature of the critical point, i.e. its \emph{index}, that 
decides what kind of feature is formed or destroyed. 
 
Generically, the adddition of an index-$p$ critical point may result in either the birth of a $p$-dimensional hole or the death of a $(p-1)$-dimensional one  
\citep{ELZ02,ZomCarl05,EdHa10}. In the situation in which the submanifolds are identified with the superlevel sets of the density field, i.e., the regions where the density is  
higher than a particular density threshold, a saddle point may merge two distinct islands in the density field. Alternatively, it may connect different ends of  
the boundary of a singular connected object. While the first will lead to the loss of one island,  
the latter will lead to the birth of a new loop. Another example is that of a cavity that gets filled up entirely and disappear as we pass through a (local) minimum.  
By establishing how the different features merge and form ever larger structural complexes as the density threshold is  
decreased, we establish a tree of hierarchically nested topological features. In a sense, this is not unlike the cosmologically more familiar merging trees  
that are defined by the dynamical evolution of dark matter halos or voids \citep{PCH2008,BWWBKP2013}.  
 
The full hierarchical embedding of topological features may subsequently be recorded and summarized in a \emph{persistence diagram} \citep{ELZ02,EdHa10} or  
\emph{persistence barcode} \citep{ZomCarl05,Carl05,Carl09}. For each ambient dimension $p=0,1,\ldots,d-1$ of a topological space, a persistence diagram records  
the \emph{birth} and \emph{death} of each topological feature or $p$-dimensional hole. For each hole $i$ it plots the function value $b_i$ at which the feature  
is created and the value $d_i$ at which it disappears. 0-dimensional diagrams record the merger events of two separate islands, 1-dimensional ones the formation  
and destruction of loops, while 2-dimensional diagrams record the birth and death of cavities or voids. The resulting persistence diagrams consists of the  
collection of points $(b_i,d_i)$, each point associated with a unique topological change in the space.  
The life-span of a topological feature, i.e. the absolute difference betweeen its death and birth values, is the \emph{persistence} value $\pi$, of the feature.  
 
Persistence diagrams contain strictly more information than the Betti numbers: the $p$-th Betti number of the superlevel set  
for threshold value $\nu$ is the number of points in the region of the  
persistence diagram delineating features that are 
created at higher function values and destroyed at lower  
function values. The important implication of this is that persistence, Betti number and Euler characteristic contain strictly  
decreasing amount of topological information about a space. Based on his observation, and taking into account that persistent  
homology is hierarchical in nature, it is evident that persistent homology entails a considerably more complete characterization of the geometry  
and topology of the cosmic mass distribution.  
 
Besides yielding a powerful statistical characterization of the topological structure of a space, the potential applications of persistence  
are numerous. One particularly interesting example in practical astronomical circumstances is that of filtering out insignificant noise features.  
In general, low-persistence features are more likely to be topological noise, while those with a high persistence values would correspond to real  
signals. In fact, persistence based filtering has the potential of substantially more profound applications. Indeed, in the context of   
complex spatial structures, such as the cosmic web, it has proven that it enables a better defined identification of individual  
features than conventional kernel filtering \citep{Gyulassy12,Sousbie1,Sousbie2,Chazal14,Shivashankar2015}. In particular noteworthy is the 
Disperse algorithm developed by Sousbie and collaborators for the identification of filaments and other structures in the large scale Universe \citep{Sousbie1,Sousbie2}. The  
concept of persistence based filtering has a rich potential for tuning it to specific problems and circumstances, as was demonstrated in the  
recent Felix algorithm for filament detection in different weblike environments, such as voids or around rich cluster nodes  
\citep{Shivashankar2015}.  
 
\begin{figure*} 
 \begin{center} 
  \includegraphics[width=\textwidth]{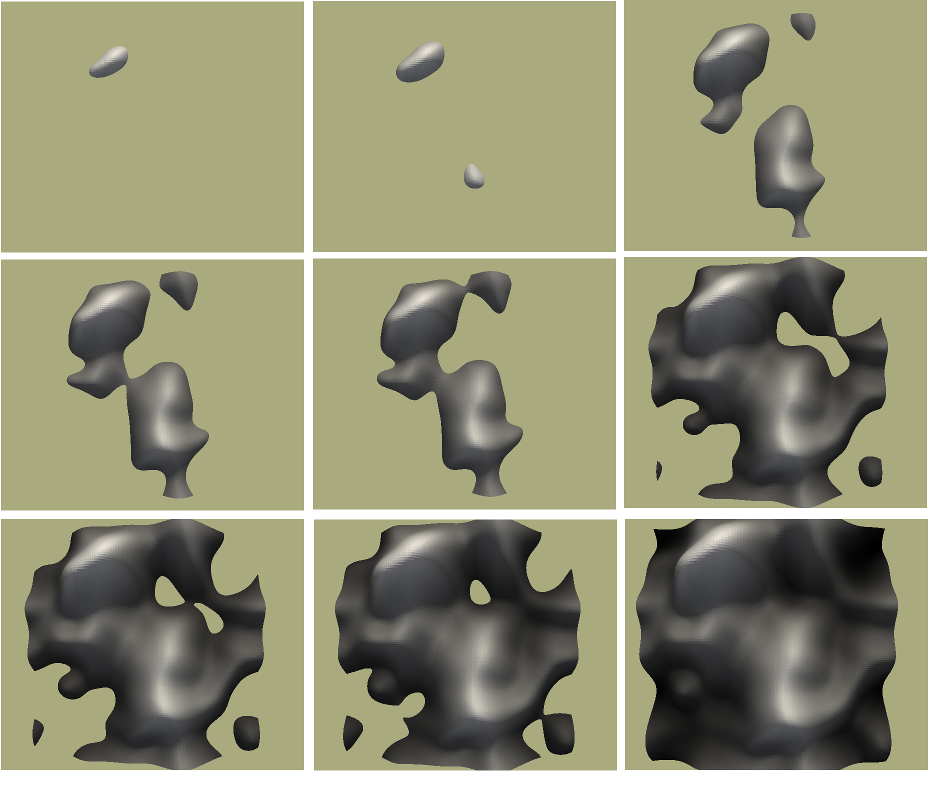}\\ 
 \end{center} 
 \caption{Topology and field singularity structure. The figure illustrates - from topleft to  
bottom right panel - the changing topology of the superlevel sets of a 2D random field as we lower  
the corresponding density threshold. The figure shows the regions of the topological space that are included  
in the superlevel set. Panel (a) starts with a single island. Panels (b) and (c) witness the birth  
of two more islands. In panel (d) two of the islands merge, so that two islands remain. In panel (e), there  
is another merger of two isolated islands, followed by the emergence of the first 1-dimensional hole  
or a loop in (f). It has the appearance of a lake surrounded by land. In panel (g), the loop splits into two,  
after which one of the loops get entirely filled up in panel (h) and disappears. In panel (i), all holes are  
filled up, with the superlevel set consisting of the entire topological space.} 
 \label{fig:superlevelmorse} 
\end{figure*} 
 
\subsection{Persistent Topology of the Cosmic Web} 
 
The obvious aim of our work is the application of homology and persistence measures for analyzing the observed spatial  
distribution of galaxies and matter on Megaparsec scales. The ultimate purpose is to develop and further our understanding  
and appreciation for the spatial connectivity aspects of the cosmic web. The expectation is that it will help us to uncover aspects  
of spatial clustering that have hitherto remained unexplored in cosmological research. To be able to interpret the quantitative results  
obtained by such an analysis, it is necessary to have a guidance for the significance of the obtained measurements.  
 
The complex reality of the observed galaxy distribution or that of a full-fledged computer simulation is the result of the  
intricate interplay between a range of physical processes. It manifests itself in a complex superposition, over a wide range of  
scales, of a rich variety of morphological features. In this respect, we encounter the complication that as yet there is no real  
insight or understanding for the expected behaviour of homology and persistence in complex spatial patterns such as the cosmic web.  
Almost without any exception there are no realistic physical situations and configurations for which exact analytical results for the  
corresponding measures are available. Even for the cosmologically canonical reference configuration of Gaussian random fields, there are  
no exact results for Betti numbers and persistence diagrams \citep[but see][]{feldbruggeBachelor}.  
 
Instead of directly analyzing full-fledged realistic cosmological situations, such as the outcome of N-body computer simulations of  
structure formation in the concordance $\Lambda$CDM cosmology \citep[see e.g.][]{springel2005,IRMP13,Illustris2014,schaye2015}, we therefore first  
need to design a baseline reference. For the understanding and interpretation of the obtained homology and persistence measures, and  
to have the ability to obtain insight into their significance, we will need to  equip ourselves with reference templates of these  
measures. The principal aim of this paper is exactly this. The reference templates will be the outcome of the topological analysis  
for a well-defined set of heuristic spatial models that each single out one particular characteristic aspect of the Cosmic Web.  
Each of the templates should provide insight and information on the impact of specific and well-defined spatial configurations on the values  
of Betti numbers and behavior of persistence diagrams. Armed with these templates, we will have the ability to interpret  
and understand the topological measures obtained for the considerably more complex reality of the real, or simulated,  
universe. The full homology and persistence analysis of the mass, halo, and galaxy distribution in cosmological  
simulations will be the subject of a series of upcoming works (for the first
results, see e.g. \cite{nevenzeel2013}). An additional  
study would involve the analysis of a set of mock galaxy catalogues that incorporate galaxy biasing effects as well as survey  
selection effects of known galaxy redshift surveys. 

We use Voronoi clustering  
models \citep{weyicke1989,weygaert94,weygaert2002,aragon2010} for investigating the manifestation of weblike and/or void-dominated  
configurations in topological measures. For the impact of the multiscale aspects of the clustering of galaxies, we  
use the fractal-like point distributions of the Soneira-Peebles model \citep{SP78}.  
 
The Voronoi clustering models are a versatile and useful class of models for the anisotropic and void-dominated nature of  
the Megaparsec mass distribution \citep{weygaert94,weygaert2002}. They use Voronoi tessellations as a spatial template for  
the weblike distribution of mass and galaxies, by a stochastic process of distributing particles in the  
various elements of the tessellations, ie. in the nodes, edges, planar faces and cell interiors of the tessellations   
\citep{weyicke1989,weygaert1991,weygaert2002,aragon2010}. The Voronoi clustering models are flexible and can be tuned to  
represent a network of interconnected filaments, or a cellular distribution dominated by walls, or a pattern of  
massive compact cluster nodes or any combination of these. In turn, this enables us to calibrate and assess quantitatively   
the way in which such configurations manifest themselves in the topological measures obtained \citep[see e.g.][]{Shivashankar2015}.  
 
The Soneira-Peebles model \citep{SP78} produces fractal-like point distributions that allow a systematic exploration  
of the influence of the multiscale clustering of galaxies and mass. It involves the nested embedding of a sequence of nodes in a  
hierarchical treelike structure. The spatial clustering of the resulting fractal point distribution can be tuned quantitatively  
by means of a few defining parameters \citep[also see][]{Sch07,weyschaap2009}. We should note that while the Soneira-Peebles model  
represents a versatile and useful heuristic model for exploring the effects of the multiscale spatial clustering, the observed  
galaxy distribution is certainly not fully fractal \citep[see e.g.][]{MJ1990}. 
 
\subsection{The Computational Formalism} 
The second major aim of this paper concerns the presentation of the computational formalism for calculating homology measures and  
persistence diagrams.  
The mathematical primer on topology in Section~\ref{sec:computation} therefore also includes extensive discussion of the computational  
machinery that we use to compute persistence diagrams and Betti numbers. The cosmological context defines a range of practical issues.  
 
The principal issue is the fact that the density field is sampled by a discrete set of points, either particles in a computer  
simulation or galaxies in observational circumstances. Most of the topological studies in cosmology depend on some sort of user-specific  
smoothing and related threshold to specify surfaces of which the topology may be determined. In cosmological studies, this usually concerns  
isodensity surfaces and/or density superlevel and sublevel sets defined on a Gaussian filter scale. Given that we do not have available the  
fully continuous density field on the topological space, we need to define a strategy to infer the topological measures from the discrete  
point set.  
 
Assuming the point sample is a representative and unbiased sample of the underlying continuous field, we may follow different  
strategies. Instrumental in this is the attempt to retain the optimal signal probing the underlying multiscale topology. The immediate  
implication for this is that we should refrain from the use of artificial filtering scales that beset so many conventional cosmological  
studies. Instead, we apply more natural filters that exploit fundamental concepts from computational geometry and computational  
topology \citep{Okabe2000,EdHa10}. These are based on the use of simplicial complexes - eg. the Delaunay tessellations that have been used 
in astronomical applications - that form the natural format for the translation of a discrete point distribution into a continuous volume-filling field that  
retains all aspects of shape, morphology over the entire spectrum of scales.  
 
A well-known strategy is the evaluation of the topological characteristics directly from the point sample distribution, on the basis of  
the distances between the sample points. A direct means of obtaining this information is via the construction  
of a simplicial complex. This is a geometric assembly of faces, edges, nodes and cells marking a discrete spatial map of the volume  
containing the point set. The edge lengths of such a complex would represent a selective sampling of the corresponding distance field. A well-known  
and topologically highly informative complex is that of \emph {Alpha Shapes}. They are subsets of a Delaunay triangulation that describe the intuitive  
notion of the shape of a discrete point set. They are one of the principal concepts from the field of Computational Topology \citep{dey1999,vegter2004, 
ZomCarl05}. Introduced by Edelsbrunner and collaborators \citep{edelsbrunner1983,EM1994}, these simplicial complexes constitute an  
ordered sequence of nested subsets of the Delaunay tessellation \citep{Okabe2000,weyicke1989,weygaert1991,EdHa10}. As they are homotopy equivalent to the distance field, they are an excellent tool for assessing the topological structure of a discrete point distribution. Instead of the  
cosmologically familiar filtration in terms of sublevel or superlevel sets defined by a density threshold, alpha shape topology is based  
on a distance filtration defined by the ``scale''  factor $\alpha$. Our earlier preliminary studies of Betti number properties in a range of  
cosmological configurations, reported in \citep{eldering05,WPVE10,ISVD10}, were based on the use of alpha shapes.  
 
In this study we follow a different strategy, and evaluate the topological measures via a density value filtration of  
a reconstruction of the density field. To this end, we translate the discrete point distribution into a volume-filling  
density field reconstruction, using the Delaunay Tessellation Field Estimator or DTFE \citep{Sch00,weyschaap2009,cautun2011}. It produces a piecewise linear continuous field of density values defined on the Delaunay triangulation  
generated by the distribution of sample points. The latter function as the vertices of the tessellation. 
 
The core of our computational formalism is that of the subsequent homology calculation. We follow a technique  
that computes the homology measures directly from the continuous DTFE density field representation on the simplicial elements of  
the Delaunay tessellation $K$, ie. on the vertices, edges, triangular faces, and the tetrahedral cells. Instrumental in the algorithm  
are the density values at the vertices of the tessellation, and the increase or decrease in density towards the vertices to which  
they are connected in the tessellation. For a given density filtration, the calculation involves the determination of the  
boundary matrix (see Section~\ref{sec:computation}, which identifies for each simplex in the superlevel filtration the simplices in its boundary. The reduction of the boundary matrix directly yields the birth-death pairs of the different $p$-dimensional persistence  
diagrams \citep[see e.g.][]{EdHa10,BEK10,BRK13}.  
 
A third computational aspects is the introduction of \emph{persistence intensity maps}. These are designed for  
the practical purpose of evaluating and analyzing the intricate topological aspects of cosmological mass distributions. The  
intensity maps are continuous maps that represent an empirical probabilistic description of persistence diagrams. They are obtained via  
the averaging of persistence diagrams for a set of realizations of the  
same stochastic process, and are supposed to converge asymptotically to a stable average. Besides forming a continuous representation  
of persistence diagrams, they form a practical condensation of the topological character of a (density) field. They facilitate  
the comparison between different spatial distributions, outline and summarize their global topological properties while simultaneously  
allowing the detection of unique topological details that otherwise would have remained hidden. The latter would surface as the grid-wise difference between 
intensity map of a specific spatial mass distribution with respect to that for a set of reference  
morphologies.  
 
\subsection{This Study} 
The first two sections of this paper introduce the necessary mathematical concepts and background. Following a short discussion and  
definition of scalar fields and of Morse theory in Section~\ref{sec:fields}, in the subsequent Section~\ref{sec:topology} we follow  
with a reasonably detailed introduction to the principal aspects of algebraic topology. This mathematical primer also includes  
an extensive and detailed presentation, in Section~\ref{sec:computation} of the computational machinery to compute persistence diagrams and Betti numbers.  
Before proceeding towards the topological analysis of clustered point distributions, Section~\ref{sec:random} establishes the base  
reference. The section presents the results obtained in terms of Betti numbers and persistence diagrams for the random, featureless point  
distributions generated by a Poisson point process. Subsequently, Section~\ref{sec:single_scale} presents the results of the topological analysis  
of pure Voronoi element models, while Section~\ref{sec:multi_scale} analyzes the topology of the multi-scale fractal Soneira-Peebles model  
\citep{SP78}. Finally, an impression of the possible time evolution of the topology of the weblike cosmic mass distribution is obtained  
in Section~\ref{sec:dynamic}, where we analyze the homology and persistence diagrams of elaborate and complex Voronoi evolution models. These are  
Voronoi clustering models that seek to emulate the morphological evolution of the cosmic web \citep{weygaert2002}. The concluding  
Section~\ref{sec:summary} presents a summary and discussion of our results and on the prospects for the application of homology and persistence  
measures for a quantitative characterization of the connectivity and morphological properties of the cosmic web.

 
 
\section{Scalar Fields and Morse theory} 
 
\label{sec:fields} 
In this study we seek to analyze the homology of cosmological density fields.  
The mass distribution in the Universe is described by the density perturbation field, 
\begin{equation} 
f(x,t)\,=\,\frac{\rho(x,t)-\rho_u(t)}{\rho_u(t)}\,, 
\end{equation} 
which describes the fractional over- or under-density at position $x$ with  
respect to the universal mean cosmological density $\rho_u(t)$.

\subsection{Stochastic Random Fields} 
 
We start with the assumption that the cosmic density perturbation field is a realization of a stochastic random field.  
A {\it random field}, $f$, on a spatial volume assigns a value, $f(x)$, to each  
location, $x$, of that volume. The fields of interest are smooth and continuous \footnote{In this  
section, the fields $f(x)$ may either be the raw unfiltered field or, without loss  
of generality, a filtered field $f_s(x)$. A filtered field is a convolution with  
a filter kernel $W(x,y)$, $f_s(x)=\int \diff y f(y) W(x,y)$.}. The stochastic  
properties of a random field are defined by its \emph{$N$-point joint probabilities},  
where $N$ can be any arbitrary positive integer. To denote them, we write  
$\xx=(x_1,x_2,\cdots,x_N)$ for a vector of $N$ points and $\ff = (f_1, f_2, \ldots, f_N)$ for a vector  
of $N$ field values. The joint probability is 
\begin{eqnarray} 
  \Probability [f(x_1)=f_1,\ldots,f(x_N)=f_N] &=& \ProbDensity_{\XX}( \ff) \diff \ff\,, 
  \label{eqn:probability} 
\end{eqnarray} 
which is the probability that the field $f$ at the locations $x_i$ has values in the range $f_i$ to $f_i+\diff f_i$,  
for each $1 \leq i \leq N$.  
 
In cosmological circumstances, we use the \emph{statistical cosmological principle}, which states that  
statistical properties of e.g. the cosmic density distribution in the Universe are uniform throughout the Universe. It  
means that the distribution functions and moments of fields are the same in each direction and at each location. The  
latter implies that ensemble averages depend only on one parameter, namely the distance between the points.  
 
Important for the cosmological reality is the validity of the {\it ergodic principle}. The Universe is unique, and its  
density distribution is the only realization we have of the underlying probability distribution. The ergodic principle  
allows us to measure the value of ensemble averages on the basis of spatial averages. These will be equal to  
the expectations over an ensemble of Universes, something which is of key significance for the ability to  
test theoretical predictions for stochastic processes like the cosmic mass distribution with observational  
reality.

\begin{figure} 
 \begin{center} 
  \includegraphics[width=7cm]{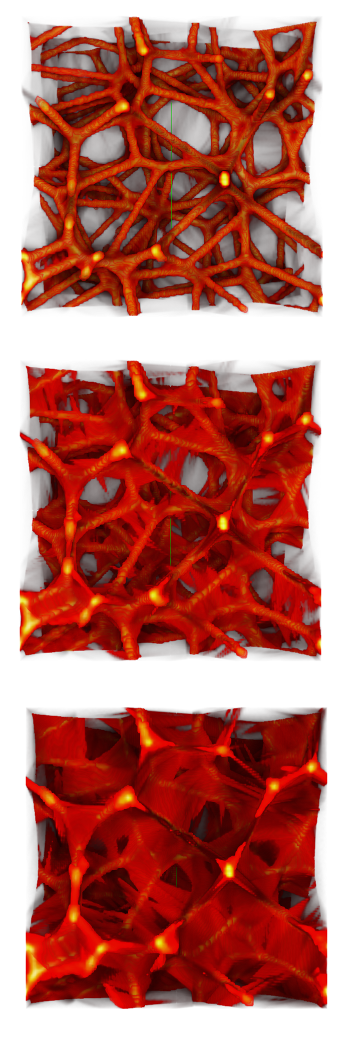}\\ 
 \end{center} 
 \caption{Density rendering of the superlevel set of the pure filamentary models. From top to bottom: three snapshots for growing superlevel sets.} 
 \label{fig:superlevelvisuals} 
\end{figure} 
 
\subsection{Superlevel Sets and Sublevel Sets} 
 
When assessing the mass distribution by a continuous density field, $f({x})$, a common practice is to  
study the sublevel or superlevel sets of the field smoothed on a scale $R_s$: 
\begin{eqnarray} 
  f_s (x)  &=&  \int f(y) W_s(y-x) \diff y , 
\end{eqnarray} 
where $W_s({x}-{y})$ is the smoothing kernel. Writing ${\mathbb M}$ for the entire space, we define the \emph{superlevel sets} of this field  
as the regions 
\begin{align} 
      {\mathbb M}_\nu  &=  \{ x \in {\mathbb M}  \mid  f_s(x) \geq \nu \} \\ 
                       &=  f_s^{-1} [\nu, \infty) . 
\end{align} 
In other words, they are the regions where the smoothed density is greater than or equal to the threshold value. 
 
The \emph{sublevel set} is the complimentary topological space of the superlevel set. the \emph{sublevel set} ${\mathbb M}^\nu$ is defined as 
\begin{align} 
 {\mathbb M}^\nu  &=  \{ x \in {\mathbb M}  \mid  f_s(x) \leq \nu \} \\ 
                  &=  f_s^{-1} (-\infty, \nu] . 
\end{align} 
     
Since both superlevel set and sublevel set are closed, they intersect 
in the \emph{level set} 
\begin{equation} 
    f^{-1} (\nu) = {\mathbb M}_\nu \cap {\mathbb M}^\nu. 
\end{equation}

\subsection{Filtrations} 
 
When addressing the topology of a mass or point distribution, a rich source of information is the topological  
structure of a filtration. Given a space $\Manifold$, a \emph{filtration} is a nested sequence  
of subspaces: 
\begin{equation} 
  \emptyset  = \Manifold_0 \subseteq \Manifold_1 \subseteq \ldots \subseteq \Manifold_m = \Manifold . 
\end{equation} 
The nature of the filtrations depends, amongst others, on the representation of the mass distribution.  
When assessing the topology of a scalar field, the filtration usually consists of the nested sequence of  
sublevel or superlevel sets. It is the evolving topology as we pass through the filtration seqencce  
which represents a rich source of information on the topological complexity of the field.  
 
A typical example of superlevel  
sets of a density field is that shown in figure~\ref{fig:superlevelvisuals}. It provides a telling illustration of a density-defined filtration of  
a weblike spatial pattern. It concerns a model of the cosmic web consisting exclusively of filaments. It shows a sequence of three growing superlevel  
sets of the weblike density field, along a sequence of decreasing density thresholds. The top panel  
corresponds to the highest density threshold. It reveals the high density regions that outline the  
underlying skeleton. The additional panels reveal complementary information on the mannner in which  
matter has distributed itself over the various structural components, revealing how the lower density  
mass elements connect up and fill in the intersticial regions of the network.  
 
The illustration shows how the sequence of filtration steps establish the connectivity of  
the cosmic mass distribution, and entails its topological structure.

\subsection{Piecewise Linear Scalar Fields}  
 
In many practical circumstances, whether it concerns the spatial distribution of  
galaxies in redshift surveys or particles in cosmological N-body simulations, we are  
dealing with datasets consisting of discrete particle positions. 
 
There are various ways in which the topology of such a discrete particle  
dataset can be analyzed. One option is to define a filtration on  
the point distribution itself. The most direct way to achieve this is that via  
a simplicial complex generated by the point distribution. Well-known examples are  
that of the alpha-complex and the  
Cech complex \citep[see][]{EdHa10}, invoking the distance function and  
a corresponding distance parameter to define the filtration.

In our study we follow a different approach. The topological analysis in our  
study is based on a density value based filtration of a piecewise linear  
density field. The latter is computed from the discrete particle  
distribution itself. The usual strategy for this is to compute a triangulation on the given discrete particle set. The density function is first calculated on the vertices of this triangulation, and subsequently extrapolated to the higher dimensional simplices, yielding a piece-wise linear function. More details on this can be found in the subsection on piece-wise linear functions, as well as Section~\ref{sec:computation}. The filtration consists of density superlevel sets.  
 
The determination of a piecewise linear density field from a discrete particle  
distribution involves a few key steps. The first step involves an estimate of  
the density at each of the sample points. Usually, the particles define the  
point sample, but in principle one may define alternatives. The second step involves  
the determination of a tessellation on the basis of the point sample. In each tetrahedron of the tessellation, the gradient can be uniquely determined from its four vertices. 
 
For a sample of $N$ points, with density value estimates $f({x}_{j})$ ($j=1,2,\ldots,N$),  
the density value $f({x})$ at a location $x$ is uniquely determined  
from the density gradient of the tetrahedron 
in which it is located, and the density value at one of its vertices, $x_i$,  
\begin{equation} 
f(x)\,=\,f(x_i)\,+\,{\nabla f}\,\cdot\,(x-x_i)\,. 
\label{eqn:linintp} 
\end{equation} 
  
One key element of a procedure to construct a linear piecewise density field is  the nature of the estimate of the density at each sample point. A second key element  
is the nature of the triangulation. For our results, In most of our results, we use the  
Delaunay Tessellation Field Estimator, DTFE \citep{Sch00,weyschaap2009,cautun2011}.  
It is based on local density estimates. The density at a particular vertex is the inverse of the volume of the delaunay star associated with it. The density is then interpolated to higher dimensional simplices, to yield a piece-wise linear field.

 
\subsection{Morse Theory}    
 
In Morse theory, we consider a compact topological space $\Mspace$, and a  
generic smooth function on this topological space. In the context of this paper,  
the topological space is the $3$-torus \footnote{In the cosmological context, the data is usually specified in a cubic box. Gluing opposite ends of the cube converts it into a 3-torus. This has the advantage of converting the data into a periodic form. This is reasonable, also from the assumptions of the cosmological principle, stating that there are no preferred locations in the Universe. Converting the data into a periodic form mimics this principle.} and the function is a density  
distribution, $f: \Mspace \to \Rspace$. Assuming $f$ is  
smooth, we can take derivatives, and we call a point $x \in \Mspace$  
\emph{critical} if all partial derivatives vanish, i.e., 
\begin{equation} 
\nabla f|_x = 0. 
\end{equation} 
 
 Correspondingly, $f(x)$ is a \emph{critical value} of the  
 function. All points of $\Mspace$ that are not critical are  
 \emph{regular points}, and all values in $\Rspace$ that are not the  
 function value of critical points are \emph{regular values}. Finally,  
 we call $f$ \emph{generic} if all critical points are  
 non-degenerate in the sense that they have invertible Hessians, which  
 is defined as the matrix of the partial double derivatives 
 \begin{equation} 
 H_{ij} = \left(\frac{\partial f}{\partial x_i \partial x_j}\right)_{i = 1,...,3; j = 1,...,3}, 
 \end{equation} 
 restricting to a 3-dimensional space. In this case, critical points are isolated from each other, and since $\Mspace$ is compact, we have only finitely many critical points and therefore only finitely many critical values. The \emph{index} of a non-degenerate critical point is the number of  
negative eigenvalues of the Hessian. Since $\Mspace$ is $3$-dimensional, we have $3$-by-$3$ Hessians and therefore only four possibilities for the index. A \emph{minimum} of $f$ has index $0$, a \emph{maximum} has index $3$, and there are two types of \emph{saddles}, with index $1$  
and $2$.

A major result of Morse theory states that the topology of a space changes only when the level set passes a critical point of the function. The change in topology is dictated by the index of the critical point. The significance of the critical points and their indices becomes apparent when we look at the sequence of growing superlevel sets: $\Mspace_\nu = f^{-1} [\nu, \infty)$, for $0 \leq \nu < \infty$. If $\nu > \mu$ are regular values for which $[\mu, \nu]$ contains no critical value then $\Mspace_\nu$ and $\Mspace_\mu$ are topologically the same, the second obtained from the first by diffeomorphic thickening all around. 
If $[\mu, \nu]$ contains the critical value of exactly one critical point, $x$, then the difference between the two superlevel sets depends only on the index of $x$. If $x$ has index $3$ then $\Mspace_\mu$ has one more component than $\Mspace_\nu$, and that component is a topological ball. If $x$ has index $2$, then $\Mspace_\mu$ can be obtained from $\Mspace_\nu$ by attaching an arc at its two endpoints and thickening all around. This extra arc can have one of two effects on the homology of the superlevel set. If its endpoints belong to different components of $\Mspace_\nu$, then $\Mspace_\mu$ has one less component, while otherwise $\Mspace_\mu$ has one more loop. If $x$ has index $1$, then $\Mspace_\mu$ can be obtained from $\Mspace_\nu$ by attaching a disk, which has again one of two effects on the homology groups. Finally, if $x$ has index $0$ then $\Mspace_\mu$ is obtained by attaching a ball. In all cases but one, this ball fills a void, the exception being the last ball that is attached when we pass the global minimum of $f$. At this time, the superlevel set is completed to $\Mspace_\mu = \Mspace$.

 
 
\section {Topology} 
 
\label{sec:topology} 
 
 
 
In this section, we introduce the topological concepts we use to analyze particle distributions. The main new methods for cosmological applications are Betti numbers and persistence, which we will relate to the more traditional notions of Minkowski functionals, Euler characteristic, and genus. 
 \begin{figure} 
   \centering 
   \includegraphics[width=9cm]{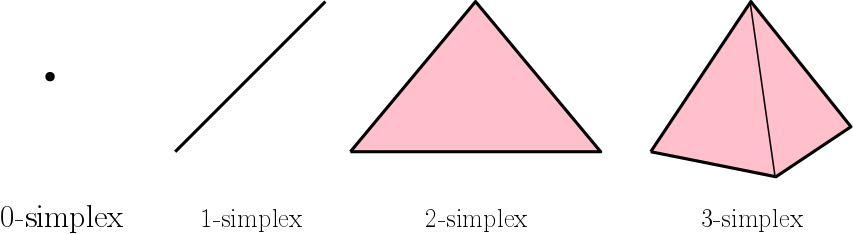} 
   \caption{From left to right: 0, 1-, 2- and 3- simplex.} 
  \label{fig:simplices} 
 \end{figure}  

\subsection{Euler Characteristic and Genus}  
Let us have a solid body $\Mspace$. Suppose now that we have the boundary of $\Mspace$ triangulated, using $v$ vertices, $e$ edges, and $t$ triangles. The vertices, edges triangles and tetrahedra are also referred to as \emph{simplices}. A vertex is a 0-dimensional simplex, an edge is a 1-dimensional simplex, a triangle is a 2-dimensional simplex, and a tetrahedron is a 3-dimensional simplex. Figure~\ref{fig:simplices} presents an illustration of simplices in dimensions up to 3.

Named after Leonhard Euler \citep{Eul58},  
the \emph{Euler characteristic} of the surface --- traditionally denoted as $\chi$ --- is the alternating sum of the number of simplices:  
\begin{equation} 
\chi  =  v - e + t\,.  
\label{eq:eulerpolyh} 
\end{equation} 
It does not depend on the triangulation, only on the surface. For example, we can triangulate the sphere with $4$ vertices, $6$ edges, and $4$ triangles, like the boundary of the tetrahedron, which gives $\chi = 4-6+4 = 2$. Alternatively, we may triangulate it with $6$ vertices, $12$ edges, and $8$ triangles, like the boundary of the octahedron, which  
again gives $\chi = 6-12+8 = 2$. 
 
Generalizing this to a orientable connected closed surface $S$ \footnote{An \emph{orientable} surface in Euclidean space is a surface for which  it is possible to make a consistent choice of surface normal vector at every point. A \emph{closed} surface is a surface which is compact and without boundary.}, with $h \geq 0$ handles the Euler characteristic is equal to $2$ minus twice the number of  
handles, is $\chi = 2 - 2h$. For example, the sphere has $\chi = 2$ and the torus has $\chi = 0$. If the boundary of $\Mspace_\nu$ consists of $k$ components with a total of $h$ holes,  
then we have $\chi = 2(k-h)$. To make this more concrete, we formalize the number of holes of a closed, connected surface to its \emph{genus}, denoted as $g = h$. It is defined  
as the maximum number of disjoint closed curves we can draw on the surface such that cutting along them leaves the surface in a single connected piece. For example, for a sphere we have  
$g=0$, and for a torus we have $g=1$. If we now drop the assumption that the surface is connected, we get the Euler characteristic and the genus by taking the sum over  
all components. Since $\chi_i = 2 - 2g_i$ for the $i$-th component, we have  
 
\begin{equation} 
\chi = \sum_{i=1}^k \chi_i = \sum_{i=1}^k (2 - 2g_i) = 2k-2g\,. 
\end{equation} 
 
We see that a minimum amount of topological information is needed to translate between Euler characteristic and genus. This is  
different from what the cosmologists have traditionally called the genus, which is defined as $\tilde{g} = - \frac{1}{2} \chi$ \citep{GDM86,HGW86}.  
Relating the two notions, we get $g = k+\tilde{g}$. We will abandon both in this paper, $\tilde{g}$ because it is redundant, and  
$g$ because it is limited to surfaces. Indeed, the Euler characteristic can also be defined for a $3$-dimensional body, taking  
the alternating sum of the simplices used in a triangulation, while the genus has no satisfactory generalization beyond $2$-dimensional  
surfaces.

\subsection{Minkowski Functionals}  
Suppose we have a solid body, $\Mspace$, whose boundary is a smoothly embedded surface in $\Rspace^3$. This surface may be a sphere or have holes, like the torus, and it may consist of one or several connected components, each with its own holes. Similarly, we do not require that $\Mspace$ is connected. Write $\Mspace^r$ for the set of points at distance $r$ or less from $\Mspace$. For small values of $r$, the boundary of $\Mspace^r$ will be smoothly embedded in $\Rspace^3$, but as $r$ grows, it will develop singularities and self-intersections. Before this happens, the volume of $\Mspace^r$ can be written as a degree-$3$ polynomial in $r$,  
\begin{equation} 
  \Volume{\Mspace^r}  ~~=~~  Q_0 + Q_1 r + Q_2 r^2 + Q_3 r^3 \,. 
\label{eq:minkvol} 
\end{equation} 
The $Q_i$ are known as the \emph{Minkowski functionals} of $\Mspace$, which are important concepts in integral geometry.  

Minkowski functionals were first introduced as measures  
of the spatial cosmic mass distribution by \cite{Mecke94} and have become an important measure of clustering of mass and galaxies  
\citep{schmalzing1997,schmalzing1999,sahni1998}.  
 
In terms of their interpration in the 3-dimensional context, following Equation~\eqref{eq:minkvol}, we see that  
$Q_0$ is the volume of $\Mspace$, $Q_1$ is the area of its boundary, $Q_2$ is the total mean curvature, and $Q_3$ is one third  
of the total Gaussian curvature of the boundary. These interpretations suggest that the Minkowski functionals are essentially  
geometric in nature, and they are, but there are strong connections to topological concepts as well. The key connection is  
established via the Euler characteristic.

\subsection{Geometry and Topology: Gauss--Bonnet Theorem}  
The key connection between the geometric Minkowski functionals and topology is established via the Euler characteristic, $\chi(S)$,  
of a surface $S$. The connection between the topological characteristics of a space and its geometrical properties  
is stated by the famous Gauss--Bonnet theorem. For a connected closed surface $S$ in $\Rspace^3$, the Gauss--Bonnet theorem asserts that  
the total Gaussian curvature is $2 \pi$ times the Euler characteristic $\chi(S)$,  
\begin{equation} 
\chi(S)\,=\,{\displaystyle 1 \over \displaystyle 2\pi}\,\oint\,\left({\displaystyle 1 \over \displaystyle R_1 R_2}\right)\,dS\,, 
\label{eq:euler} 
\end{equation} 
where $R_1$ and $R_2$ are the principal radii of curvature at each point of the surface. Note that the Gauss--Bonnet theorem only holds for smooth surfaces, meaning surfaces for which at least the second derivative is well-defined. For the situation sketched above,  
a boundary of space $\Mspace$ consisting of $k$ components with a total of $h$ holes, it tells that the total Gaussian  
curvature will be equal to $4 \pi (k-h)$. For example, the Gaussian curvature of a sphere with radius $r$ is $1 / r^2$ at every point. Multiplying with the area,  
which is $4 \pi r^2$, we get the total Gaussian curvature equal to $4 \pi$, which is independent of the radius. This agrees  
with $\chi = 4 \pi (k-h)$ given above since $k-h = 1$ in this case.  
 
The Gauss--Bonnet theorem (eq.~\ref{eq:euler}) underlines the key position of the Euler characteristic at the core of the topological and geometric  
characterization of topological spaces. The Euler characteristic establishes profound and perhaps even surprising links between seemingly widely different areas of mathematics.  
While in simplicial topology Euler's polyhdron formula states that it is the alternating sum of the number of $k$-dimensional simplices  
of a simplicial complex (eq.~\ref{eq:eulerpolyh}), its role in \emph{algebraic topology} as the alternating sum of Betti numbers is  
expressed by the Euler--Poincar\'e formula (see eq.~\ref{eq:eulerpoincare} in the next subsection). Even more intricate is the  
connection that it establishes between these topological aspects and the singularity structure of a field, which is the realm of  
\emph{differential topology}. In particular interesting is the relation established by Morse theory of the Euler characteristic  
being equal to the alternating sum of the number of different field singularities, ie. of maxima, minima and saddle points. Finally,  
its significance in \emph{integral geometry} is elucidated via Crofton's formula, which establishes the fact  
that Minkowski functionals are integrals over the Euler characteristic of affine cross-sections.  
 
\begin{figure*} 
  \centering 
  \includegraphics[width=18cm]{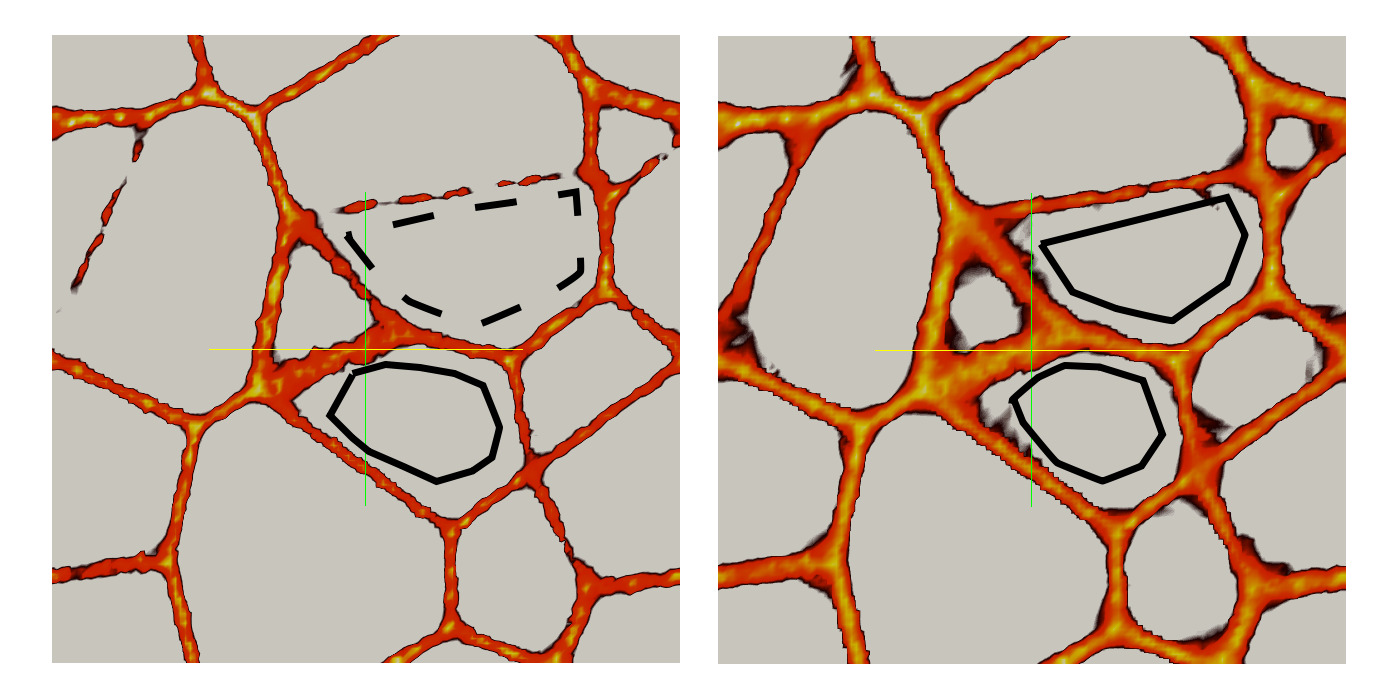} 
  \caption{Chains and Cycles. Density rendering of the superlevel set of a 2-dimensional cross section of a Voronoi wall model. The lefthand  
frame corresponds to a higher density threshold value than that in the righthand frame. Particularly attention concerns the cells in which we  
have marked the outline by black lines. For a high threshold value, the superlevel structure traced by the dashed closed curve does not form a loop: the  
multiple broken segments are \emph{chains}. At a lower threshold value, the superlevel structure becomes continuous, and individual segments  
merge together to form a loop: a 1-dimensional \emph{cycle}.} 
 \label{fig:chains_cycles} 
\end{figure*}

\subsection{Homology and Betti Numbers}  
While the Euler characteristic can distinguish between connected, closed surfaces in $\Rspace^3$, it has no discriminative power if applied to $3$-manifolds, which is the most direct generalization of surfaces to the next higher dimension. Indeed, Poincar\'{e} duality implies $\chi = 0$ for all $3$-manifolds. Fortunately, we can write the Euler characteristic as an alternating sum of more descriptive topological invariants named after Enrico Betti \citep{Bet71}. To introduce them, we find it convenient to generalize the space $\Mspace$ by dropping most limitations, such as that it be embedded or even embeddable in $\Rspace^3$. Letting the intrinsic dimension of $\Mspace$ be $d$, we get $d+1$ possibly non-zero \emph{Betti numbers}, which traditionally are denotes as $\beta_0, \beta_1, \ldots, \beta_d$. The relationship to the Euler characteristic is give by the Euler--Poincar\'{e} Formula: 
\begin{equation} 
  \chi ~~=~~ \beta_0 - \beta_1 + \beta_2 - \ldots (-1)^d \beta_d . 
\label{eq:eulerpoincare} 
\end{equation} 
This relation holds in great generality, requiring only a triangulation of the space, and even this limitation can sometimes be lifted. In this paper, we only consider subspaces of the $3$-torus, $\Mspace$. For this case, only $\beta_0$, $\beta_1$, $\beta_2$, and $\beta_3$ are possibly non-zero, and we have $\beta_3 \neq 0$ only if $\Mspace$ is equal to the $3$-torus, in which case $\beta_3 = 1$. The first three Betti numbers have intuitive interpretations: $\beta_0$ is the number of \emph{components}, $\beta_1$ is the number of \emph{loops}, and $\beta_2$ is the number of \emph{shells} in $\Mspace$. Often, it is convenient to consider the complement of $\Mspace$, which shows $\beta_0 - 1$ \emph{gaps} between the components, $\beta_1$ \emph{tunnels} going through the loops, and $\beta_2$ \emph{voids} enclosed by the shells. 
 
A formal definition of the Betti numbers requires the algebraic notion of a homology group. While a serious discussion of this topic is beyond the scope of this paper, we provide a simplified exposition and refer to texts in the algebraic topology literature for details \citep[see e.g.][]{munkres1984elements}.  
 
For simplicity, we assume a triangulated space and we use the coefficients $0$ and $1$ and addition, modulo $2$. A \emph{p-chain} is a formal sum of the $p$-simplices in the triangulation, which we may interpret as a subset of all $p$-simplices, namely those with coefficients $1$. The \emph{sum} of two $p$-chains is again a $p$-chain. Interpreted as sets, the sum is the symmetric difference of the two sets. Note that each $p$-simplex has $p+1$ $(p-1)$-simplices as faces. The \emph{boundary} of the \emph{p-chain} is then the sum of the boundaries of all $p$-simplices in the chain. Equivalently, it is the set of $(p-1)$-simplices that belong to an odd number of $p$-simplices in the chain. We call the $p$-chain a \emph{p-cycle} if it is the boundary of a $(p+1)$-chain. Importantly, every $p$-boundary is a $p$-cycle. The reason is simply that the boundaries of the $(p-1)$-simplices in the boundary of a $p$-simplex contain all $(p-2)$-simplices twice, meaning the boundary of the boundary is necessarily empty. To get homology, we still need to form \emph{classes}, which we do by not distinguishing between two $p$-cycles that together form the boundary of a $(p+1)$-chain.  
 
To get the group structure, we add $p$-cycles by taking their symmetric difference or, equivalently, by adding simplices modulo $2$. Homology classes can now be added simply by adding representative $p$-cycles and taking the class that contains the sum. The collection of classes together with this group structure is the \emph{$p$-th homology group}, which is traditionally denoted as $H_p$. Finally, the \emph{$p$-th Betti number} is the rank of this group, and since we use modulo $2$ arithmetic to add, this rank is the binary logarithm of the order: $\beta_p = \log_2 |H_p|$. We note that modulo $2$ arithmetic has multiplicative inverses and therefore forms what in algebra is called a \emph{field} \footnote{The algebraic concept of field is not to be confused with the physical notion of
(scalar density) field that also plays a prominent role in this paper.}. For example, arithmetic with integers is not a field. Whenever we use a field to construct homology groups, we get vector spaces. In particular, the groups $H_p$ defined above are vector spaces, and the $\beta_p$ are their dimensions, as defined in standard linear algebra. 
 
In our study, we forward Betti numbers for the characterization of the topological aspects of the cosmic mass  
distribution.  

\begin{figure} 
  \centering 
  \includegraphics[width=0.48\textwidth]{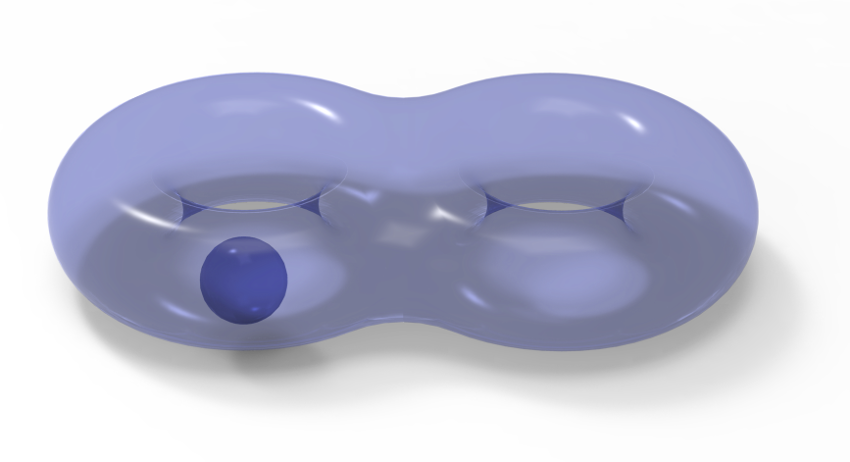} 
  \caption{Running example of nontrivial topology: a solid double-torus containing an empty bubble. The boundary surface of this double-donut with small void inside consists  
 of two parts, a double-torus on the outside and a sphere encapsulating the bubble. For further explanation see sect.~\ref{sec:topology}.} 
 \label{fig:running-example} 
\end{figure} 
 
 \subsection{Running Example}  
 We begin with an example, which we use to illustrate the geometric and topological concepts, ahead of formally defining them. For this purpose, let $\Mspace$ be a solid double-torus with an empty bubble, that is: a double-donut with a small void inside; see Figure \ref{fig:running-example}. Its boundary, denoted as $\partial \Mspace$ consists of two surfaces: a double-torus on the outside and a sphere bounding the bubble. 
 
 The \emph{Minkowski functionals} are the volume of $\Mspace$, the area, the total mean curvature, and the total Gaussian curvature of $\partial \Mspace$. These are geometric properties, but they are not independent of the purely topological concepts we will introduce next. 
 
 The \emph{Euler characteristic} is the alternating sum of the number of simplices of different dimensions needed to triangulate a space. Applied to $\partial \Mspace$, the number of vertices minus the number of edges plus the number of triangles needed to triangulate the double-torus gives $-2$, and for the sphere we get $+2$. It follows that the Euler characteristic of $\partial \Mspace$ is $\chi = 0$. There are many other $2$-dimensional topological spaces that have the same Euler characteristic, the torus being one, the union of two tori being another.  
 
 Indeed, the total Gaussian curvature of the sphere is $4 \pi$, no matter how large it is, and the Euler chatacteristic of the same is $2$. The \emph{genus} of $\partial \Mspace$ is $2$, namely $2$ for the double-torus plus $0$ for the sphere. For a connected closed surface, the genus equals $1$ minus half the Euler characteristic. More generally, the genus of a $2$-manifold that is the union of disjoint closed surfaces is therefore 
 
 \begin{equation} 
 g = \sum_i g_i = \sum_i \left( 1 - \frac{\chi_i}{2} \right) = \#{\rm components} - \left( \frac{\chi}{2} \right), 
 \end{equation} 
  
 where we write $\chi_i$ and $g_i$ for the Euler characteristic and the genus of the $i$-th component. The reader may check that this relation holds for $\partial \Mspace$. We get a refinement of the concepts by introducing \emph{Betti numbers}. Formally, they are ranks of homology groups, one for each dimension (more on homology and homology groups later). We have

 \begin{eqnarray} 
   \beta_0  &=&  \# \rm{components},           \nonumber               \\ 
   \beta_1  &=&  \# \rm{independent~loops},                            \\ 
   \beta_2  &=&  \# \rm{independent~closed~surfaces}. \nonumber 
 \end{eqnarray} 
 
 \noindent For $\partial \Mspace$, we have $\beta_0 = 2$, $\beta_1 = 4$, $\beta_2 = 2$. Indeed, we have two components and two closed surfaces: the double-torus and the sphere. To see the four loops, draw one around each hole of the double-torus and another one around each handle. We get the Euler characteristic by taking the alternating sum: $\chi = \beta_0 - \beta_1 + \beta_2$, which for $\partial \Mspace$ gives $0$, as required. 
 
 Suppose now that $\Mspace$ is the portion of the Universe at which the local density exceeds some threshold, $\nu$. What if we decrease $\nu$ by some small but positive amount? Decreasing the threshold enlarges the portion at which the density threshold is exceeded. It may be that the bubble fills up. Assuming that nothing else changes, $\partial \Mspace$ is now a double-torus, with $\beta_0 = 1$, $\beta_1 = 4$, $\beta = 1$. The sphere and the bubble have gone.

\subsection{Persistent Homology}  
In Morse theory, we learned that passing a critical point either increases the rank of a homology group by one, or it decreases the rank of another group by one. Equivalently, it gives birth to a generator of one group or death to a generator of another group. Our goal is to pair up births with deaths such that we can talk about the subsequence in the filtration over which a homology class exists. This is precisely what persistent homology accomplishes. 
 
Recall that between two consecutive critical values, the homology of the superlevel sets is constant. It therefore suffices to pick one regular value within each such  
interval. Writing $r_0 > r_1 > \ldots > r_n$ for these regular values, this induces a sequence of inclusions
\begin{equation}
\Mspace_0 \to \Mspace_1 \to \cdots \to \Mspace_i \to \cdots \to \Mspace_n,
\end{equation}
where $\Mspace_i$ is the manifold defined by the superlevel set $r_i$.

The inclusion $\Mspace_{i-1} \to \Mspace_i$ maps a $p$-cycle in $\Mspace_{i-1}$ to a $p$-cycle in $\Mspace_i$, and a p-boundary in $\Mspace_{i-1}$ to a
$p$-boundary in $\Mspace_i$. Therefore, it induces a map 
$H_p(\Mspace_i) \to H_p(\Mspace_{i+1})$, which is a homomorphism
since it preserves the group structure.
So we have d+1 sequences 
\begin{equation}
H_p(\Mspace_0) \to H_p(\Mspace_1) \to \cdots \to H_p(\Mspace_n),
\end{equation}
for $p = 0,1,\ldots, d$.

Assuming coefficients in a field \footnote{The betti numbers might depend on the choice of the field. For example, $\beta_2$ of the projective plane is 1, if the field is $Z_2$, and 0 for the field of rational numbers. However, such considerations do not apply if the surfaces are orientable, which is the case that we deal with.}, as before, we have a sequence of vector spaces with linear maps between them. These maps connect the groups by telling us where to find the cycles of a homology group within later homology groups. Sometimes, there are new cycles that cannot be found as images of incoming maps, and sometimes classes merge to form larger classes, which happens when we get chains that further wash out the difference between cycles.  
 
To simplify notation, we will assume a particular dimension, $p$,
so that we can suppress the subscript.  Instead, we write
$H_i = H_p ({\Mspace}_i)$, effectively indexing the homology
groups with the position along the filtration.  We can now be
specific about the persistence of homology classes. Letting $\gamma$ be a class in $H_i$, we say $\gamma$ is \emph{born} at $H_i$ and \emph{dies entering} $H_j$ if 
\begin{itemize} 
  \item  $\gamma$ is not in the image of $H_{i-1}$ in $H_i$; 
  \item  the image of $\gamma$ is not in the image of $H_{i-1}$ in $H_{j-1}$, but it is in the image of $H_{i-1}$ in $H_j$. 
\end{itemize} 
 
\noindent Letting $r_{i-1} > \nu_i > r_i$ and $r_{j-1} > \nu_j > r_j$ be the critical values in the relevant intervals, we represent $\gamma$ by $(\nu_i, \nu_j)$, which we call a \emph{birth-death pair}. Furthermore, we call $\persistence{\gamma} = \nu_i - \nu_j$  the \emph{persistence} of $\gamma$, but also of its birth-death pair.  
 
To avoid any misunderstanding, we note that there is an entire coset of homology classes that are born and die together with $\gamma$, and all these classes are represented by the same birth-death pair. Calling the image of $H_i$ in $H_{j-1}$ a \emph{persistent homology group}, we note that its rank is equal to the number of birth-death pairs $(\nu_b, \nu_d)$ that satisfy $\nu_b \geq \nu_i > \nu_j \geq \nu_d$. They represent the classes that are born at or before $H_i$ and that die entering $H_j$ or later. 

Finally, for whom this description of persistence and homology is not immediately clear, we refer to Section~\ref{sec:triangle_filtration} for a concrete example of filtration and persistent homology.

\subsection{Intensity Maps} 
\label{sec:intensity_map}
 
This paper concerns itself with the topology of stochastic point  
processes, and density field computed on them. In the context of the  
Universe, both the cosmic microwave background and the density  
distribution in the Universe are examples of spatial stochastic  
processes. 
It is a universal property of stochastic processes that the  
expectation value of the quantities defined on them converge over many  
realizations. Our conjecture is that this must also be true for the  
birth-death events, as reflected in the persistence diagrams, if  
averaged over many realizations. While a rigorous attempt at deriving a probabilistic and statistical  
description of persistence topology is beyond the scope of this  
paper, we provide an emperical description and test, as proof of the  
hypothesis, by introducing the intensity maps. 
 
We are interested in  
the statistical description of persistence diagrams, as an average over  
many realization, of the stochastic process $f$. To this end, we construct the \emph{intensity map}, which is the function $p:\Rspace^2 \to \Rspace$ in the  
\emph{mean density-persistence} plane \footnote{This is a plane defined by the mean-density of the features on the horizontal axis, which is the mean of birth and death values of the features. The vertical axis is defined by the persistence value of the features.}, whose integral over every  
region $R \subset \Rspace^2$ is the expected number of points in $R$. Let $\langle N_{tot} \rangle$ be representative of the \emph{total intensity} of the map. We discretize the intensity  
map into a number of regular grid-cells in the plane, and define  
the bin-wise intensity, for the grid-cell $(i,j)$ as 
\begin{equation} 
	I_{ij}= \frac{\langle N_{ij} \rangle}{\langle N_{tot} \rangle}, 
	\label{eqn:intensity_map} 
\end{equation} 
where, $\langle N_{ij} \rangle$ is the expected intensity in the grid-cell ($i,j$), and $\langle N_{tot} \rangle$ is the expected total intensity, over many realizations of the same random experiment.
 
The total \emph{intensity} of the maps is  
proportional to the average number of total dots in the persistence  
diagrams. For each grid cell, the intensity function represents the  
fraction of the total intensity of  
the map. Since the intensity in each bin is normalized by the total  
intensity of the map, the integral of the intensity function over  
$\Rspace^2$ always evaluates to $1$, irrespective of the model in  
question. In the limit of the size of the grid-cells going to zero, the  
discretized intensity function approximates the  probability  
density function. At this point, we only have emperical  
evidence that if $f$ arises from a stochastic process and 
is tame (all the derivatives well defined), the intensity maps are well  
defined . As we will show shortly, the intensity  
maps are highly sensitive to the  
parameters of the model, and capture local variations in  
topology across the whole range of function value. As such, we propose their use to characterize  
and discriminate between various models.

 
 
\section{Computation} 
 
\label{sec:computation} 
 
 
 
The geometric and topological concepts outlined in Sections \ref{sec:fields} and \ref{sec:topology} have all matured to a stage at which we have fast software to run on simulated and observed data. In this section, we describe the principles of these algorithms, and we provide sufficient information  
for the reader to understand the connection between the mathematics, the data, and the computed results. 
 
The computational framework of our study involves three components. The first component concerns the definition and calculation of the  
density field on which we apply the field's filtration. This is described in Section~\ref{sec:comp_density}. A directly related issue is the representation of the density  
field in the homology calculation, ie. whether we retain its representation by density estimates at the original  
sampling points or whether we evaluate it on the basis of a density image on a regular grid. The second component of the computational pipeline, is the algorithm used for computing persistent homology. This involves building a filtration, described in Section~\ref{sec:comp_filtration}, and the subsequent computation of persistent homology on this filtration, which is described in Section~\ref{sec:pers_hom}. The third aspect concerns the representation of  
the results of the homology and persitent homology computation. The principal products consist of {\it intensity maps} and {\it Betti numbers} 
of the analyzed samples, which form the visual representation and summary 
of persistent homology and homology. The construction of intensity maps is described in detail in Section~\ref{sec:intensity_map}, as well as Section~\ref{sec:pers_hom}.

\subsection{Density Reconstruction from Point Sample}
\label{sec:comp_density}  
We use DTFE \citep{Sch00,weyschaap2009,cautun2011} to construct a piecewise linear scalar-valued density field from a  
particle distribution. The DTFE formalism involves the computation of  
the Delaunay tessellation of the particles in $\Mspace$, the determination of tessellation based density estimates, and the  
subsequent piecewise linear interpolation of the density values at the  
Delaunay vertices, ie. the sample points, to the higher dimensional  
simplices, yielding a field $f: \Mspace \to \Rspace$. Figure~\ref{fig:simplices} presents an illustration of simplices in spatial dimensions up to 3. 
 
For the calculation of the Delaunay tessellation, we use software in the {\sc Cgal} library. We use the $3$-torus option  
of {\sc Cgal}, which is the periodic form of the original data set in a cubic box obtained by identifying  
opposite faces of the box.

\begin{table} 
 \begin{center} 
  \begin{tabular}{r|rrrr} 
   Model   &   \# particles  & \# simplices   & Del. (s) & Pers. (s)  \\ \hline 
   Poisson &      500,000    &   14,532,164   &  10.15   &    6414.16       \\ \hline 
   Cluster &      262,144    &    7,491,308   &  81.48   &      12.58       \\ \hline 
  Filament &      262,144    &    7,346,712   &  77.76   &     402.36       \\ \hline 
      Wall &      262,144    &    7,345,520   &   5.26   &     555.46       \\ \hline 
   Voronoi &		     &                &          &                  \\ 
 Kinematic &      262,144    &    7,409,364   &   5.93   &     125.33       \\ 
   Stage 3 &                 &                &          &                  \\ \hline 
  Soneira-- &                 &                &          &                  \\ 
  Peebles  &      531,441    &   14,300,836   & 162.42   &     168.15       \\ 
$\zeta=9.0$ &                 &                &          &                  \\ \hline 
  \end{tabular} 
 \end{center} 
 \caption{Parameters of computation for the various models described in this paper. All computations are performed on an Intel(R) Xeon(R) CPU @ 2.00GHz. Columns 1 \& 2 present the models described in the later sections, and the number of particles used for the computation. Column 3 gives the total number of simplices of the Delaunay tessellation. Columns 4 \& 5 give the time required to compute the tessellation and persistence respectively, in seconds.} 
 \label{tab:comp_param} 
\end{table} 
 
In a second step, we compute the DTFE density value for each vertex, $u$, of the Delaunay tessellation. The DTFE density value  
at the vertices is the inverse of the volume of its \emph{star}. The  star consists of all simplices that contain $u$ as a vertex (see  
Figure~\ref{fig:star_vertex} for an illustration),  
and we assign one over this volume as the density value to $u$. Finally, we use piece-wise linear interpolation to define  
$ f: \Mspace \to \Rspace$. 

The particular nature of the discretely sampled density field involves a complication. Because the number density of the sample points  
represents a measure of the value of the density field itself, the DTFE density field has a much higher spatial resolution in  
high density regions than in low density regions. This might be a source of a strong bias in the retrieved topological  
information, given that most of this will focus on the topological structure of the high-density regions. To alleviate a density bias 
towards the highly sampled regions, one may invoke a range of strategies. An option that is often followed is to sample the density field 
on a regular grid. In other words, to create an image of the DTFE density field reconstruction. It has the advantage 
of representing a uniformly sampled density field,  with a uniform spatial resolution dictated by the voxel size of the image. However, 
following this option involves the loss of resolution in the high density regions. On the other hand, it retains the DTFE 
advantage of sampling the low density - void - regions well. In the context of homology analysis, we should also note 
that the use of a gridbased image involves a few extra complications. The details of this are extensively discussed 
in the follow-up study analyzing the homology and persistence of Gaussian random fields 
\citep[see][]{pranav2016a,pranav2016b}.  
 
Dependent on the region of interest one may therefore choose to follow the full formal DTFE procedure or to use 
the alternative option of a grid-sampled one DTFE field. In the context of our study, we follow the formal DTFE definition. 

Another strategy to moderate the bias towards high-density regions is to use the singularity structure of the  
piecewise linear density field, and use the persistence of singularity pairs to remove insignificant topological  
features. This natural feature-based smoothing of the density field has been described extensively and has been  
applied in studies of cosmic structures by \cite{Sousbie1} and  
by \cite{Shivashankar2015}.

Table~\ref{tab:comp_param} presents the noteworthy parameters of computations for a single realization of the different models  
used in the results section of this paper. Naming the models in Column 1, we see the number of particles and simplices in the  
Delaunay tessellation in Columns 2 and 3 \citep[also see][]{Okabe2000,weygaert94}, and the number of seconds needed to compute  
the Delaunay tessellation and the persistence pairs in Columns 4 and 5. Apparently, the number of particles is not strongly  
correlated with the time it takes to construct the Delaunay tessellation. Indeed, the algorithm is also sensitive to other  
parameters --- such as the number of simplices in the final simplicial complex or ever constructed and destroyed during the runtime  
of the algorithm --- that depend on how the particles are distributed in space.   
 
\subsection{Critical Values and Filtration} 
\label{sec:comp_filtration}
\begin{figure} 
\begin{center} 
  \includegraphics[width=8cm]{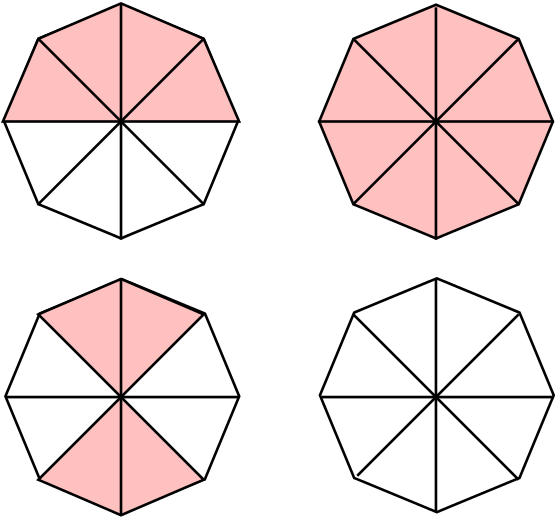}\\ 
  \caption{Figure illustrating the upper star of a regular vertex,  
  minimum, saddle and maximum, respectively in top-left, top-right,  
  bottom-left, bottom-right panels. The star of a vertex is consists  
  of all the simplices incident to it. The shaded simplices in pink  
  have a function value higher than the vertex.} 
  \label{fig:star_vertex} 
\end{center}  
\end{figure}

As mentioned in the paragraph on Morse theory, the superlevel set does not change topology as long as $\nu$ does not pass a critical value of the function, and this is also true for piecewise linear functions, except that we need to adjust the concept of critical point. Here we do the obvious, looking at how $ f$ varies in the \emph{link},  of a vertex. The link consists of all faces of simplices in the star, that do not themselves belong to the star \citep[Chapter VI]{EdHa10}. Indeed, the topology can change only when $\nu$ passes the value of a vertex, so it suffices to consider only one (regular) value between any two contiguous vertex values. To describe this, we let $n$ be the number of vertices in the tessellation, and we assume $\nu_i =  f (u_i) < \nu_{i+1} =  f (u_{i+1})$ for $1 \leq i < n$.\footnote{It is unlikely that the estimated density values at two vertices are the same, and if they are, we can pretend they are different, eg.\ by simulating a tiny perturbation that agrees with the ordering of the vertices by index; see eg.\ \citep[Section I.4]{Ede01}.} We thus consider superlevel sets  
at the regular values in the sequence 
$$ 
  r_0 > \nu_1 > r_1 > \nu_2 > \ldots > \nu_n > r_n . 
$$ 
Constructing these superlevel sets and computing their homology individually would be impractical for the data-sets we study in this paper. Fortunately, there are short-cuts we can take that speed up the computations while having no effect on the computed results. 
The first short-cut is based on the observation that $\Mspace_\nu$ has  
the same homotopy type as the subcomplex $K_\nu$ of the tessellation  
$K$ of $\Mspace$ that consists of all vertices with $ f (u_i) \geq  
\nu$ and all simplices connecting them. There is a convenient  
alternative description of $K_\nu$. Define the \emph{upper star} of a  
vertex $u$ as the collection of simplices in the star for which $u$ is  
the vertex with smallest density value (see Figure~\ref{fig:star_vertex}  
for the upper star of a regular vertex, a $1$-saddle, a $2$-saddle an  
a maximum). Then $K_\nu$ is the union of the upper stars of all  
vertices with $ f (u_i) \geq \nu$. This description is  
computational convenient because it tells us that $K_{r_{i+1}}$ can be  
obtained from $K_{r_i}$ simply by adding the simplices in the upper  
star of $u_{i+1}$. We say the superlevel sets can be computed  
\emph{incrementally}, and we will be careful to follow this paradigm in  
every step of our computational pipeline. This incremental  
construction of the superlevel sets is equivalent to constructing the  
upper-star filtration, which is an essential pre-cursor to computing  
persistence homology. 
 
\begin{figure*} 
 \begin{center} 
\subfloat[]{\includegraphics[width=0.49\textwidth]{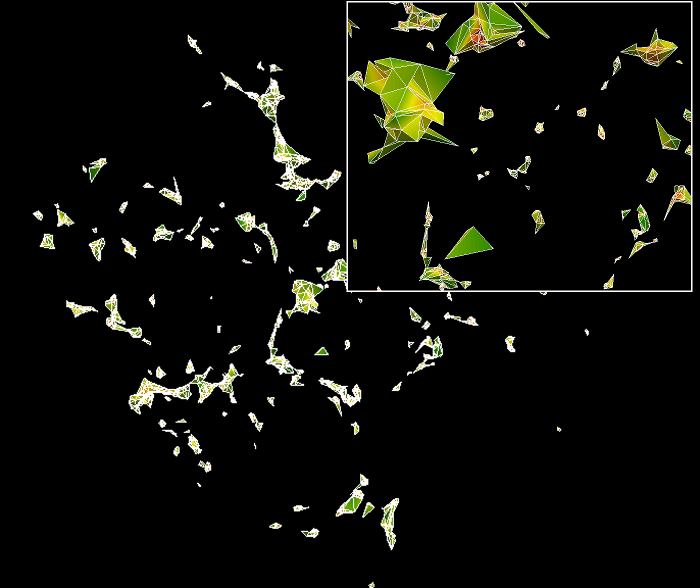}} 
   \hspace{0.01\textwidth}    
   \subfloat[]{\includegraphics[width=0.49\textwidth]{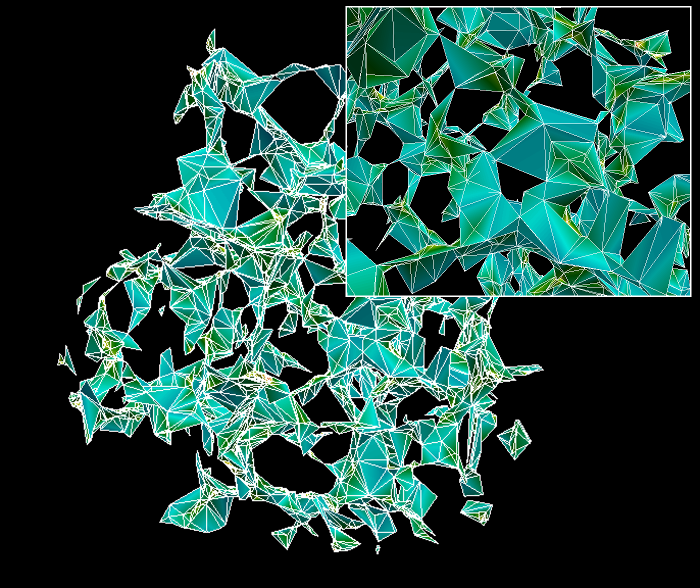}}\\ 
  \subfloat[]{\includegraphics[width=0.49\textwidth]{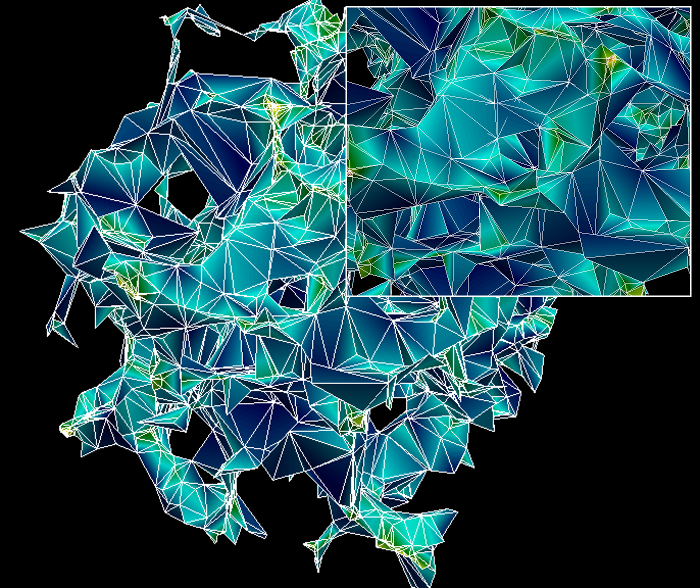}} 
\end{center} 
  \caption{From panel (a) to panel (c): Growing superlevel sets of a filtration of a simplicial complex constructed on a discrete pointset. The insets present a zoomin. The density threshold decreases from left to right. As the density threshold decreases, more and more simplices from the underlying triangulation get included in the simplicial complex defined by the superlevel set corresponding to the density threshold.} 
   \label{fig:simplicialFiltration} 
\end{figure*}

To give a visual impression of superlevel sets of tessellations constructed in the practical circumstances, Figure~\ref{fig:simplicialFiltration} presents an illustration of the growing superlevel sets of a filtration of a simplicial complex constructed on a pointset obtained from a typical cosmological simulation.   
 
\subsection{Persistent Homology}  
\label{sec:pers_hom} 
Next, we sketch the algorithm that computes the persistent homology of the sequence of superlevel sets. We begin with a linear ordering of the simplices in $K$ that contains all $K_\nu$ as prefixes. To describe it, let $u_i = \sigma_{j_i}, \sigma_{j_i+1} , \ldots, \sigma_{j_{i+1}-1}$ be the simplices in the upper star of $u_i$, sorted in increasing order of dimension. Setting $j_1 = 1$ and $m = j_{n+1}-1$, this linear ordering of the simplices is $\sigma_1, \sigma_2, \ldots, \sigma_m$. It has the property that each simplex is preceded by its faces, which implies that every prefix, $K_j = \{\sigma_1, \sigma_2, \ldots, \sigma_j\}$, is a simplicial complex. We require this property so that every step of our incremental algorithm is well defined. It should be clear that $K_{\nu_i} = K_j$ for $j = j_{i+1}-1$.  
 
\begin{algorithm} 
  \caption{\sc Matrix Reduction}   
  \begin{algorithmic}[1] 
  \label{alg:matrix_reduce} 
  \STATE R = $\Delta$ 
  \FOR{ $j = 1$ to $m$}   
		\WHILE{there exists $j_{0} < j$ with $low(j_{0}) = low(j)$} 
			\STATE{add colum $j_{0}$ to column $j$} 
		\ENDWHILE 
  \ENDFOR 
\end{algorithmic} 
\end{algorithm} 
 
The persistence algorithm is easiest to describe as a matrix reduction  
algorithm, with the input matrix being the ordered boundary matrix of  
$K$.\footnote{We hasten to mention that storing this matrix explicitly  
is too costly for our purposes.  Instead, we use the tessellation as a  
sparse matrix representation, and we implement all steps of the matrix  
reduction algorithm accordingly.  However, for the purpose of  
explaining the algorithm, we maintain the illusion of an explicit  
representation of the matrix.} Specifically, this is the $m \times m$  
matrix $\Delta$ whose rows and columns correspond to the simplices in  
the mentioned linear ordering. Specifically, the $j$-th column records  
the boundary of $\sigma_j$, namely $\Delta_{i,j} = 1$, if $\sigma_i$ is  
a face of $\sigma_j$ and the dimension of $\sigma_i$ is one less than  
that of $\sigma_j$, and $\Delta_{i,j} = 0$, otherwise. Symmetrically,  
the $i$-th row records the star of $\sigma_i$. The persistence  
algorithm transforms $\Delta$ into \emph{reduced form}, in which every  
row contains the lowest non-zero entry of at most one column. Making  
sure that we do not permute rows, and we add columns strictly from left  
to right, the lowest non-zero entries in the reduced matrix correspond  
to the birth-death pairs of the density field --- precisely the  
information we are after. To describe the transformation, we write  
$\low{j} = i$ if $i$ is the maximum row index of a non-zero entry in  
column $j$, and we set $\low{j} = 0$ if the entire column is $0$.  
Algorithm~\ref{alg:matrix_reduce} presents the algorithm for such a  
reduction. Section~\ref{sec:triangle_filtration} illustrates these  
concepts and steps through an example. 
 
The search for the fastest algorithm to reduce an ordered boundary  
matrix is an interesting question of active research in the field of  
computational topology. Most known algorithms use row and column  
operations, like in Gaussian elimination, which takes time proportional  
to $m^3$ in the worst case. A fortunate but largely not understood  
phenomenon is the empirical observation that some of these algorithms  
are significantly faster than cubic time for most practical input data.  
This is lucky but also necessary since we could otherwise not compute  
the results we present in this paper. The time to compute the  
persistence pairs for different models is displayed in Column $5$ of Table~\ref{tab:comp_param}.

\subsubsection{Persistence Diagrams}  
 
\begin{figure} 
 \begin{center} 
  \includegraphics[width=8.5cm]{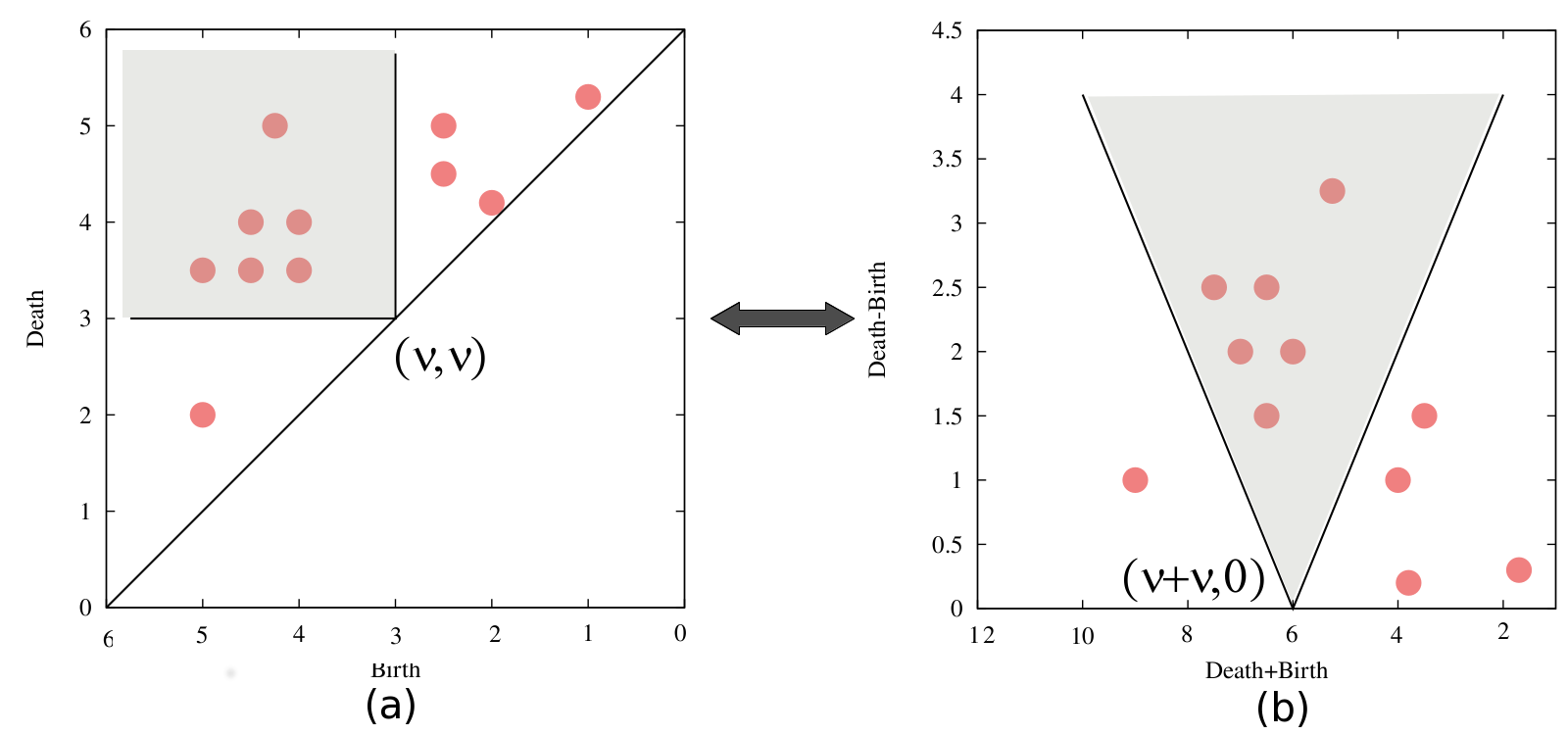}\\ 
 \end{center} 
 \caption{Figure illustrating the transition from the \emph{birth-death} to the \emph{mean age-persistence} plane. If the coordinates of a point in panel (a) are $(b,d)$, the coordinates in panel (b) are $(d+b,d-b)$. The Betti numbers can be read off from the persistence diagrams. The contribution to the Betti numbers for a level set $\nu$ comes from all the persistent dots that are born before $\nu$ and die after $\nu$ --- in other words, the shaded region in panel (a) anchored at ($\nu,\nu$). The shaded region transforms in panel (b) to a V-shaped region anchored at ($\nu+\nu,0$). The arms of the V have slope $-1$ and $1$ respectively.} 
 \label{fig:dgm_axis_transform} 
\end{figure}

Given the reduced boundary matrix, we generate the birth-death pairs of $\varrho$ from the lowest non-zero entries in the columns. Specifically, for every non-zero $i' = \low{j'}$, the addition of $\sigma_{i'}$ gives birth to a homology class that dies when we add $\sigma_{j'}$. If $\sigma_{i'}$ is in the upper star of $u_i$, and $\sigma_{j'}$ is in the upper star of $u_j$, then we get $(\nu_i, \nu_j)$ as the corresponding birth-death pair. It is quite possible that $i = j$, namely if both simplices belong to the same upper star, in which case we talk of a \emph{still-birth}. We draw this birth-death pair as the point $(\nu_i,\nu_j)$ in the \emph{birth-death} plane. Alternatively, we can also draw them as $(\nu_i+\nu_j, \nu_j-\nu_i)$ in the plane. This amounts to a scaling by a factor of $\sqrt{2}$ and a rotation of coordinates by 45 degrees clock-wise. This is our preferred representation of the persistence diagrams throughout this paper. An illustration of the transformation is depicted in Figure~\ref{fig:dgm_axis_transform}. Drawing all points representing $p$-dimensional homology classes gives the \emph{$p$-th persistence diagram} of $ f$, which we denote as $\Ddgm{p}{ f}$. Recall that the second coordinate is the persistence, and because a still-birth has zero persistence, it is drawn right on the horizontal axis. The persistence is a measure of significance of the feature represented by a birth-death point, and still-births are artifacts of the representation of $ f$ and have indeed no significance. The first coordinate is the sum of birth- and death-values, and we refer to half that coordinate as the \emph{mean age}. It gives information about the range of density values the corresponding feature is visible.\footnote{Almost every homology class that is ever born will also die at finite time, but there are eight exceptions, namely the classes that describe the $3$-torus itself. They are not relevant for the study in this paper, and we do not  
draw them in the diagrams.} 
 
Persistence diagrams contain more information than the Betti numbers.  
Indeed, we can read the $p$-th Betti number of the superlevel set for  
$\nu$ as a number of points of $\Ddgm{p}{ f}$. The contribution  
to the Betti numbers for the superlevel set at $\nu$ comes from all the dots in  
the persistence diagram corresponding to cycles that are born before  
$\nu$ and die after $\nu$ --- in other words, the shaded region in panel  
(a) anchored at ($\nu,\nu$) in Figure~\ref{fig:dgm_axis_transform}. The  
shaded region transforms appropriately in panel (b) to a V-shaped  
region anchored at ($\nu,0$) on the horizontal axis. The arms of the V have slope $-1$ and $1$ respectively.  Another useful property is the stability of the diagram under small perturbations of the input. Specifically, the diagram of a density function, $ f'$, that differs from $ f$ by at most $\varepsilon$ at every point  
of the space, has bottleneck distance at most $\varepsilon$ from $\Ddgm{p}{ f}$; see \citep{CEH07}. This implies that every point of $\Ddgm{p}{ f'}$ is at a distance at most $\varepsilon$ from a point in $\Ddgm{p}{ f}$ or from the horizontal axis.

\subsubsection{Intensity Map} 
 
Our preferred visual presentation of a diagram is averaged over a  
number of realizations of the same random experiment; see Figure  
\ref{fig:poisson_dgm_poisson_dens_alldim_avg_ver1}, which shows the  
plots for the data generated as described in Section \ref{sec:random}. To construct it, we superimpose the diagrams of the different realizations, we discretize $\Rspace^2$ using a grid of $100 \times 100$ squares, and we form the histogram by counting the points in each square. The result is a real-valued function on the plane, which we denote as the \emph{averaged persistence diagram} or the \emph{intensity map} of the diagram.  
 
\subsection{Example: Persistent Homology of a Triangle} 
\label{sec:triangle_filtration} 
 
\begin{figure*} 
\begin{center} 
  \includegraphics[width=18cm]{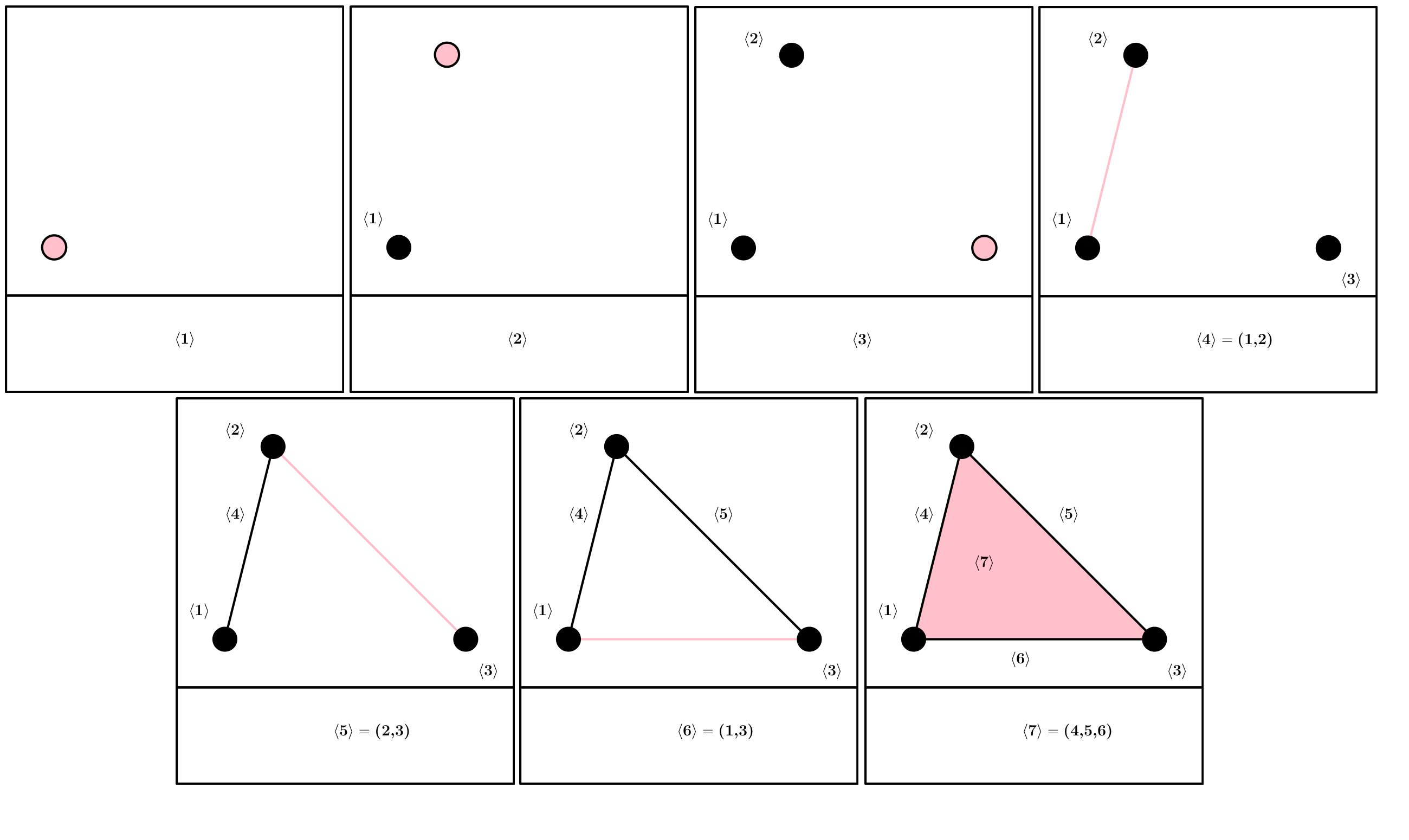}\\ 
  \caption{Figure illustrating the order in which the simplices of the  
  triangle appear in the filtration.} 
  \label{fig:triangle_filtration} 
\end{center}  

\begin{center} 
  \includegraphics[width=18cm]{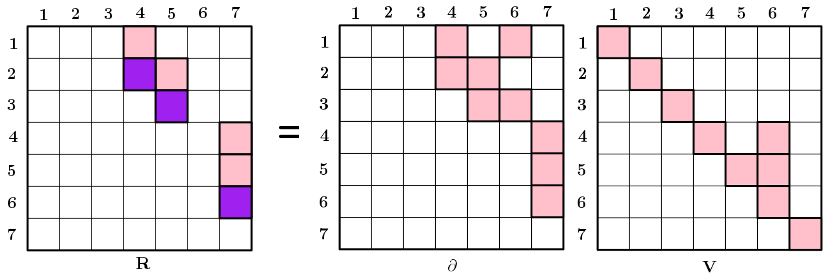}\\ 
  \caption{Figure illustrating reduction of the boundary matrix. $R$  
  is the reduced matrix, $\partial$ is the original boundary matrix  
  and $V$ is the matrix whose column $j$ encodes the columns of  
  $\partial$ that add up to give the column $j$ of $R$. The shaded  
  entries in the matrices denote $1$. All other entries are zero.} 
  \label{fig:reduction_proc} 
\end{center}  
\end{figure*} 
 
In this section, we illustrate the construction of filtration and  
boundary matrix, and the subsequent reduction of the boundary matrix  
through an example. We take a triangle as our input simplicial complex. 
 
\subsubsection{Filtration} 
\label{sec:filt_proc_example} 
We assume there is a function defined on the simplices that constitute  
the triangle. The function is such that it induces an ordering of the  
simplices, from the lowest to the highest dimension.  
Figure~\ref{fig:triangle_filtration} depicts such an ordering and the  
order in which the simplices appear in the filtration. We examine the  
filtration now, while simultaneously keeping track of the birth and  
death events.  
 
First the vertex $\langle 1 \rangle$, appears in the filtration. this  
corresponds to the birth of a $0$-dimensional hole, or an isolated  
object. Subsequently, vertices $\langle 2 \rangle$ and $\langle 3 \rangle$  
appear, in that order, taking the number of isolated objects to $3$. 
Thereafter, the edge $\langle 4 \rangle$ appears, merging  
the vertices $\langle 1 \rangle$ and $\langle 2 \rangle$ into a single  
component. We have a death of a $0$-dimensional hole here. According  
to \emph{elder rule} \citep[page 150,]{EdHa10}, the component that forms early lives, and the younger  
component dies. In other words: the edge $\langle 4 \rangle$ kills the  
vertex $\langle 2 \rangle$, and  
\{$\langle 2 \rangle, \langle 4 \rangle$\}  
form a birth-death persistence pair in the corresponding to a  
$0$-dimensional hole. Thereafter  
comes edge $\langle 5 \rangle$, merging the vertex $\langle 3 \rangle$  
with the connected component $\langle 1 \rangle$ (note that, since  
$\langle 2 \rangle$ is dead, the connected component resulting from  
the merger of $\langle 1 \rangle$ and $\langle 2 \rangle$ has the same  
index as $\langle 1 \rangle$). 
 
The first topological hole in $1$-dimension is born when the edge  
$\langle 6 \rangle$ appears in the filtration. This completes the  
boundary of the triangle, forming a loop. This $1$-dimensional hole dies when the   
triangle appears in the final phase of the filtration, patching up the loop that had  
formed due to the introduction of the edge $\langle 6 \rangle$. In  
other words, \{$\langle 6 \rangle, \langle 7 \rangle$\} form a  
birth-death persistence pair in $1$-dimension. 
 
In summary, there are three birth-death pairs in the filtration of the  
triangle : two corresponding to isolated components --- 
\{$\langle 2 \rangle$, $\langle 4 \rangle$\} \& 
\{$\langle 3 \rangle$,$\langle 5 \rangle$\}, and one corresponding to  
the loop --- \{$\langle 6 \rangle$, $\langle 7 \rangle$\}. 
 
From the point of view of need to construct the boundary matrix, we  
also enumerate the simplices and their boundaries here. The boundary  
of the edges constitutes of the vertices --- for example, the boundary of the edge  
$\langle 4 \rangle$ consists of the vertices $\langle 1 \rangle$ and  
$\langle 2 \rangle$. The boundary of the triangluar face $\langle 7  
\rangle$ consists of the edges $\langle 4 \rangle$,  
$\langle 5 \rangle$ and $\langle 6 \rangle$. 
 
\subsubsection{Boundary Matrix and its Reduction} 
 
We construct the boundary matrix, $\partial$ of the filtration of the  
triangle. Since the  
number of simplices in the filtration is $7$ (three vertices, three edges,  
and one triangle), the size of the boundary matrix is $7 \times 7$. If the  
simplex $i$ is in the boundary of the simplex $j$, the ($i,j$)-th  
element of the matrix is $1$. All other elements are $0$. We reduce  
the boundary matrix to $R$, using Algorithm~\ref{alg:matrix_reduce},  
to the form detailed in Section~\ref{sec:pers_hom}.  
Figure~\ref{fig:reduction_proc} illustrates this operation in the form  
of the matrix multiplication notation $R = \partial\cdot V$, where $R$  
and $\partial$ are the reduced matrix and the original boundary matrix  
respectively. One may verify that the shaded entries in the $\partial$  
matrix of Figure~\ref{fig:reduction_proc} indeed correspond to the  
simplices of the triangle, and its boundary  
(Figure~\ref{fig:triangle_filtration} and  
Section~\ref{sec:filt_proc_example}). 
 
\subsubsection{Persistence Diagrams} 
 
\begin{figure} 
\begin{center} 
  \includegraphics[width=8.5cm]{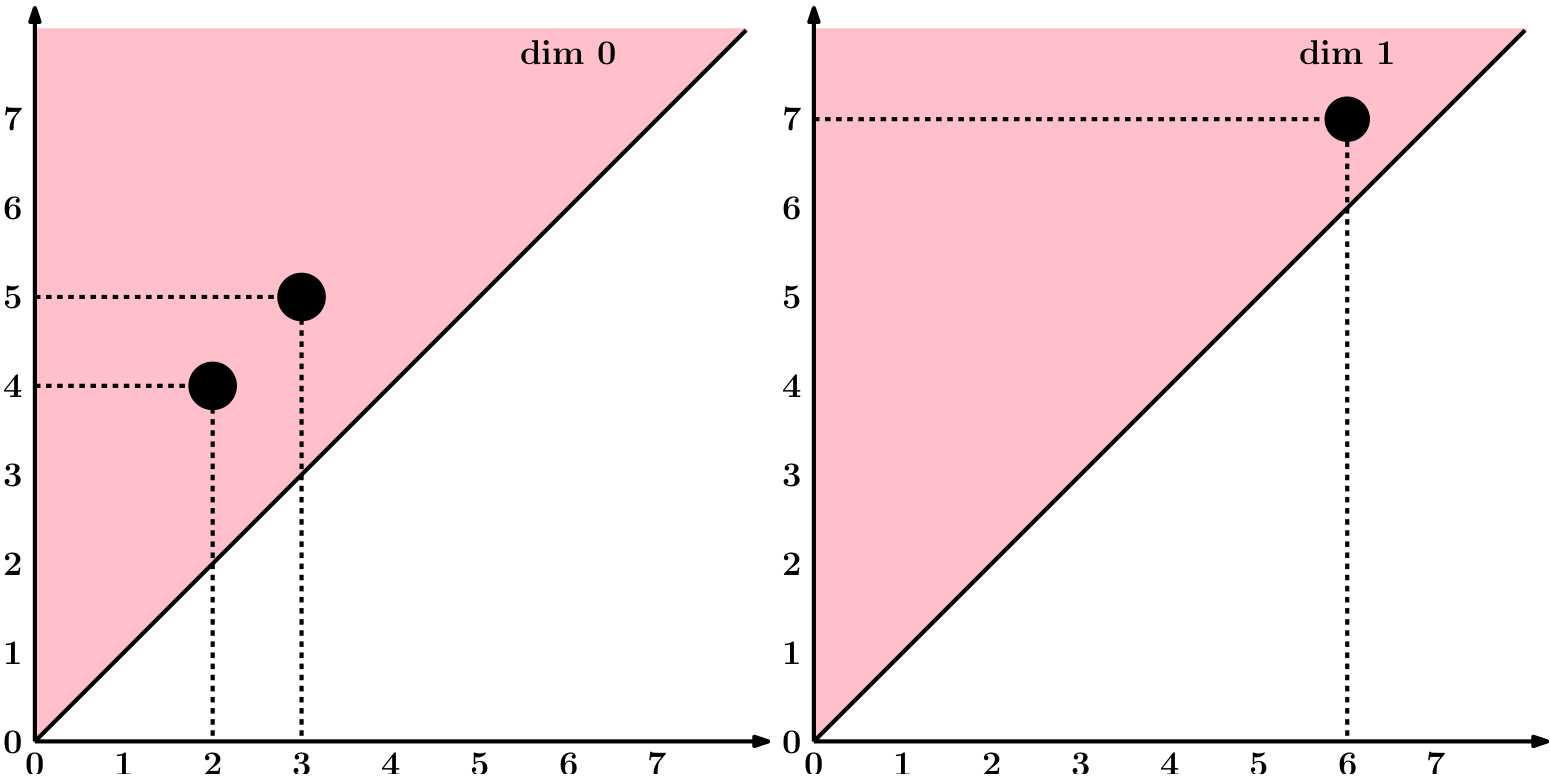}\\ 
  \caption{Persistence diagrams corresponding to the birth-death pairs  
  in the filtration of a triangle. Left panel presents the  
  $0$-dimensional persistence diagram, corresponding to birth-death or  
  merger events of isolated objects. Right panel presents the  
  $1$-dimensional persistence diagram, corresponding to birth-death  
  events of loops.} 
  \label{fig:triangle_diagrams} 
\end{center}  
\end{figure}

It is easy to read off the persistence diagrams from the reduced  
matrix $R$. In Figure~\ref{fig:reduction_proc}, the matrix $R$ is the  
reduced matrix corresponding to the persistence homology computation  
of the filtration of a triangle. The shaded entries in this matrix  
have a value $1$. Moreover, the entries in a deeper shade of  
pink denote the lowest row of a column whose entry is $1$. The lowest $1$'s indicate the  
birth-death persistence pair. In this example, the lowest $1$ entry  
indices correspond to the pairs $(i,j) \in \{(2,4), (3,5), (6,7)\}$.  
The first entry in the pair is the index of the simplex that gives birth  
to a topologcal hole. The second entry is the index of the simplex  
that kills that particular topological hole. One can verify that the  
indices of these pairs indeed correspond to the birth-death pairs, as  
enumerated in Section~\ref{sec:filt_proc_example}. 
Figure~\ref{fig:triangle_diagrams} presents the information of  
birth-death pairs in the filtration of a triangle in the form of  
persistence diagrams.  
 
\subsection{Points of Caution}  
The methods employed in this paper are perhaps on the more sophisticated end of the spectrum of cosmic web analyzes. It is therefore important to make sure that each step is rational and reliable, and the results are not contaminated by side-effects. There are indeed a few subtleties we need to keep in mind, and we list them here to avoid possible pit-falls. 
 
\begin{itemize} 
  \item{\sc Periodic tiling.} Instead of the $3$-dimensional Euclidean space as a model of the Universe, we use the $3$-torus, which has non-trivial homology, with Betti numbers $\beta_0 = 1$, $\beta_1 = 3$, $\beta_2 = 3$, and $\beta_3 = 1$. These numbers interfere with our statistical analysis of the topology of superlevel sets, but they are barely noticeable in the midst of usually thousands for ranks we observe. 
 
  \item{\sc Density Field Estimation.} Among the many possible density field estimators, we rely mostly on the DTFE as it naturally adapts to the particle distribution. It has the side-effect of forming high density spikes above particles that are completely and tightly surrounded by others. 
 
  \item{\sc Symbolic perturbation and superlevel sets.} We use the technical tools of symbolically perturbing the density values at the vertices, and retracting each superlevel set to the subcomplex above the threshold. Both techniques simplify the computation but have otherwise no effect. In particular, they give precisely the same persistence diagrams and intensity plots. 
 
  \item{\sc Intensity maps.} The averaged diagrams are meant to approximate the underlying distribution from which the persistence diagrams are sampled. We have no proof that they exist, other than the visual evidence that the diagrams for statistically similar particle distributions appear similar. We draw these plots by counting points within each square of a $100 \times 100$ grid, which implies that small shifts of the grid would give (slightly) different plots. \item{\sc Perturbations and stability.} Recalling the Stability Theorem for persistence diagrams \citep{CEH07}, we note an $\varepsilon$-perturbation of the density function can lead to the addition or removal of points at distance at most $\varepsilon$ from the horizontal axis. As a consequence, the intensity plots may change an arbitrary amount near the horizontal axis, but not at a distance larger than $\varepsilon$. 
 
\end{itemize} 
 
 
 
\section{Random Topology} 
 
\label{sec:random} 
 
 
 
Random processes play a crucial role in many aspects of life. In this paper, the analysis of random data provides a baseline for comparison, training the eye to pay attention to features that are not accidental, caused by inevitable random configurations in the data. We create this baseline by picking particles in space uniformly at random. 
 
\begin{figure*} 
  \includegraphics[width=18cm]{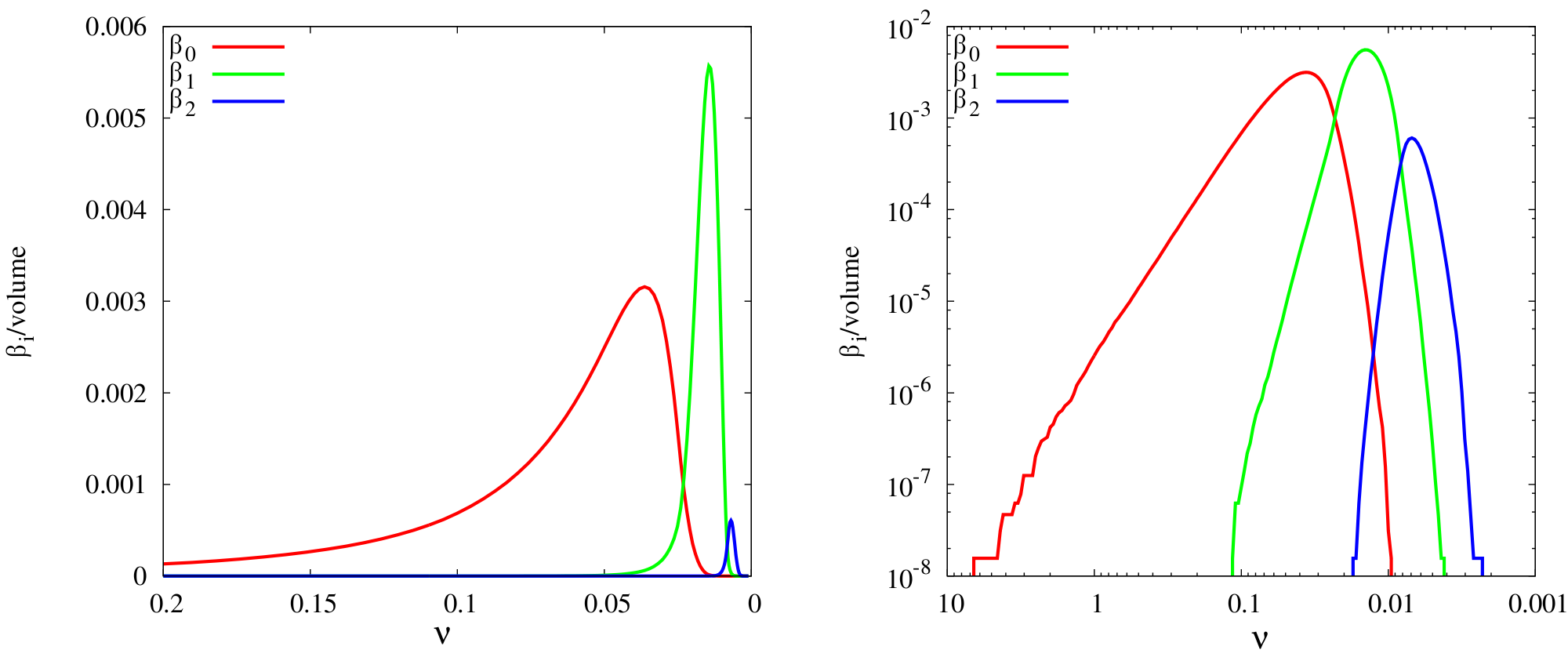}\\ 
  \caption{Left:  The three Betti numbers of the superlevel sets of a density function on the $3$-torus. The threshold, $\nu$, decreases from left to right, and the numbers of components, tunnels, and voids increase from bottom to top. Generating $500,000$ particles in a Poisson process, we get the density with the DTF estimator as explained in Section \ref{sec:fields}. The graphs are averaged over ten realizations. Right:  the same graphs in log-log scale.} 
\label{fig:poisson_dens_betti_ver1} 

  \includegraphics[width=18cm]{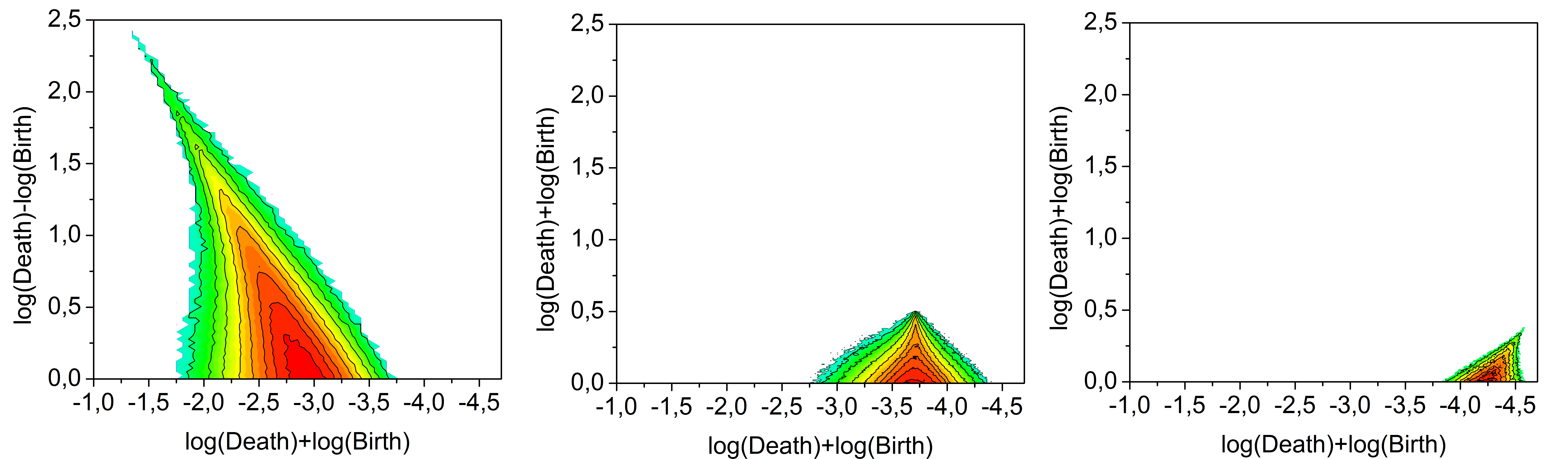}\\ 
  \caption{From left to right:  the intensity maps of the persistence diagrams for dimensions $0, 1, 2$, averaged over ten realizations. The sum of the logarithms of birth- plus death-values decreases from left to right, while the logarithm of the persistence increases from bottom to top.} 
  \label{fig:poisson_dgm_poisson_dens_alldim_avg_ver1} 
\end{figure*} 
 
\begin{figure} 
  \centering 
  \includegraphics[width=8cm]{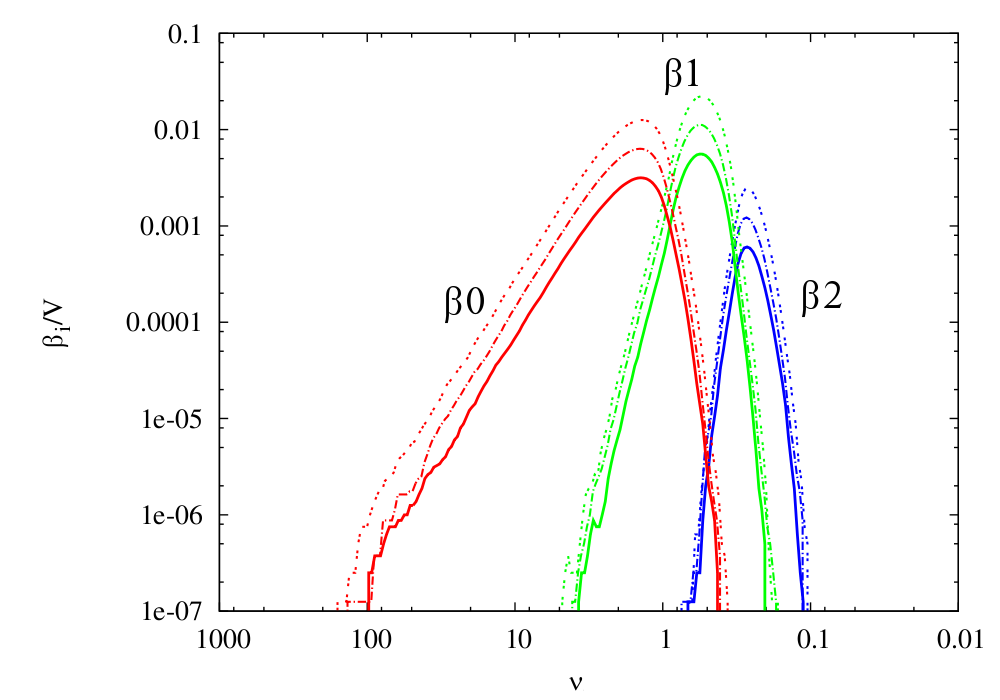} 
  \caption{Betti numbers for the uniform distribution with $\lambda$, the parameter of distribution varying. For each realization, the levelset values on the horizontal axis are normalized by the standard deviation of that particular realization. In the representation of normalized horizontal axis, the peak positions for realizations with different $\lambda$ are coincident. The lowest peak-amplitude corresponds to $\lambda=0.25$, followed by $\lambda = 0.125$ and $0.0625$ respectively.} 
 \label{fig:poisson_betti_scaled} 
\end{figure}

\subsection{Poisson Point Process}  
Recall that our model of the Universe is the $3$-dimensional cube with opposite faces glued to each other to create a periodic tiling of space. We call this the \emph{$3$-torus model}, denoting it by $\Mspace$. We choose the length unit such that each edge is $200 \Mpch$ long. Within this cube, we pick $n = 500,000$ particles in a \emph{Poisson point process}\footnote{The Poisson process depends on a density parameter that determines the expected number of particles. We slightly rig the process such that the number of chosen particles is precisely the expected number.}.  

For practical purposes, the particles are thus chosen from a \emph{uniform distribution} over the $3$-torus. This forms a  
reasonable approximation of a Poisson point process.

\subsection{Graphs of Betti Numbers}  
To get a feeling for the DTF estimator of the particle sample, we compute the Betti numbers of the superlevel sets. Writing $ f: \Mspace \to \Rspace$ for the estimated density function, we plot the $p$-the Betti number of $ f^{-1} [\nu, \infty)$ as a function of $\nu$, for $p = 0, 1, 2$. Drawing $\nu$ decreasing from left to right, we superimpose the graphs of the Betti numbers for ease of comparison; see Figure \ref{fig:poisson_dens_betti_ver1}. We observe that the graph of $\beta_0$ peaks first, at a density threshold of $\nu \approx 0.04$. As expected, the graph of $\beta_1$ peaks second, at $\nu \approx 0.015$, and the graph of $\beta_3$ peaks last, at $\nu \approx 0.007$. This suggests that loops are formed preferably by merging clusters into filaments, as opposed to growing horns that eventually meet. Similarly, voids are formed preferably by merging clusters and filaments into walls that eventually meet to completely enclose junks  
of empty space. In addition to the clear order, we observe that each of the three graphs has a clean shape with a clearly defined single mode. These properties are indicative of the data following a single, well-defined distribution.  
 
\subsection{Averaged Persistence Diagrams}  
As explained in Section~\ref{sec:topology}, persistence diagrams contain strictly more information than the graphs of the Betti numbers. Figure \ref{fig:poisson_dgm_poisson_dens_alldim_avg_ver1} shows the intensity maps of the density function, $ f: \Mspace \to \Rspace$, again in log-log scale. To compare these plots with the curves in Figure \ref{fig:poisson_dens_betti_ver1} on the right, we observe that the number of birth-death pairs, $(\nu_b, \nu_d)$, with $\nu_b \geq \nu > \nu_d$ gives the Betti numbers for the superlevel set for threshold $\nu$.\footnote{This relation may be violated   by the $8 = 1+3+3+1$ essential homology classes of the $3$-torus, which are not drawn in our diagrams. Their number is too small to be notices in our figures.} Since we draw the diagrams as intensity maps, we need to compare the integral over the V-shaped region anchored at the point $(\log \nu + \log \nu, 0)$ with the Betti number at $\log \nu$. When doing this,  
note that the horizontal axes in Figure \ref{fig:poisson_dens_betti_ver1} are labeled with values of $\nu$, while the horizontal axes in Figure \ref{fig:poisson_dgm_poisson_dens_alldim_avg_ver1} are labeled with twice the logarithm to the base $10$ of $\nu$. Similar to the graphs in Figure \ref{fig:poisson_dens_betti_ver1}, the diagrams of $\beta_0$, $\beta_1$, $\beta_2$ are ordered along the horizontal axis. In addition, the persistence, which we see as the vertical distance from the horizontal axis, decreases from $\beta_0$ to $\beta_1$, and then again from $\beta_1$ to $\beta_2$. This is a reflection of the DTF estimator, which tends to form spikes of high density at clusters. The height of these spikes is measured by the persistence of dots in the diagram of $\beta_0$, and these spikes are visible even after taking the logarithm of the density. In contrast, the depth of voids is measured by the persistence of the dots in the diagram of $\beta_2$, which is much milder, as seen in Figure \ref{fig:poisson_dgm_poisson_dens_alldim_avg_ver1}. Finally, we point out the characteristic ``pointed hat'' shape of the diagrams, and more specifically the sideways leaning tips for $\beta_0$ and $\beta_2$.These shapes seem related to heavily studied but difficult questions in percolation theory, and in particular to threshold phenomena, which are characteristic of this field. 
\begin{figure*} 
  \centering 
  \includegraphics[width=18cm]{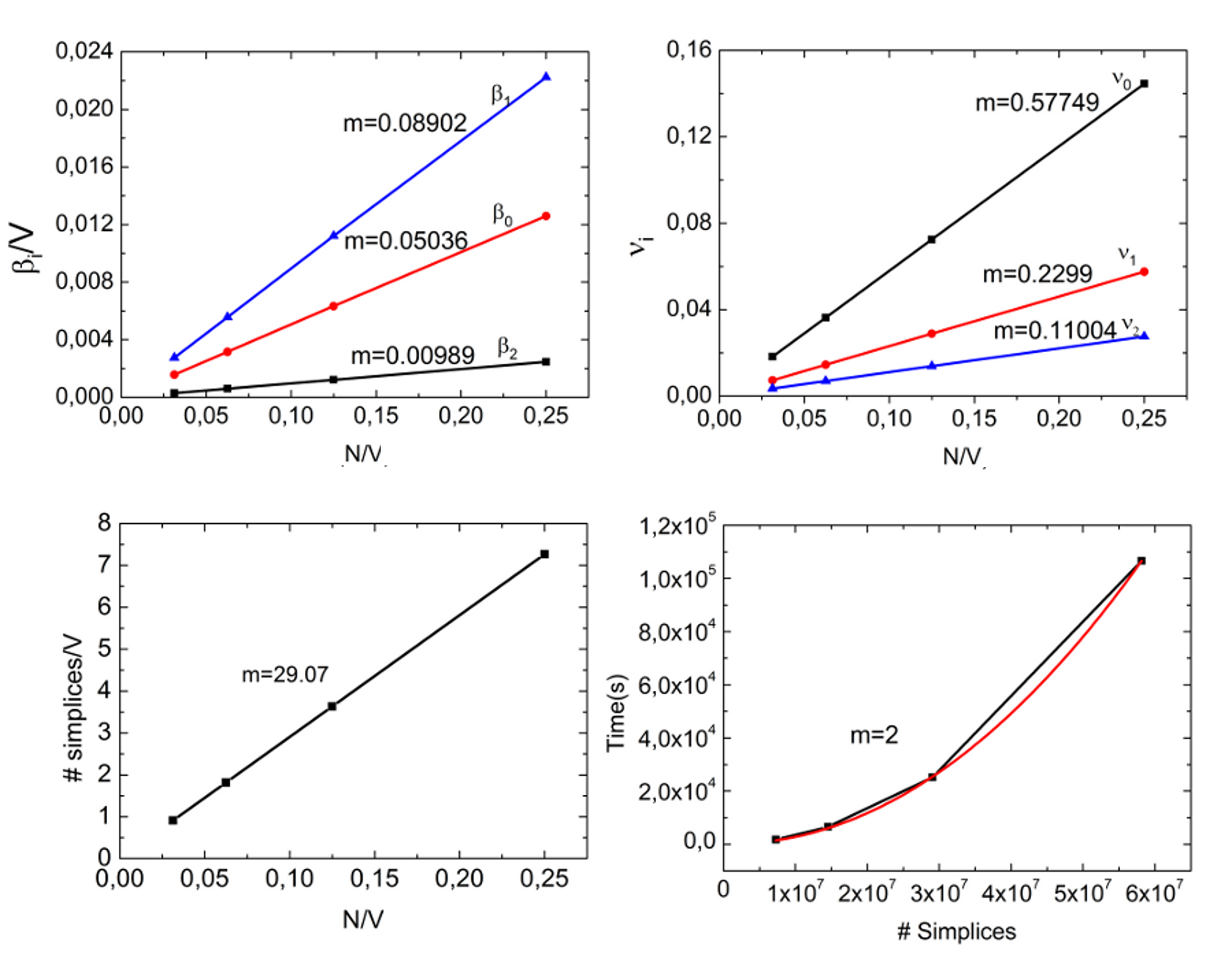} 
  \caption{Scaling relations for different quantities for the uniform distribution.  The quantities on the vertical axis (except the bottom-right panel) are per-unit-volume. Top-left panel : Scaling of peak-amplitude of $\beta_0$, $\beta_1$ and $\beta_2$, with number of particles per unit volume. Top-right: scaling of un-normalized (with the standard deviation) peak-position (on the horizontal axis), with the mean number of particles per unit volume. Bottom-left: scaling of number of simplices with $\lambda$. This can be translated to the scaling of number of simplices with the number of particles in the box. Bottom-right : scaling of time required to compute persistence with the number of simplices. The quantities on vertical axis scale linearly with quantities on horizontal axis in the top-left, top-right and bottom-left panel. The scaling in bottom-right panel has a power-law form. the slope of scaing is denoted by ``m'' in the first three panels. In the fourth panel, m is the index of the power-law distribution.} 
 \label{fig:poisson_ampl_nu_smplx_time_scaling} 
\end{figure*} 
 
\subsection{Scaling Relations of Poisson Topology} 
In order to probe the scaling relations of various quantities for the uniform distribution, we constuct realizations with different mean inter-particle separation $\lambda = 0.0625, 0.125$, and $ 0.25$. Keeping the box size same, this amounts to an increased number of particles with decreasing $\lambda$. Figure~\ref{fig:poisson_betti_scaled} plots the Betti numbers for realizations with different $\lambda$, where the horizontal axis (density threshold) is scaled with the variance of density. The $\beta_i$s for different $\lambda$'s have the same peak positions after scaling. Peak positions are well separated, denoting that topology is predominantly either ``meatball-like'', ``sponge-like'' or ``cheese-like'' at different values of $\nu$. $\beta_0$ peaks at $\nu \approx 1.8$, $\beta_1$ at $\nu \approx 0.6$ and $\beta_2$ at $\nu \approx 0.3$. The coincidence of peak-positions suggest a functional form of Betti numbers as a function of density threshold.  
 
In addition to the scaling of peak positions with normalized density threshold values, the peak amplitudes and the location of the peak of the $\beta_i$ also scale with $\lambda$. This scaling is shown in the top-left and top-right panels of Figure~\ref{fig:poisson_ampl_nu_smplx_time_scaling}. Peak amplitudes of $\beta_0$, $\beta_1$ and $\beta_2$ scale linearly with $\lambda$, with different slopes. $\beta_1$, the number of loops rises the sharpest with $\lambda$, with a slope of $m=0.08902$, followed by $\beta_0$ ($m=0.05036$) and $\beta_2$ ($m=0.00989$). The non-normalized (with respect to variance) peak positions on the horizontal axis also scale with $\lambda$. However, the trend is not the same as the peak amplitudes. In this domain, $\nu_0$, the peak position for $\beta_0$ rises the sharpest with increasign $\lambda$, with a slope of $m=0.57749$, followed by $\nu_1$ ($m=0.2299$) and $\nu_2$ ($m=0.11004$), in that order. The number of simplices per unit volume also scales linearly with $\lambda$ and has  
a slope of $m=29.07$. This is presented in the bottom-left panel of Figure~\ref{fig:poisson_ampl_nu_smplx_time_scaling}. The bottom-right panel of Figure~\ref{fig:poisson_ampl_nu_smplx_time_scaling} presents the scaling of time required to compute persistence for the uniform distribution with respect to the number of simplices in the tessellation. The time required to compute persistence seems to follw a power-law with respect to the number of simplices. We fit a power-law of the form $f(x)=\emph{a}x^{\emph{b}}$ where \emph{b} is the index of the power-law. The fitted curve to the data points gives the value of  the index $b = 2$. 
 
\begin{figure*} 
 \begin{center} 
  \includegraphics[width=18cm]{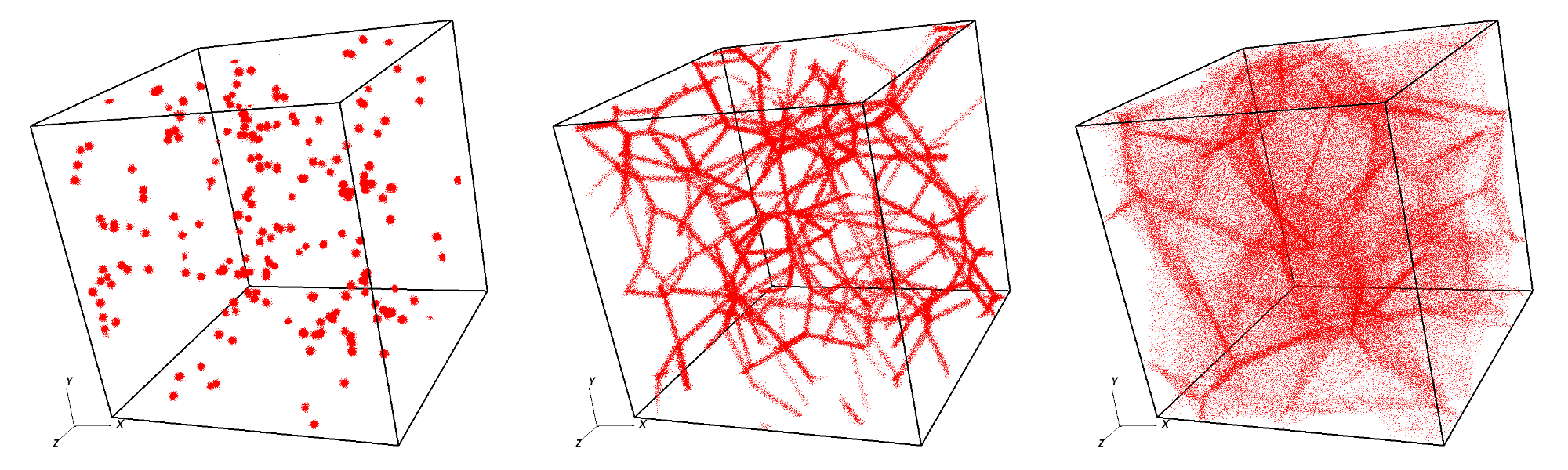}\\ 
 \end{center} 
 \caption{Top row, from left to right: particle distribution in the three pure Voronoi element models corresponding to clusters, filaments, and walls. Each data set consists of $262,144$ particles inside a periodic box of side length $200 \Mpch$. Bottom row, from left to right : density rendering of the same.} 
 \label{fig:vorelmvisuals} 
  \rotatebox{-90}{\includegraphics[height=18cm]{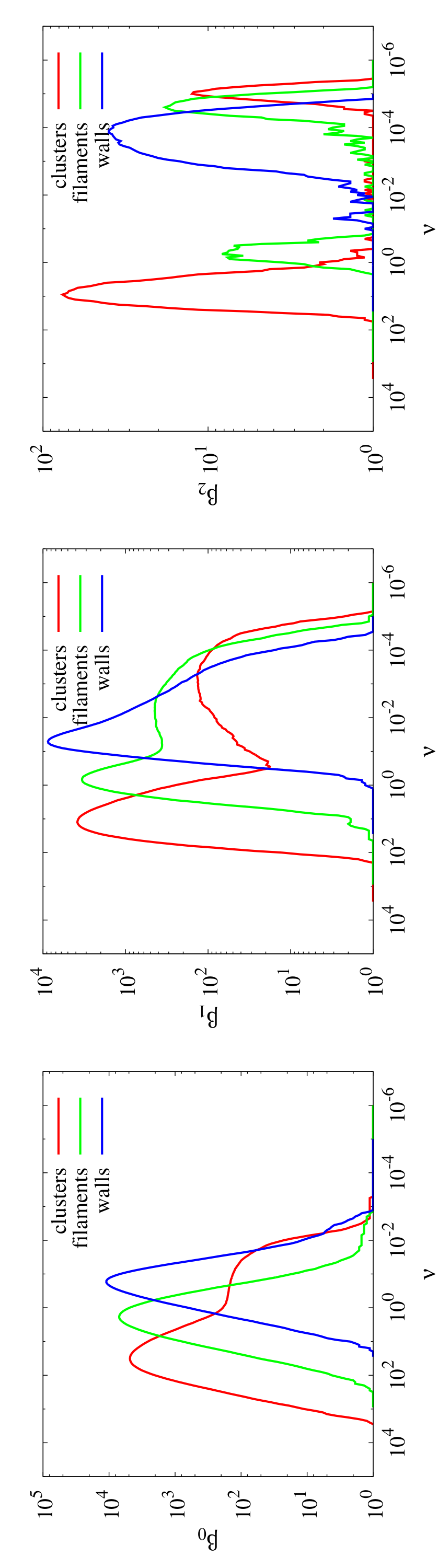}} 
  \caption{The Betti numbers of the superlevel sets of the density function for pure Voronoi element models as functions of the threshold. From left to right: $\beta_0$ $\beta_1$, $\beta_2$.} 
 \label{fig:voronoi_density_betti} 
\end{figure*} 
 
 
 
\section{Single-Scale Topology} 
\label{sec:single_scale} 
 
 
 
In this section, we consider a random process that produces particle distributions near the elements of a  
fixed Voronoi tessellation. While heuristic in nature, these distributions mimic the structural patterns observed  
in the Universe: the clusters, filaments, and walls in the Cosmic Web.  
 
In these Voronoi clustering models, a geometrically fixed  
Voronoi tessellation defined by a small set of nuclei is complemented with a heuristic prescription for  
the location of particles or model galaxies within the tessellation \citep{weyicke1989,weygaert1991,weygaert2007}.  
We distinguish two classes of Voronoi models: the \emph{pure Voronoi element models} 
and the \emph{Voronoi evolution models}. Both are obtained by moving an initially random  
distribution of $N$ particles toward the faces, lines, and nodes of the Voronoi tessellation. 
The pure Voronoi element models do this by a heuristic and user-specified mixture of projections onto the various geometric  
components of the tessellation. The Voronoi evolution models accomplish this via a gradual motion  
of the galaxies from their initial, random locations towards the boundaries of the cells. 
 
\begin{figure*}    
  \includegraphics[width=18cm]{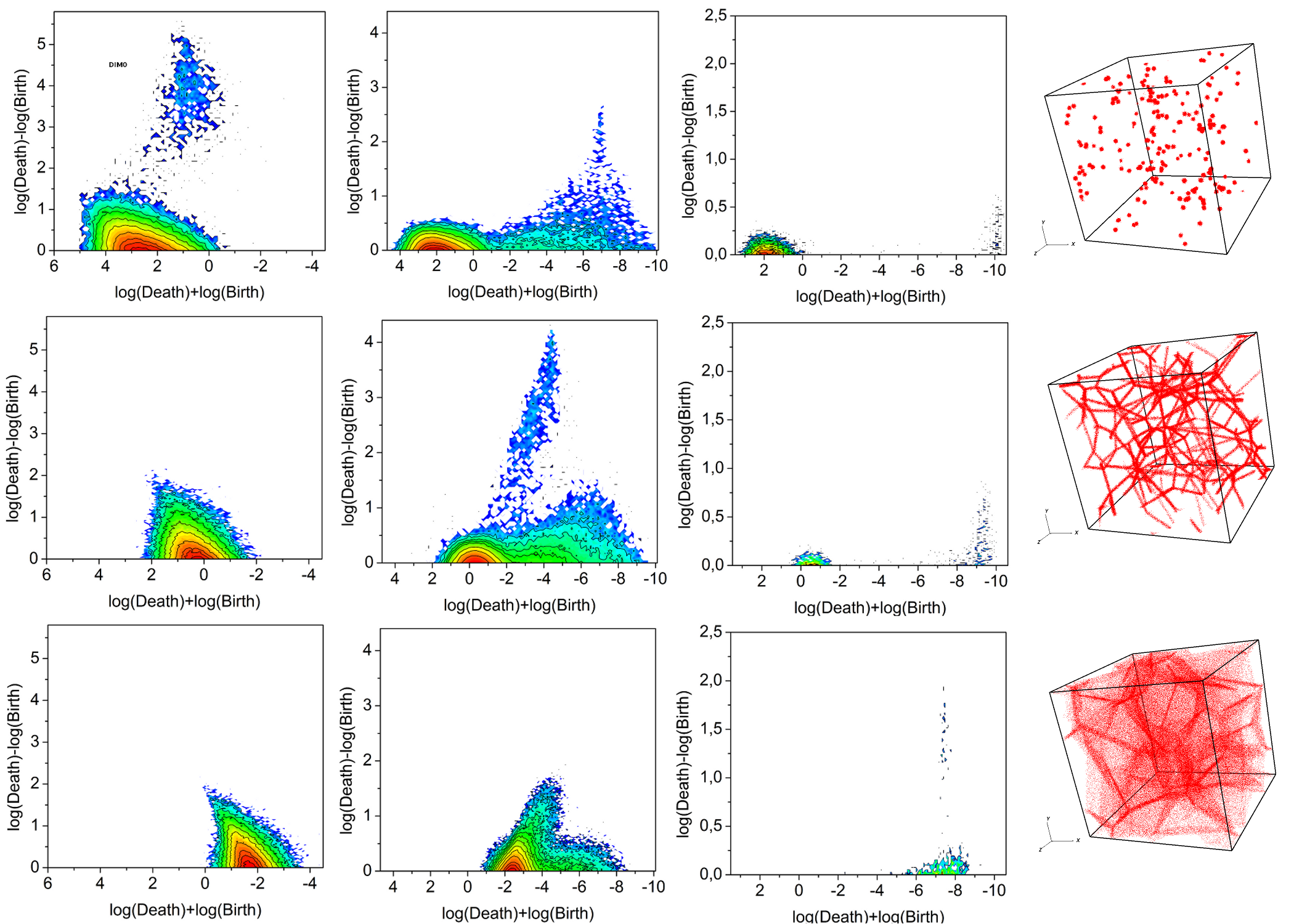} 
  \caption{The averaged persistence diagrams of the density functions for pure Voronoi element models. From top to bottom, we show the intensity for clusters, filaments, walls, and from left to right for classes of dimension $0, 1, 2$.} 
\label{fig:voronoi_dgm_dens} 
\end{figure*}

\subsection{Pure Voronoi Element Models}  
Recall that a Voronoi tessellation in space has four types of elements:  vertices, edges, faces, and cells. Constructing and fixing a diagram for only $32$ nuclei within a periodic box with sides of length $200 \Mpch$, we consider three random processes that generate particles near the vertices, edges, and faces. With each realization, we get $262,144$ particles distributed uniformly along and with a Gaussian spread of $1 \Mpch$ around the elements of the Voronoi skeleton; see Figure \ref{fig:vorelmvisuals}. The first process generates the particles in \emph{clusters} around the vertices, the second forms \emph{filaments} along the edges, and the third creates \emph{walls} following the faces. Since each process focuses on the elements of a single dimension, we call  
the resulting distributions \emph{pure Voronoi element models}.

\subsection{Graphs of Betti Numbers}  
We begin our analysis by looking at the Betti numbers of the superlevel sets of the estimated density field. Figure \ref{fig:voronoi_density_betti} shows the numbers as functions of the threshold. All results are averaged over eight realizations. The number of particles being the same in all three models, the average density in the clusters is higher than along the filaments, which in turn is higher than inside the walls. This is reflected by the graphs of $\beta_0$, in which the density threshold of the maximum is highest for clusters, between the extremes for filaments, and lowest for walls. The value at the maximum (the number of components) follows an opposite trend. 
 
Note the prominent shoulder in the graph of $\beta_0$ for clusters, which we do not see in the graphs for filaments and voids. The shoulder is a reflection of the merging process, which first consolidates the particles into clusters and second merges the clusters into one connected whole. We thus observe a transition from \emph{intra-cluster} to \emph{inter-cluster} merging, with the parameters of the shoulder identifying the density values at which this transition happens. In the filament and wall models, we have a single connected component as soon as all filaments and walls have been consolidated, which explains the absence of shoulders. Nevertheless, we observe a transition from a focus on intra- to inter-structural connectivity as a function of the density threshold. Indeed, the graph for $\beta_1$ has a shoulder, both for clusters and for filaments, and the explanation is similar.  
 
\begin{figure*} 
  \begin{center} 
    \includegraphics[width=18cm]{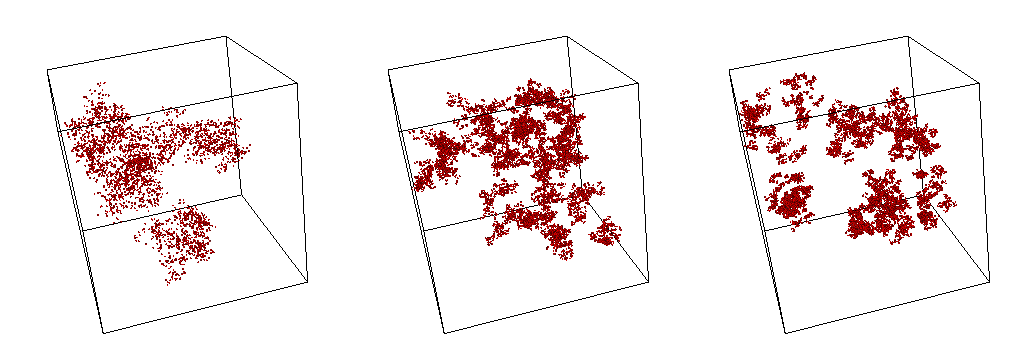}\\ 
  \end{center} 
  \caption{Particle distributions generated with the Soneira--Peebles process. Fixing the height to $\eta = 6$ and the branching factor to $\psi=9$, we vary the concentration from left to right as $\zeta=5.0, 7.0, 9.0$. There are $6^9$ particles in each dataset. The apparent low number of particles to the naked eyes is due to the high concentration factor. Zooming into a particular region shows similar structure at higher levels of hierarchy. Density rendering of the distribution is not feasible due to high concentration.} 
  \label{fig:soneiravisuals} 

  \rotatebox{-90}{\includegraphics[height=18cm]{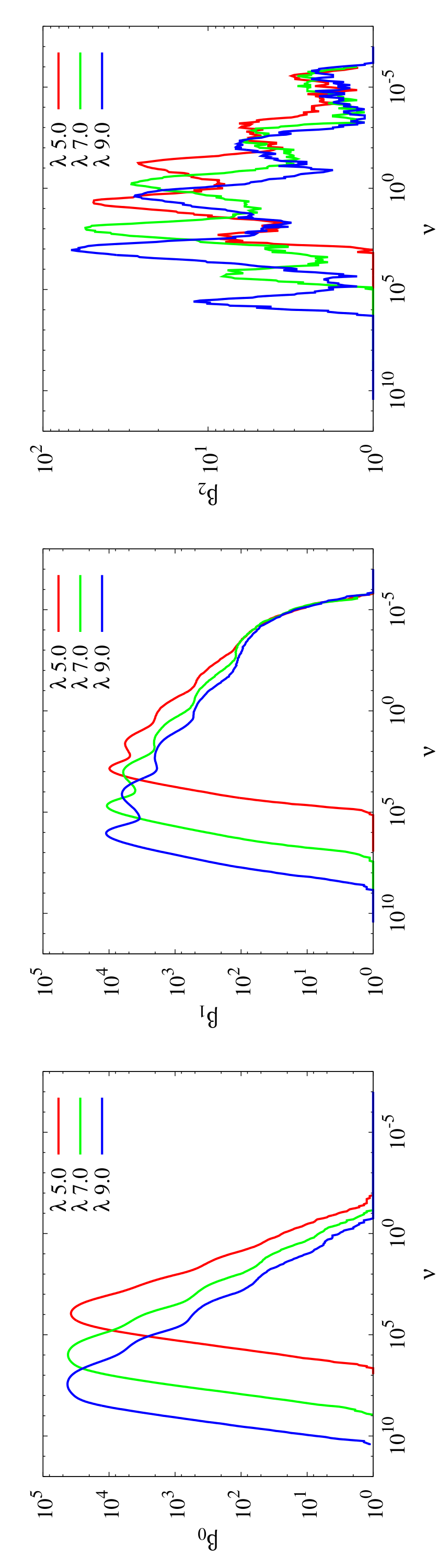}}\\ 
  \caption{From left to right:  the $0$-th, $1$-st, $2$-nd Betti numbers of the superlevel sets of the density function for the Soneira--Peebles particle distributions plotted on a logarithmic scale. Fixing the height to $\eta = 5$ and the branching factor to $\psi=9$, we vary the concentration as $\zeta=5.0, 7.0, 9.0$.} 
  \label{fig:betti_dens_soneira_log} 
\end{figure*} 
 
Continuing the trend, the graph for $\beta_2$ has two clear modes for clusters and filaments, and a hint of two modes for  
voids. A comparison with the intensity maps shows that this hint is a fluke, and while the separation into two populations of voids is real, it is not visible in the graph. 
More about this shortly. Returning to the graphs of $\beta_2$, we note that the left modes reflect the consolidation of the particles sampling the Voronoi elements, and the second modes reflect the filling up of the global, inter-structural voids. We see that the ordering of the left modes from clusters to filaments to walls is reversed for the right modes, remembering that $\beta_2$ for walls does not distinguish between the two populations and combines the left and right modes into one. The reversal of order makes geometric sense, since we are talking about the same voids in all three models, but these voids are shallower and appear at lower density values 
for clusters than for filaments, and more so for walls. 
 
\begin{figure*} 
  \includegraphics[width=18cm]{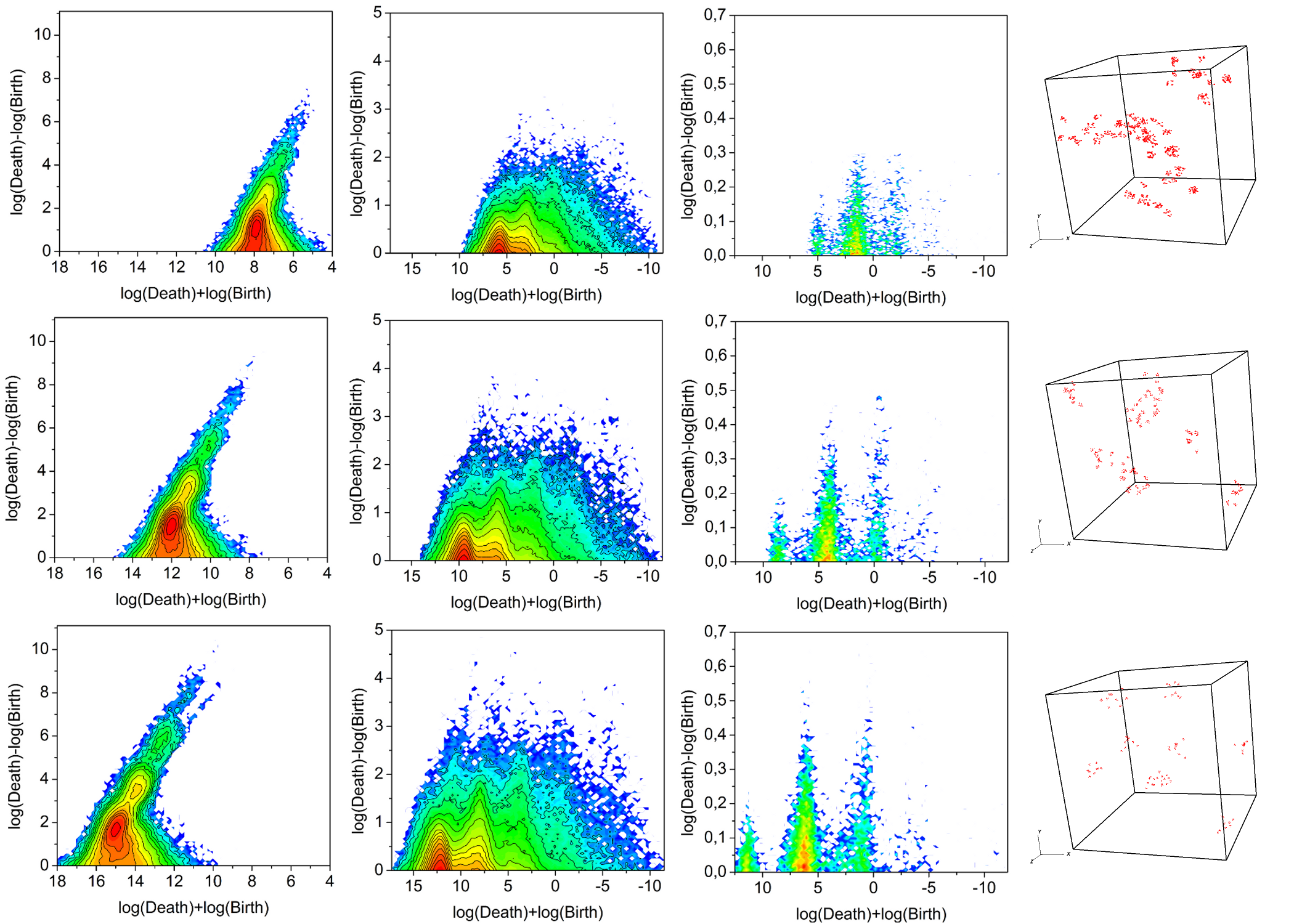}\\ 
  \caption{From left to right:  the $0$-, $1$-, $2$-dimensional averaged persistence diagrams of the density functions obtained from the Soneira--Peebles particle distributions. Fixing the height to $\eta = 5$ and the branching factor to $\psi=9$, we vary the concentration from top to bottom as $\zeta=5.0, 7.0, 9.0$.} 
  \label{fig:sp_dgm_dens_alldim} 
\end{figure*}

\subsection{Averaged Persistence Diagrams}  
The intensity maps for the pure Voronoi element models display features the graphs of the Betti numbers fail to capture, primarily because the maps distinguish between significant and insignificant features. For example, each realization of the filament model has a large number of tiny loops inside the filaments, but also a smaller number of larger loops that are carried by the filaments themselves. The $1$-st averaged persistence diagram distinguishes between these two populations.  
 
More generally, Figure \ref{fig:voronoi_dgm_dens} shows the intensity maps of all diagrams for all pure Voronoi element models: from top to bottom for clusters, filaments, voids, and from left to right for $\beta_0$, $\beta_1$, $\beta_2$. To a first degree of approximation, all diagrams contain a red and green high-intensity region and a blue low-intensity region. For the six diagrams in the upper-right triangle of the $3 \times 3$ array, the second region forms a \emph{island}, by  
which we mean a hill that is completely surrounded by a ring of zero intensity. As before, the high-intensity regions reflect the intra-structural  consolidation, while the low-intensity regions consist of points that represent large topological structures each carried by several clusters, filaments, or walls. For components, the two populations are clearly separated in the upper-left diagram for clusters.  
 
Similar to the graphs, we see no separation into the two populations of components in the diagrams for filaments and walls. For loops, the two populations are most clearly separated in the center diagram of Figure \ref{fig:voronoi_dgm_dens}, which plots the intensity for filaments. The two populations of loops are less clearly separated in the top diagram for clusters, and not at all separated in the bottom diagram for walls. Nevertheless, that map has a tongue suggesting a population of loops emigrating from the bulk. The geometric interpretation of this phenomenon is that the walls meet in filaments,  
which are therefore more densely sampled, so that global loops can form before the walls are completely filled.  
 
For voids, the separation into two populations is clearly visible in all three diagrams; see the third column in Figure \ref{fig:voronoi_dgm_dens}. Most noteworthy is the separation in the bottom diagram, in which the two populations have roughly the same mean age but very different persistence. Such populations cannot be separated by V-shapes, which is the reason the function of Betti numbers is oblivious to this difference.

 
\section{Multi-Scale Topology} 
 
\label{sec:multi_scale} 
 
 
One of the major features of the matter distribution at large scales is the presence of a hierarchy of substructures, with a large dynamic range in density and spatial scale. As a result, we see a multi-scale distribution, with interesting features at every scale.

\subsection{The Soneira--Peebles Model}  
Soneira--Peebles is a random point process with adjustable parameters that generates a fractal distribution of particles \citep{SP78}. Both the two-point correlation function and the fractal dimension of these particle sets are well understood analytically. The parameters can be chosen such that the correlation function of the particle distribution mimics that of the galaxies in the sky. It is used to explain the clustering statistics of the galaxy distribution, taking into account the fact that they display strong self-similarity. The placement of the particles is controlled by three parameters, each responsible for tuning a different aspect of the hierarchy:  
 
\begin{description} 
 \item[$\eta$:]    the \emph{height}, equal to the number of levels minus $1$; 
 \item[$\zeta$:]    the \emph{concentration}, equal to the ratio between consecutive radii; 
 \item[$\psi$:]    the \emph{branching factor}, equal to the number of children. 
\end{description} 
 
We start the construction with a unit sphere at level $0$, inside which we place the centers of $\psi$ level-$1$ spheres, each with radius ${1}/{\zeta}$ at random positions. The next iteration places the centers of $\psi$ level-$2$ spheres with radius $1/\zeta^2$ inside each level-$1$ sphere. We continue the process until we reach level $\eta$, with a total of $\psi^\eta$ spheres of radius $1/\zeta^\eta$. Finally, we pick a particle at the center of each level-$\eta$ sphere. Figure \ref{fig:soneiravisuals} shows three sample distributions with fixed height and branching factor, but with varying concentration. 
 
While this produces a pure \emph{singular} Soneira--Peebles model, 
it is common to superimpose a number of them to produce a somewhat 
more realistically looking model of the galaxy distribution. 
 
The Soneira--Peebles model involves a hierarchy of structures of 
varying densities and characteristic scales, with the higher level 
spheres corresponding to high density structures of small scale and 
the lower level spheres corresponding to low density structures of 
large scale.  As each sphere is constructed in the same way, the  
resulting point distribution is self-similar, forming a bounded fractal. 
The fractal geometry of a point set is often characterized by the 
fractal dimension, $D$, which is defined as 
\begin{align} 
     D  &=  \frac{\log N(r)}{\log (1/r)} , 
\end{align} 
in which $N(r)$ is the number of non-empty cells in a partition of constant cell size $r$.  If the Soneira--Peebles model would contain an infinite number of levels, the resulting point distribution would have 
fractal dimensions $D = \log \psi / \log \zeta$. 
One important manifestation of the self-similarity is reflected in the 
power-law two-point correlation function.  For $3$ dimensions, it is 
given by $r^{- \gamma}$, with 
\begin{align} 
   \gamma  &=  3 - \frac{\log \psi}{\log \zeta}, 
\end{align} 
for $1 / {\zeta^{\eta-1}} < r < 1$. 
The parameters $\psi$ and $\zeta$ may be adjusted such that they yield 
the desired value for the correlation slope, $\gamma$.

\subsection{Graphs of Betti Numbers}  
We study particle distributions generated with height $\eta = 6$, branching factor $\psi = 9$, and three different concentrations, $\zeta = 5.0, 7.0, 9.0$. For each parameter triplet, we average the results over eight realizations. Figure \ref{fig:betti_dens_soneira_log} shows the Betti numbers as functions of the threshold defining the superlevel set of the density functions defined by the particle distributions. Evidence of modularity\footnote{The term ``modularity'' is used for particle distributions with distinguishable levels in the hierarchy. A \emph{modular} distribution is hierarchical in nature.} 
is present in the curves for all chosen values of $\zeta$. For $\beta_0$, it manifests itself as ripples on the right side of the mode, when the number of components decreases after reaching a maximum. For $\beta_1$ and $\beta_2$, the evidence can be seen in the number of modes. Higher concentration results in a more clearly defined modular distribution. Indeed, the number of distinct ripples in the graphs for $\beta_0$ is the largest for $\zeta = 9.0$, while they are barely visible for $\zeta = 5.0$. 
 
The peak amplitude for $\beta_0$ is the same for all three distributions. The reason may be trivial, namely the fact that $\eta$ and $\psi$ are the same for all three experiments, implying that all data sets contain the same number of particles, namely $\psi^\eta = 9^5$. However, the peaks occur at different density thresholds, reflecting the varying local density of the distributions generated for different concentrations. Indeed, more concentrated particle distributions have higher density peaks, and as a result we see the mode at higher thresholds. We observe the same trend in the curves for $\beta_1$, and even for $\beta_2$, although the latter curves a much rougher, reflecting overall smaller numbers and more noise. The number of levels in the hierarchy is reflected in the number of peaks in the graph of $\beta_1$. We see five distinct peaks, while the number of levels in the distribution is six. It seems that the lowest level has too few components to be visible in the graphs. While the graphs of $\beta_2$  
are noisy, they also exhibit five distinct peaks. 
 
\begin{figure*} 
 \begin{center} 
   \includegraphics[width=18cm]{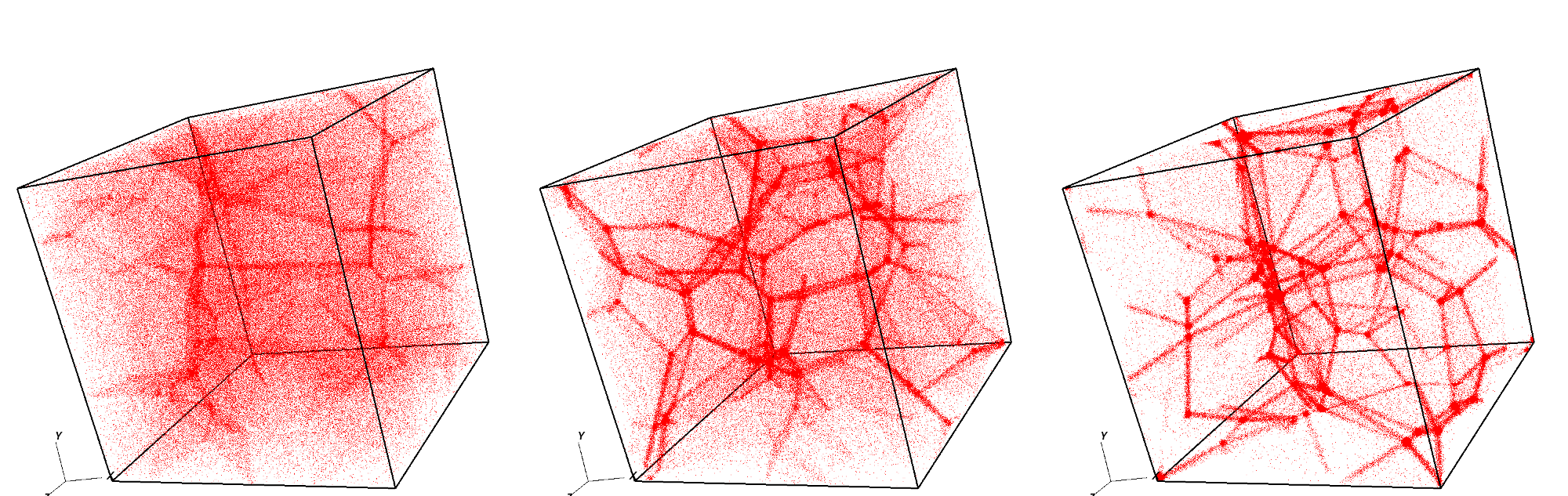}\\ 
 \end{center} 
 \caption{Snap-shots in the Voronoi evolution time-series. Top row, from left to right: particle distribution at the least, medium, most evolved stage. Bottom row, from left to right: volume rendering of the same.} 
 \label{fig:kinvisual} 

 \rotatebox{-90}{\includegraphics[height=18cm]{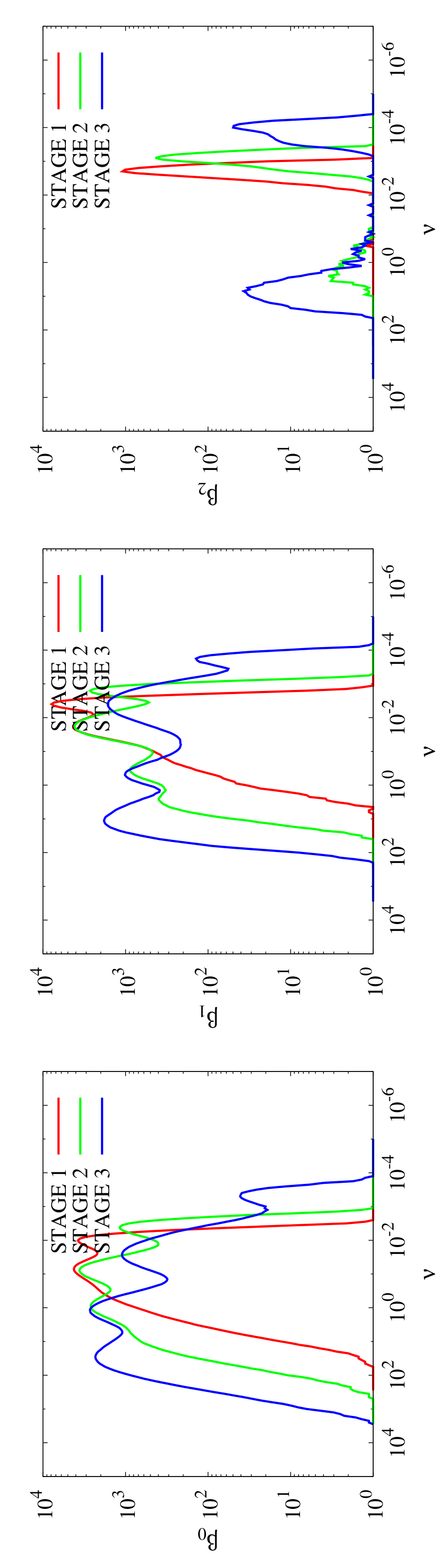}}\\ 
 \caption{The graphs of the Betti numbers computed for the density function of evolving particle distributions. From left to right: $\beta_0, \beta_1, \beta_2$ at different stages of the evolution. Stages 1, 2, 3 progress from least, to medium, to most evolved.} 
 \label{fig:betti_kin_dens} 
\end{figure*}

\subsection{Averaged Persistence Diagrams}  
The intensity maps of the particle distributions described above are shown in Figure \ref{fig:sp_dgm_dens_alldim}, for $\zeta = 5.0, 7.0, 9.0$ from top to bottom, and for dimension $0$, $1$, $2$ from left to right. The features in the diagrams show a clear transition as a function of the concentration, with evidence of modularity present in all diagrams. In particular, we notice \emph{hills} in the intensity, which we define as the neighborhood of a local maximum away from the horizontal axis. Note that these are different from \emph{tongues} in the intensity maps, which are local persistence maxima. 
 
Hills seem rather unusual features as the intensity usually decreases monotonically from bottom to top. For the $0$-dimensional diagrams, we notice an increase in the number of hills when we increase the concentration: there is a single hill for $\zeta = 5.0$, we see the hint of a second hill for $\zeta = 7.0$, and there are three clear hills for $\zeta = 9.0$. In words, we get progressively more evidence for modularity as the concentration increases, which is hardly surprising. Interestingly, the hills come in sequence, from bottom to top, so that later hills represent birth-death pairs of higher persistence. Furthermore, the intensity of the hills decreases from bottom to top. This makes sense since lower levels in the construction contain fewer clusters with lower persistence. Indeed, the highest level in the hierarchy generates the densest regions with the largest number of particles. Physically this means that many tiny clusters form at high density thresholds. These clusters are short lived, and as we go  
down from the highest level, a large number of tiny clusters merge together to form fewer but larger clusters. These larger clusters are of higher persistence and correspond to the low-intensity, high-persistence hills in the diagrams. The bias of the higher persistence hills towards the lower density values, is interesting, as it counters the higher density leaning pointy hat shape we see for the uniformly distributed particles in Figure \ref{fig:poisson_dgm_poisson_dens_alldim_avg_ver1}. 
 
Progressively better defined modularity as a function of increased concentration is also evident in the $1$-dimensional intensity maps. Here, we see tongues that correspond to the hills in the $0$-dimensional maps. Larger concentration corresponds to smaller filling rate, which results in bigger patches of empty space. This is reflected in the $2$-dimensional intensity maps, which record the information for the voids or empty regions: we see three or perhaps four grainy tongues, which are fuzzy for $\zeta = 5.0$, and progressively better defined for $\zeta = 7.0$ and $9.0$. 
 
 
 
\section{Dynamic Topology} 
 
\label{sec:dynamic} 
 
 
 
In this section, we consider particle distributions that change over time, similar to the matter in the Cosmos. Under the influence of gravity, the relatively uniform distribution at early epochs accumulates in the potential wells, evolving into galaxies and clusters. These clusters seem connected by filaments and walls. 
 
\begin{figure*} 
  \includegraphics[width=18cm]{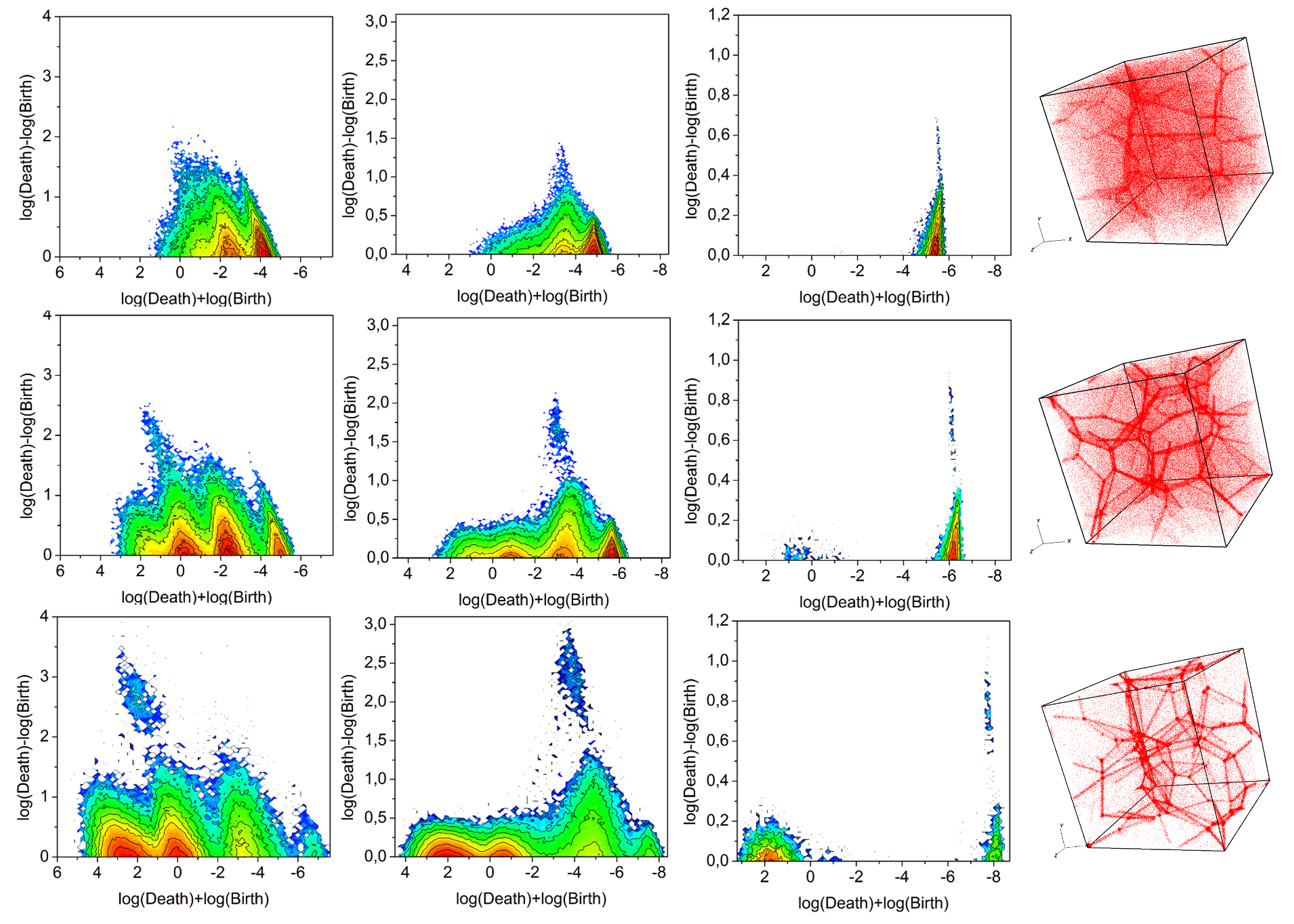}\\ 
  \caption{The averaged persistence diagrams of the density function for the Voronoi evolution models. From top to bottom, we show the intensity maps for least, medium, most evolved stages, and from left to right for classes of dimension $0, 1, 2$.} 
  \label{fig:dgm_dens_voronoi_evolution} 
\end{figure*}

\subsection{Voronoi Evolution Models}  
Starting with a random distribution of particles over the entire volume, \emph{Voronoi evolution} generates a time-series of  
particle distributions driven by slow drifts from higher- to lower-dimensional elements of an underlying Voronoi tessellation. They attempt to  
provide weblike galaxy distributions that reflect the outcome of realistic cosmic structure formation  
scenarios. They are based upon the notion that voids play a key organizational role in the development  
of structure and makes the Universe resemble a soapsud of expanding bubbles \citep{weyicke1989}.  
While the galaxies move away from the void centres, and stream out of the voids towards the sheets,  
filaments and clusters in the Voronoi network the fraction of galaxies in the voids (cell interior),  
the sheets (cell walls), filaments (wall edges) and clusters (vertices) is continuously changing  
and evolving. The details of the model realization depends on the time evolution specified by the  
particular Voronoi Evolution Model.  
 
Within the class of Voronoi Evolution Models the most representative and most frequently used are the  
{\it Voronoi kinematic models}. They form the idealized and asymptotic description of the outcome of  
hierarchical gravitational structure formation process, with single-sized voids forming around depressions in  
the primordial density field. This is translated into a scheme for the displacement of initially randomly  
distributed galaxies within the Voronoi skeleton. Within a void, the mean distance between galaxies increases  
uniformly in the course of time. When a galaxy tries to enter an adjacent cell, the velocity component  
perpendicular to the cell wall disappears. Thereafter, the galaxy continues to move within the wall, until it  
tries to enter the next cell; it then loses its velocity component towards that cell, so that the galaxy  
continues along a filament. Finally, it comes to rest in a node, as soon as it tries to enter a fourth  
neighbouring void.  
 
We have sampled the time-series at three moments in time, called \emph{stages}, and we show the results for these, emphasizing the continuous change that becomes visible by comparing the graphs and diagrams. To parametrize the stages, we keep track of the percentage of particles that lie in the interior of cells, faces, edges, and vertices of the Voronoi diagram; see Table \ref{tbl:kinstat} for the percentages at the chosen stages. Stage 1 is the least evolved particle distribution, with the highest percentage of particles in cells, while Stage 3 is the most evolved distribution, with the highest percentage at and around the vertices. 
 
Figure \ref{fig:kinvisual} shows the three stages as point clouds, going from left to right in the evolution. 
 
\begin{table} 
 \begin{center} 
  \begin{tabular}{r|rrrr} 
           &    cell  &  wall   & filament & cluster  \\ \hline 
   Stage 1 &  49.93\% & 38.52\% &  10.46\% &  1.08\%  \\ 
   Stage 2 &   5.03\% & 23.50\% &  41.26\% & 30.22\%  \\ 
   Stage 3 &   2.00\% & 14.72\% &  39.81\% & 43.47\% 
  \end{tabular} 
 \end{center} 
 \caption{The relative abundance of particles in each structural element throughout the course of evolution. Stage 1 is the least evolved, with almost half the particles residing in cells, while Stage 3 is the most evolved, with almost half the particles residing in clusters.} 
 \label{tbl:kinstat} 
\end{table}

\subsection{Graphs of Betti Numbers}  
We show the graphs of the Betti numbers as functions of the threshold defining the superlevel set in Figure \ref{fig:betti_kin_dens}. The graphs are significantly different from the ones we see for the single-scale Voronoi models in Figure \ref{fig:voronoi_density_betti}. The graphs for $\beta_0$ show a gradual transition from two to four peaks. The four peaks in Stage 3 reflect the fact that we have a non-trivial number of particles populating each of the four morphological features (clusters, filaments, walls, and the space in between) so that each population contributes its own peak to the graph. As before, the contributions are ordered from left to right as the clusters are densest and merge first, and so on. In contrast to Stage 3, Stage 1 has most particles near the walls and in the space between them, so that there are only two modes in the graph. 
 
A similar trend is also seen in the graphs for $\beta_1$. The particle distribution gets progressively more segregated into the morphological features, each with its own density, which explains the clear four peaks we see for Stage 3. The signal we get from $\beta_2$ is different while consistent with our explanation. We see one peak at Stage 1 and two peaks each at Stages 2 and 3. As before, the difference is between intra- and inter-structural consolidation, and the second one barely exists in Stage 1, at which time a large fraction of the particles populates the space between the walls.

\subsection{Averaged Persistence Diagrams}  
The evolution of the particle distribution is well visible in the averaged persistence diagrams, which we show separated for the three stages and the different dimensions in Figure \ref{fig:dgm_dens_voronoi_evolution}. Each intensity map is obtained by averaging eight realizations. While the evolution flows from top to bottom, we show the results for the components, loops, and voids from left to right. 
 
Recall that Stage 1 is dominated by particles distribution near the walls and in the space between the walls. Corresponding to the two peaks of the graph for $\beta_0$, we see two tongues in the upper-left intensity map, which shows the averaged diagram for the components. Note that the tongue with higher intensity is on the right hand side, where the mean age is larger. Indeed, the density in the space between the walls is smaller while the population there is larger. Two things happen when we go from Stage 1 to Stage 3: the number of tongues increases to four, and the order of the tongues by intensity is reversed. Similar to the graphs of the Betti numbers, we contribute the four tongues at Stage 3 to a clean segregation of the particles into four morphological elements. The change in order is of course due to the trend to put larger populations of particles into lower-dimensional elements. We point out that the two phenomena are related to each other. The percentage of particles in a morphological  
component dictates its average density, which, in turn, drives the segregation. 
 
Note also the formation of a low-intensity island in the intensity maps, which breaks from the bulk and migrates towards high persistence values as the model evolves. We see this phenomenon in all three dimensions. The underlying reason is that the cells deplete of particles during the evolution, and the created empty space favors the appearance of inter-structural consolidation --- a manifestation of the structure of the underlying Voronoi skeleton itself ---  which is represented by the islands. 
 
 
 
\section{Summary and Discussion} 
 
\label{sec:summary} 
 
In this study we have described and introduced a multiscale topological description  
of the Megaparsec cosmic matter distribution. Emanating from algebraic topology and  
Morse theory, Betti numbers and topological persistence \citep{EdHa10} offer a powerful  
means of describing the rich connectivity structure the Cosmic Web. They represent a major  
extension and deepening of the cosmologically familiar topological genus measure, and the  
related geometric Minkowski functionals, and are more tuned towards the analysis of the  
complex spatial weblike and multiscale arrangement of matter and galaxies in the cosmic web.  
 
With the intention to use Betti numbers and topological persistence to  
analyze the large scale galaxy and matter distribution, this study is a first  
in a series of publications towards this goal. The present paper has three  
aims. The first is the presentation of the mathematical foundation. The second aim is  
the presentation and discussion of the algorithms for computing Betti  
numbers and persistence diagrams for a given spatial distribution of points, galaxies  
or simulation particles or objects. The third aspect concerns a systematic exploration  
of the imprint of different weblike morphologies and different levels and patterns of  
multiscale clustering in the computed Betti numbers and persistence diagrams.   
 
The specific formalism from algebraic topology that we use to describe the  
topological structure of the cosmic mass distribution is  
known as \emph{homology}. This is the mathematical formalism for the quantitative  
characterization of the connectivity of space by assessing the presence and identity  
of holes in a topological space, usually via the descripton of the boundaries of these  
holes. For a given superlevel set of the cosmic density field, Betti numbers are  
topological invariants that quantify the presence of isolated islands, tunnels  
and cavities or enclosed void regions. They have a direct relation to the  
more conventionally known Euler characteristic, but extend its description of the  
global topology as it entails, for a 3D density field, 3 independent numbers.  
 
The details of the spatial connections between the various topological spaces,   
holes or boundaries leads to the concept of \emph{persistence} \citep{EdHa10}.  
Persistence formalizes topology as a hierarchical concept, and represents a  
major extension of the available topological machinery to characterize  
the cosmic mass distribution. Using the singularity structure of a density field,  
and the realization that the topology of a space is entirely --- and only --- determined by its critical points, persistence maps the changes in topology  
that occur at these points. By identifying the formation of new topological  
features and the destruction of existing features at each of the critical  
points, persistence produces a quantitative characterization of the multiscale  
topological structure of the cosmic web in terms of the \emph{birth} and  
\emph{death} of topological features. These are summarized in a \emph{persistence diagram},  
one for each class of $p$-dimensional topological holes \citep{EdHa10}. In our study we  
introduce and use \emph{persistence intensity maps}, continuous maps representing an empirical  
probabilistic description of persistence diagrams.  
 
As for the computational formalism, a major complication enters via the fact of having to  
deal with a discrete point sample of an underlying density field, while the underlying  
theory is based on a continuous density field. To this end, we translate the sample  
point distribution into a piecewise linear continuous density field reconstruction  
by means of the DTFE algorithm \citep{Sch00}. On the basis of its  
representation on the corresponding Delaunay tessellation, the boundary  
relations between its simplices --- tetrahedral cells, triangular faces, edges and  
vertices --- are transformed into a boundary matrix, using the density value estimates  
at the vertices to evaluate which simplices belong to the density superlevel set  
at a given density threshold level. Reduction of the boundary matrix translates  
directly into the set of corresponding pairs of birth-death pairs of topological  
features, or merger events of separate features, in the persistence diagrams.  
 
An important aspect of the present study is the development of an understanding of  
the impact of various key characteristics of the cosmic web on the statistics of  
Betti numbers and persistence. This forms a necessary step in the application of  
these to the observed reality of galaxy surveys or fully fledged cosmological N-body  
simulations. Because analytical expressions for Betti numbers and persistence  
do not exist for any cosmologically representative situation, not even for  
Gaussian random fields (but see \cite{feldbruggeBachelor}), we use a set of heuristic  
models of spatial clustering to investigate the influence of a range of  
morphological features on topological measures.  
 
The first reference template is that of Betti numbers and persistence for uniform distributions sampled from a Poisson point process. The topological imprint of such random featureless distributions  
also informs us of the contribution by shot noise in generic features sampled by  
discrete points. Subsequently, we invoke a set of Voronoi clustering  
models \citep{weyicke1989,weygaert2002} to study the topology measures in a range of  
weblike galaxy distributions, each differing in prominence of wall-like planes,  
elongated filaments, cluster nodes or underdense void regions. The influence of the  
multiscale mass distribution, the result of the hierarchical buildup of cosmic  
structure, is explored on the basis of the fractal-like Soneira--Peebles model \citep{SP78}.  
 
We find that the dominant presence of the various morphological features in  
the Voronoi clustering models is clearly reflected in the persistence intensity  
maps. The presence of prominent filamentary structures is particularly strongly manifest  
in the $1$-dimensional persistence diagrams in the form of high persistence cloud.  
A wall-like distribution, which in Voronoi models goes along with the presence of large voids,  
induces isolated high persistence clouds in the $2$-dimensional persistence diagrams. In the  
situation wherein most particles are concentrated in and around cluster nodes, we find  
high persistence clouds in $0$-dimensional persistence maps. However, in all  
situations we find that the discrete nature of the point distributions in the  
various components of the cosmic web generates a prominent and extended  
base of low persistence features, that is: features of a low topological significance.  
In the situation of a multiscale matter distribution, modelled by the fractal  
Soneira--Peebles model, we find as well a clear manifestation of the clustering  
properties in the persistence maps and the Betti numbers. Different levels in  
a nested hierarchy of point clusters reflect themselves in the presence of  
a sequence of concentrations in a persistence diagram.  
 
In two upcoming studies, we direct the presented topological measures to  
more realistic cosmological mass distributions. The topology of the dark matter 
distribution will be addressed in the context of a few large N-body simulations 
of cosmic structure formation. The relation between the topological  
characteristics of the dark matter field and the corresponding dark halo distribution  
is adressed in the same study. It will highlight the expected impact of halo bias on the  
recorded topological measures, as halos of different masses and assembly epoch trace  
different parts of the cosmic web. In \cite{nevenzeel2013} the first results of this study have been presented,
pertaining to the topology of the dark matter distribution in cosmologies with
a varying nature of dark energy. Also within the context of a large cosmological simulation, a second study combines the dark matter and dark halo topology with that  
of gas that settled in the cosmic web, and galaxies that emerged in different  
cosmic environments. 
 
The application of our topological toolbox to the observational reality offers  
substantial challenges. For the analysis of galaxy surveys we have to deal with 
measurement errors, selection effects, systematic biases and errors, substantial shot  
noise effects, and a range of other practical effects. An important exercise towards this will be assessing the  
impact of such effects on the topological measurements on the basis of mock galaxy catalogues extracted  
from standard N-body simulations like the Millennium and Millenium-2 simulations.  
 
Our principal motivation is the understanding of the complex and intricate  
structure of the cosmic web, the earliest emerging and largest nontrivial structure in the  
Universe. Nonetheless, persistent topology also opens up a new perspective on the structure  
of the primordial density field. Homology and persistent topology of Gaussian random  
fields has been the subject of several insightful studies \citep{Adl10}. In  
the cosmological context it may provide a rich new characterization of the spatial  
structure and connectivity in the primordial density field. One issue of high interest  
is whether the sensitivity of persistence diagrams to slight deviations from  
Gaussianity, a direct manifestation of inflationary physics, is considerably  
larger than recorded with more conventional measures. Following a first  
numerical assessment of Betti numbers in Gaussian fields \citep{PPC13}, it forms the  
rationale behind our first theoretical paper on the subject \citep{feldbruggeBachelor}.  
Amongst others, the latter has established approximate analytical expressions describing the  
behaviour of Betti numbers in two dimensional Gaussian random fields, which may be used to  
allow the detection of non-Gaussian deviations. In two major studies \citep{pranav2016a,pranav2016b},  
we present an extensive numerical study of the topological analysis of Gaussian random fields.  
These studies present and investigate the Betti numbers, Minkowski functionals and  
persistence diagrams for Gaussian random field realizations, comprising a range of  
different power spectra, with the purpose of identifying systematic trends.  
 
In summary, while it has lasted some time before powerful concepts from the abstract  
mathematical branch of algebraic topology have become available for practical applications, major developments in  
computational topology and geometry over the past years have made them accessible for applications in  
a wide range of scientific disciplines. In turn, these were enabled by the surge in necessary computational  
resources. In this study we have demonstrated the potential for a significantly more versatile  
topological analysis of the cosmic mass distribution. It has paved the path of interesting applications towards  
a vast range of cosmologically significant issues, and opens up the possibility of answering several  
questions on the basis of the new perspectives offered by persistent topology.  
 
 
 
\section*{Acknowledgements} 
We are grateful to Bob Eldering and Nico Kruithof for important  
and contributions and discussions at the start of this project. Discussions  
with and insights obtained from Job Feldbrugge, Matti van Engelen and  
Keimpe Nevelzeel have been of key significance in shaping this paper, and  
are gratefully acknowledged. We are also very grateful to Robert Adler for  
insightful comments on this manuscript.  
 
Part of this work has been supported by the 7th Framework Programme for Research  
of the European Commission, under FET-Open grant number 255827 (CGL Computational  
Geometry Learning), and ERC advanced grant, URSAT (Understanding Random Systems via Algebraic Topology) number 320422. 
 
 
 
\bigskip 
 
 
 
\bibliographystyle{mn2e} 
\bibliography{../../references.bib} 

\begin{thebibliography}{}

\bibitem[\protect\citeauthoryear{{Abel}, {Hahn} \& {Kaehler}}{{Abel}
  et~al.}{2012}]{abel11}
{Abel} T.,  {Hahn} O.,    {Kaehler} R.,  2012, MNRAS, 427, 61

\bibitem[\protect\citeauthoryear{Adler \& Taylor}{Adler \&
  Taylor}{2010}]{Adl10}
Adler R.,  Taylor J.,  2010, Random Fields and Geometry.
Springer Monographs in Mathematics, Springer

\bibitem[\protect\citeauthoryear{{Arag{\'o}n-Calvo}, {Jones}, {van de Weygaert}
  \& {van der Hulst}}{{Arag{\'o}n-Calvo} et~al.}{2007a}]{aragon07a}
{Arag{\'o}n-Calvo} M.~A.,  {Jones} B.~J.~T.,  {van de Weygaert} R.,    {van der
  Hulst} J.~M.,  2007a, Astrophys. J. Lett., 655, L5

\bibitem[\protect\citeauthoryear{{Arag{\'o}n-Calvo}, {Jones}, {van de Weygaert}
  \& {van der Hulst}}{{Arag{\'o}n-Calvo} et~al.}{2007b}]{aragon07b}
{Arag{\'o}n-Calvo} M.~A.,  {Jones} B.~J.~T.,  {van de Weygaert} R.,    {van der
  Hulst} J.~M.,  2007b, Astron. Astrophys., 474, 315

\bibitem[\protect\citeauthoryear{Aragon-Calvo \& {Szalay}}{Aragon-Calvo \&
  {Szalay}}{2013}]{aragon2013}
Aragon-Calvo M.~A.,  {Szalay} A.~S.,  2013, MNRAS, 428, 3409

\bibitem[\protect\citeauthoryear{{Arag{\'o}n-Calvo}, {van de Weygaert} \&
  {Jones}}{{Arag{\'o}n-Calvo} et~al.}{2010}]{aragon2010}
{Arag{\'o}n-Calvo} M.~A.,  {van de Weygaert} R.,    {Jones} B.~J.~T.,  2010,
  MNRAS, 408, 2163

\bibitem[\protect\citeauthoryear{{Bauer}, {Kerber} \& {Reininghaus}}{{Bauer}
  et~al.}{2013}]{BRK13}
{Bauer} U.,  {Kerber} M.,    {Reininghaus} J.,  2013, ArXiv e-prints

\bibitem[\protect\citeauthoryear{{Behroozi}, {Wechsler}, {Wu}, {Busha},
  {Klypin} \& {Primack}}{{Behroozi} et~al.}{2013}]{BWWBKP2013}
{Behroozi} P.~S.,  {Wechsler} R.~H.,  {Wu} H.-Y.,  {Busha} M.~T.,  {Klypin}
  A.~A.,    {Primack} J.~R.,  2013, \apj, 763, 18

\bibitem[\protect\citeauthoryear{Bendich, Edelsbrunner \& Kerber}{Bendich
  et~al.}{2010}]{BEK10}
Bendich P.,  Edelsbrunner H.,    Kerber M.,  2010, IEEE Transactions on
  Visualization and Computer Graphics, 16, 1251

\bibitem[\protect\citeauthoryear{Betti}{Betti}{1871}]{Bet71}
Betti E.,  1871, Ann.\ Mat.\ Pura Appl., 2, 140

\bibitem[\protect\citeauthoryear{{Bond}, {Kofman} \& {Pogosyan}}{{Bond}
  et~al.}{1996}]{BKP96}
{Bond} J.~R.,  {Kofman} L.,    {Pogosyan} D.,  1996, Nature, 380, 603

\bibitem[\protect\citeauthoryear{{Bond}, {Strauss} \& {Cen}}{{Bond}
  et~al.}{2010}]{bond10}
{Bond} N.~A.,  {Strauss} M.~A.,    {Cen} R.,  2010, MNRAS, 409, 156

\bibitem[\protect\citeauthoryear{Carlsson}{Carlsson}{2009}]{Carl09}
Carlsson G.,  2009, Bulletin of the American Mathematical Society, 46, 255

\bibitem[\protect\citeauthoryear{Carlsson \& Zomorodian}{Carlsson \&
  Zomorodian}{2009}]{CarlZom09}
Carlsson G.,  Zomorodian A.,  2009, Discrete \& Computational Geometry, 42, 71

\bibitem[\protect\citeauthoryear{Carlsson, Zomorodian, Collins \&
  Guibas}{Carlsson et~al.}{2005}]{Carl05}
Carlsson G.,  Zomorodian A.,  Collins A.,    Guibas L.~J.,  2005, International
  Journal of Shape Modeling, 11, 149

\bibitem[\protect\citeauthoryear{{Cautun}, {van de Weygaert} \&
  {Jones}}{{Cautun} et~al.}{2013}]{Nexus}
{Cautun} M.,  {van de Weygaert} R.,    {Jones} B.~J.~T.,  2013, \mnras, 429,
  1286

\bibitem[\protect\citeauthoryear{{Cautun}, {van de Weygaert}, {Jones} \&
  {Frenk}}{{Cautun} et~al.}{2014}]{CWJF14}
{Cautun} M.,  {van de Weygaert} R.,  {Jones} B.~J.~T.,    {Frenk} C.~S.,  2014,
  \mnras, 441, 2923

\bibitem[\protect\citeauthoryear{{Cautun} \& {van de Weygaert}}{{Cautun} \&
  {van de Weygaert}}{2011}]{cautun2011}
{Cautun} M.~C.,  {van de Weygaert} R., , 2011, {The DTFE public software: The
  Delaunay Tessellation Field Estimator code}

\bibitem[\protect\citeauthoryear{{Chazal}, {Cohen-Steiner}, {Guibas}, {Mémoli}
  \& {Oudot}}{{Chazal} et~al.}{2009}]{Chazal09}
{Chazal} F.,  {Cohen-Steiner} D.,  {Guibas} L.,  {Mémoli} F.,    {Oudot} S.,
  2009, Computer Graphics Forum, 28, 1393–1403

\bibitem[\protect\citeauthoryear{Chazal \& Sun}{Chazal \& Sun}{2014}]{Chazal14}
Chazal F.,  Sun J.,  2014, in Proceedings of the Thirtieth Annual Symposium on
  Computational Geometry SOCG'14, Gromov-hausdorff approximation of filament
  structure using reeb-type graph.
ACM, New York, NY, USA, pp 491:491--491:500

\bibitem[\protect\citeauthoryear{Cohen-Steiner, Edelsbrunner \&
  Harer}{Cohen-Steiner et~al.}{2007}]{CEH07}
Cohen-Steiner D.,  Edelsbrunner H.,    Harer J.,  2007, Discrete Comput. Geom.,
  37, 103

\bibitem[\protect\citeauthoryear{{Colless}, {Peterson} \& {Jackson}}{{Colless}
  et~al.}{2003}]{Col03}
{Colless} M.,  {Peterson} B.~A.,    {Jackson} C. e.~a.,  2003, ArXiv
  Astrophysics e-prints

\bibitem[\protect\citeauthoryear{{Colombi}, {Pogosyan} \&
  {Souradeep}}{{Colombi} et~al.}{2000}]{colombi1}
{Colombi} S.,  {Pogosyan} D.,    {Souradeep} T.,  2000, Physical Review
  Letters, 85, 5515

\bibitem[\protect\citeauthoryear{Dey, Edelsbrunner \& Guha}{Dey
  et~al.}{1999}]{dey1999}
Dey T.~K.,  Edelsbrunner H.,    Guha S.,  1999, in Advances in Discrete and
  Computational Geometry Computational topology.
American Mathematical Society, pp 109--143

\bibitem[\protect\citeauthoryear{Edelsbrunner}{Edelsbrunner}{2001}]{Ede01}
Edelsbrunner H.,  2001, Geometry and Topology for Mesh Generation.
Cambridge Monographs on Applied and Computational Mathematics, Cambridge
  University Press

\bibitem[\protect\citeauthoryear{Edelsbrunner \& Harer}{Edelsbrunner \&
  Harer}{2010}]{EdHa10}
Edelsbrunner H.,  Harer J.,  2010, Computational Topology: An Introduction.
Applied mathematics, American Mathematical Society

\bibitem[\protect\citeauthoryear{Edelsbrunner, Kirkpatrick \&
  Seidel}{Edelsbrunner et~al.}{1983}]{edelsbrunner1983}
Edelsbrunner H.,  Kirkpatrick D.~G.,    Seidel R.,  1983, {IEEE} Trans.
  Information Theory, 29, 551

\bibitem[\protect\citeauthoryear{Edelsbrunner, Letscher \&
  Zomorodian}{Edelsbrunner et~al.}{2002}]{ELZ02}
Edelsbrunner H.,  Letscher J.,    Zomorodian A.,  2002, Discrete \&
  Computational Geometry, 28, 511

\bibitem[\protect\citeauthoryear{Edelsbrunner \& M\"{u}cke}{Edelsbrunner \&
  M\"{u}cke}{1994}]{EM1994}
Edelsbrunner H.,  M\"{u}cke E.~P.,  1994, ACM Trans. Graph., 13, 43

\bibitem[\protect\citeauthoryear{Eldering}{Eldering}{2005}]{eldering05}
Eldering B.,  2005, Topology of Galaxy Models.
MSc thesis, University of Groningen

\bibitem[\protect\citeauthoryear{Euler}{Euler}{1758}]{Eul58}
Euler L.,  1758, Novi Commentarii academiae scientiarum Petropolitanae, 4, 140

\bibitem[\protect\citeauthoryear{{Feldbrugge}}{{Feldbrugge}}{2013}]{feldbruggeBachelor}
{Feldbrugge} J.,  2013, {Stochastic Homology of Random Fields:Graphs towards
  Betti Numbers and Persistence Diagrams }.
Bachelor thesis, University of Groningen

\bibitem[\protect\citeauthoryear{{Forero-Romero}, {Hoffman}, {Gottl{\"o}ber},
  {Klypin} \& {Yepes}}{{Forero-Romero} et~al.}{2009}]{forero09}
{Forero-Romero} J.~E.,  {Hoffman} Y.,  {Gottl{\"o}ber} S.,  {Klypin} A.,
  {Yepes} G.,  2009, MNRAS, 396, 1815

\bibitem[\protect\citeauthoryear{Genovese, Perone-Pacifico, Verdinelli \&
  Wasserman}{Genovese et~al.}{2012}]{GPVW10}
Genovese C.~R.,  Perone-Pacifico M.,  Verdinelli I.,    Wasserman L.,  2012, J.
  Mach. Learn. Res., 13, 1263

\bibitem[\protect\citeauthoryear{{Gonz{\'a}lez} \& {Padilla}}{{Gonz{\'a}lez} \&
  {Padilla}}{2010}]{Gonzalez09}
{Gonz{\'a}lez} R.~E.,  {Padilla} N.~D.,  2010, \mnras, 407, 1449

\bibitem[\protect\citeauthoryear{{Gott} III, {Dickinson} \& {Melott}}{{Gott}
  et~al.}{1986}]{GDM86}
{Gott} III J.~R.,  {Dickinson} M.,    {Melott} A.~L.,  1986, Astrophysical
  Journal, 306, 341

\bibitem[\protect\citeauthoryear{{Guzzo} \& {The Vipers Team}}{{Guzzo} \& {The
  Vipers Team}}{2013}]{vipers}
{Guzzo} L.,  {The Vipers Team} 2013, The Messenger, 151, 41

\bibitem[\protect\citeauthoryear{Gyulassy, Kotava, Kim, Hansen, Hagen \&
  Pascucci}{Gyulassy et~al.}{2012}]{Gyulassy12}
Gyulassy A.,  Kotava N.,  Kim M.,  Hansen C.~D.,  Hagen H.,    Pascucci V.,
  2012, IEEE Transactions on Visualization and Computer Graphics, 18, 1549

\bibitem[\protect\citeauthoryear{{Hahn}, {Carollo}, {Porciani} \&
  {Dekel}}{{Hahn} et~al.}{2007}]{hahn07}
{Hahn} O.,  {Carollo} C.~M.,  {Porciani} C.,    {Dekel} A.,  2007, \mnras, 381,
  41

\bibitem[\protect\citeauthoryear{{Hamilton}, {Gott} III \&
  {Weinberg}}{{Hamilton} et~al.}{1986}]{HGW86}
{Hamilton} A.~J.~S.,  {Gott} III J.~R.,    {Weinberg} D.,  1986, Astrophysical
  Journal, 309, 1

\bibitem[\protect\citeauthoryear{{Huchra}, {Macri} \& {Masters}}{{Huchra}
  et~al.}{2012}]{HMM12}
{Huchra} J.~P.,  {Macri} L.~M.,    {Masters} K.~L. e.~a.,  2012, Apj Supl.,
  199, 26

\bibitem[\protect\citeauthoryear{Ishiyama, Makino, Portegies~Zwart, Groen,
  Nitadori, Rieder, de Laat, McMillan, Hiraki \& Harfst}{Ishiyama
  et~al.}{2013}]{IRMP13}
Ishiyama T.,  Makino J.,  Portegies~Zwart S.,  Groen D.,  Nitadori K.,  Rieder
  S.,  de Laat C.,  McMillan S.,  Hiraki K.,    Harfst S.,  2013, Astrophys.
  J., 767, 146

\bibitem[\protect\citeauthoryear{{Kauffmann} \& {White}}{{Kauffmann} \&
  {White}}{1993}]{kauffmann1993}
{Kauffmann} G.,  {White} S.~D.~M.,  1993, \mnras, 261

\bibitem[\protect\citeauthoryear{{Lacey} \& {Cole}}{{Lacey} \&
  {Cole}}{1994}]{LC94}
{Lacey} C.,  {Cole} S.,  1994, \mnras, 271, 676

\bibitem[\protect\citeauthoryear{{Libeskind}, {Hoffman}, {Knebe}, {Steinmetz},
  {Gottl{\"o}ber}, {Metuki} \& {Yepes}}{{Libeskind} et~al.}{2012}]{libeskind12}
{Libeskind} N.~I.,  {Hoffman} Y.,  {Knebe} A.,  {Steinmetz} M.,
  {Gottl{\"o}ber} S.,  {Metuki} O.,    {Yepes} G.,  2012, MNRAS, 421, L137

\bibitem[\protect\citeauthoryear{{Martinez} \& {Jones}}{{Martinez} \&
  {Jones}}{1990}]{MJ1990}
{Martinez} V.~J.,  {Jones} B.~J.~T.,  1990, MNRAS, 242, 517

\bibitem[\protect\citeauthoryear{{Mecke}, {Buchert} \& {Wagner}}{{Mecke}
  et~al.}{1994}]{Mecke94}
{Mecke} K.~R.,  {Buchert} T.,    {Wagner} H.,  1994, Astronomy \& Astrophysics,
  288, 697

\bibitem[\protect\citeauthoryear{Milnor}{Milnor}{1963}]{Mil63}
Milnor J.,  1963, Morse Theory.
Annals of mathematics studies, Princeton University Press

\bibitem[\protect\citeauthoryear{Munkres}{Munkres}{1984}]{munkres1984elements}
Munkres J.,  1984, Elements of Algebraic Topology.
Advanced book classics, Perseus Books, New York City, NY

\bibitem[\protect\citeauthoryear{Nevenzeel}{Nevenzeel}{2013}]{nevenzeel2013}
Nevenzeel K.,  2013, Triangulating the Darkness.
MSc thesis, University of Groningen

\bibitem[\protect\citeauthoryear{{Neyrinck}}{{Neyrinck}}{2008}]{Neyrinck2008}
{Neyrinck} M.~C.,  2008, \mnras, 386, 2101

\bibitem[\protect\citeauthoryear{{Neyrinck}}{{Neyrinck}}{2012}]{neyrinck12}
{Neyrinck} M.~C.,  2012, MNRAS, 427, 494

\bibitem[\protect\citeauthoryear{{Novikov}, {Colombi} \& {Dor{\'e}}}{{Novikov}
  et~al.}{2006}]{colombi2}
{Novikov} D.,  {Colombi} S.,    {Dor{\'e}} O.,  2006, \mnras, 366, 1201

\bibitem[\protect\citeauthoryear{Okabe, Boots, Sugihara \& Chiu}{Okabe
  et~al.}{2000}]{Okabe2000}
Okabe A.,  Boots B.,  Sugihara K.,    Chiu S.~N.,  2000, Spatial tessellations:
  Concepts and applications of {V}oronoi diagrams, 2nd edn.
Probability and Statistics, Wiley, NYC

\bibitem[\protect\citeauthoryear{{Park}, {Pranav}, {Chingangbam}, {van de
  Weygaert}, {Jones}, {Vegter}, {Kim}, {Hidding} \& {Hellwing}}{{Park}
  et~al.}{2013}]{PPC13}
{Park} C.,  {Pranav} P.,  {Chingangbam} P.,  {van de Weygaert} R.,  {Jones} B.,
   {Vegter} G.,  {Kim} I.,  {Hidding} J.,    {Hellwing} W.~A.,  2013, Journal
  of Korean Astronomical Society, 46, 125

\bibitem[\protect\citeauthoryear{{Parkinson}, {Cole} \& {Helly}}{{Parkinson}
  et~al.}{2008}]{PCH2008}
{Parkinson} H.,  {Cole} S.,    {Helly} J.,  2008, \mnras, 383, 557

\bibitem[\protect\citeauthoryear{Peebles}{Peebles}{1980}]{Pee80}
Peebles P.,  1980, The Large-scale Structure of the Universe.
Princeton series in physics, Princeton University Press

\bibitem[\protect\citeauthoryear{{Platen}, {van de Weygaert} \&
  {Jones}}{{Platen} et~al.}{2007}]{PWJ07}
{Platen} E.,  {van de Weygaert} R.,    {Jones} B.~J.~T.,  2007, MNRAS, 380, 551

\bibitem[\protect\citeauthoryear{{Pranav et. al.}}{{Pranav et.
  al.}}{2016a}]{pranav2016a}
{Pranav et. al.} 2016a, MNRAS, in prep.

\bibitem[\protect\citeauthoryear{{Pranav et. al.}}{{Pranav et.
  al.}}{2016b}]{pranav2016b}
{Pranav et. al.} 2016b, MNRAS, in prep.

\bibitem[\protect\citeauthoryear{Rote \& Vegter}{Rote \&
  Vegter}{2006}]{vegter2004}
Rote G.,  Vegter G.,  2006, Computational Topology: An Introduction.
Springer Berlin Heidelberg

\bibitem[\protect\citeauthoryear{Sahni, Sathyprakash \& Shandarin}{Sahni
  et~al.}{1998}]{sahni1998}
Sahni V.,  Sathyprakash B.,    Shandarin S.,  1998, Astrophys. J., 507, L109

\bibitem[\protect\citeauthoryear{Schaap}{Schaap}{2007}]{Sch07}
Schaap W.,  2007, The Delaunay Tessellation Field Estimator.
PhD Thesis, University of Groningen

\bibitem[\protect\citeauthoryear{{Schaap} \& {van de Weygaert}}{{Schaap} \&
  {van de Weygaert}}{2000}]{Sch00}
{Schaap} W.~E.,  {van de Weygaert} R.,  2000, Astronomy \& Astrophysics, 363,
  L29

\bibitem[\protect\citeauthoryear{{Schaye}, {Crain}, {Bower} \& et.
  al.}{{Schaye} et~al.}{2015}]{schaye2015}
{Schaye} J.,  {Crain} R.~A.,  {Bower} R.~G.,    et. al. 2015, \mnras, 446, 521

\bibitem[\protect\citeauthoryear{{Schmalzing} \& {Buchert}}{{Schmalzing} \&
  {Buchert}}{1997}]{schmalzing1997}
{Schmalzing} J.,  {Buchert} T.,  1997, Astrophys. J. Lett., 482, L1+

\bibitem[\protect\citeauthoryear{Schmalzing, Buchert, Melott, Sahni,
  Sathyaprakash \& Shandarin}{Schmalzing et~al.}{1999}]{schmalzing1999}
Schmalzing J.,  Buchert T.,  Melott A.,  Sahni V.,  Sathyaprakash B.,
  Shandarin S.,  1999, Astrophys. J., 526, 568

\bibitem[\protect\citeauthoryear{{Shandarin}}{{Shandarin}}{2011}]{shandarin11}
{Shandarin} S.~F.,  2011, JCAP, 5, 15

\bibitem[\protect\citeauthoryear{{Sheth} \& {van de Weygaert}}{{Sheth} \& {van
  de Weygaert}}{2004}]{shethwey2004}
{Sheth} R.~K.,  {van de Weygaert} R.,  2004, MNRAS, 350, 517

\bibitem[\protect\citeauthoryear{Shivashankar, Pranav, Natarajan, van~de
  Weygaert, Bos \& Rieder}{Shivashankar et~al.}{2016}]{Shivashankar2015}
Shivashankar N.,  Pranav P.,  Natarajan V.,  van~de Weygaert R.,  Bos E. G.~P.,
     Rieder S.,  2016, {IEEE} Trans. Vis. Comput. Graph., 22, 1745

\bibitem[\protect\citeauthoryear{{Soneira} \& {Peebles}}{{Soneira} \&
  {Peebles}}{1978}]{SP78}
{Soneira} R.~M.,  {Peebles} P.~J.~E.,  1978, \aj, 83, 845

\bibitem[\protect\citeauthoryear{Sousbie}{Sousbie}{2011}]{Sousbie1}
Sousbie T.,  2011, MNRAS, 414, 350

\bibitem[\protect\citeauthoryear{{Sousbie}, {Pichon}, {Courtois}, {Colombi} \&
  {Novikov}}{{Sousbie} et~al.}{2008}]{colombi3}
{Sousbie} T.,  {Pichon} C.,  {Courtois} H.,  {Colombi} S.,    {Novikov} D.,
  2008, \apjl, 672, L1

\bibitem[\protect\citeauthoryear{Sousbie, Pichon \& Kawahara}{Sousbie
  et~al.}{2011}]{Sousbie2}
Sousbie T.,  Pichon C.,    Kawahara H.,  2011, MNRAS, 414, 384

\bibitem[\protect\citeauthoryear{{Springel}}{{Springel}}{2005}]{springel2005}
{Springel} V.,  2005, MNRAS, 364, 1105

\bibitem[\protect\citeauthoryear{Stoica, Gregori \& Mateu}{Stoica
  et~al.}{2005}]{Stoica05}
Stoica R.,  Gregori P.,    Mateu J.,  2005, Stochastic Processes and their
  Applications, 115, 1860

\bibitem[\protect\citeauthoryear{{Stoica}, {Mart{\'{\i}}nez} \&
  {Saar}}{{Stoica} et~al.}{2010}]{Stoica10}
{Stoica} R.~S.,  {Mart{\'{\i}}nez} V.~J.,    {Saar} E.,  2010, \aap, 510, A38

\bibitem[\protect\citeauthoryear{{Sutter}, {Lavaux}, {Wandelt}, {Weinberg},
  {Warren} \& {Pisani}}{{Sutter} et~al.}{2014}]{SLWW14}
{Sutter} P.~M.,  {Lavaux} G.,  {Wandelt} B.~D.,  {Weinberg} D.~H.,  {Warren}
  M.~S.,    {Pisani} A.,  2014, \mnras, 442, 3127

\bibitem[\protect\citeauthoryear{{Tegmark}, {Strauss}, {Blanton} \& et.
  al.}{{Tegmark} et~al.}{2004}]{TSB04}
{Tegmark} M.,  {Strauss} M.~A.,  {Blanton} M.~R.,    et. al. 2004, Physica Rev.
  D., 69, 103501

\bibitem[\protect\citeauthoryear{{Tempel}, {Stoica} \& {Saar}}{{Tempel}
  et~al.}{2012}]{Tempel12}
{Tempel} E.,  {Stoica} R.~S.,    {Saar} E.,  2012, \mnras, p.~138

\bibitem[\protect\citeauthoryear{{van de Weygaert}}{{van de
  Weygaert}}{1991}]{weygaert1991}
{van de Weygaert} R.,  1991, Voids and the geometry of large scale structure.
PhD thesis, University of Leiden

\bibitem[\protect\citeauthoryear{{van de Weygaert}}{{van de
  Weygaert}}{1994}]{weygaert94}
{van de Weygaert} R.,  1994, \aap, 283, 361

\bibitem[\protect\citeauthoryear{{van de Weygaert}}{{van de
  Weygaert}}{2002}]{weygaert2002}
{van de Weygaert} R.,  2002, in {Plionis} M.,  {Cotsakis} S.,  eds, Modern
  Theoretical and Observational Cosmology Vol.~276 of Astrophysics and Space
  Science Library, {Froth across the Universe}.
p.~119

\bibitem[\protect\citeauthoryear{van~de Weygaert}{van~de
  Weygaert}{2007}]{weygaert2007}
van~de Weygaert R.,  2007, in ISVD '07: Proc. of Symp. on Voronoi Diagrams in
  Science and Engineering Voronoi tessellations and the cosmic web: Spatial
  patterns and clustering across the universe.
IEEE Computer Society, Washington, DC, USA, pp 230--239

\bibitem[\protect\citeauthoryear{{van de Weygaert} \& {Bond}}{{van de Weygaert}
  \& {Bond}}{2008}]{WeygaertLectureNotesI}
{van de Weygaert} R.,  {Bond} J.~R.,  2008, in {Plionis} M.,  {L{\'o}pez-Cruz}
  O.,   {Hughes} D.,  eds, A Pan-Chromatic View of Clusters of Galaxies and the
  Large-Scale Structure Vol.~740 of Lecture Notes in Physics, Berlin Springer
  Verlag, {Clusters and the Theory of the Cosmic Web}.
p.~335

\bibitem[\protect\citeauthoryear{{van de Weygaert} \& {Icke}}{{van de Weygaert}
  \& {Icke}}{1989}]{weyicke1989}
{van de Weygaert} R.,  {Icke} V.,  1989, \aap, 213, 1

\bibitem[\protect\citeauthoryear{{van de Weygaert}, {Platen}, {Vegter},
  {Eldering} \& {Kruithof}}{{van de Weygaert} et~al.}{2010}]{WPVE10}
{van de Weygaert} R.,  {Platen} E.,  {Vegter} G.,  {Eldering} B.,    {Kruithof}
  N.,  2010, International Symposium on Voronoi Diagrams in Science and
  Engineering, 0, 224

\bibitem[\protect\citeauthoryear{{van de Weygaert} \& {Schaap}}{{van de
  Weygaert} \& {Schaap}}{2009}]{weyschaap2009}
{van de Weygaert} R.,  {Schaap} W.,  2009, in {Mart{\'{\i}}nez} V.~J.,  {Saar}
  E.,  {Mart{\'{\i}}nez-Gonz{\'a}lez} E.,   {Pons-Border{\'{\i}}a} M.-J.,  eds,
  Data Analysis in Cosmology Vol.~665 of Lecture Notes in Physics, Berlin
  Springer Verlag, {The Cosmic Web: Geometric Analysis}.
pp 291--413

\bibitem[\protect\citeauthoryear{van~de Weygaert, Vegter, Edelsbrunner, Jones,
  Pranav, Park, Hellwing, Eldering, Kruithof, Bos, Hidding, Feldbrugge, ten
  Have, van Engelen, Caroli \& Teillaud}{van~de Weygaert et~al.}{2011}]{ISVD10}
van~de Weygaert R.,  Vegter G.,  Edelsbrunner H.,  Jones B. J.~T.,  Pranav P.,
  Park C.,  Hellwing W.~A.,  Eldering B.,  Kruithof N.,  Bos E. G. P.~P.,
  Hidding J.,  Feldbrugge J.,  ten Have E.,  van Engelen M.,  Caroli M.,
  Teillaud M.,  2011, Transactions on Computational Science, 14, 60

\bibitem[\protect\citeauthoryear{{Vogelsberger}, {Genel}, {Springel}, {Torrey},
  {Sijacki}, {Xu}, {Snyder}, {Bird}, {Nelson} \& {Hernquist}}{{Vogelsberger}
  et~al.}{2014}]{Illustris2014}
{Vogelsberger} M.,  {Genel} S.,  {Springel} V.,  {Torrey} P.,  {Sijacki} D.,
  {Xu} D.,  {Snyder} G.~F.,  {Bird} S.,  {Nelson} D.,    {Hernquist} L.,  2014,
  ArXiv e-prints

\bibitem[\protect\citeauthoryear{Zomorodian \& Carlsson}{Zomorodian \&
  Carlsson}{2005}]{ZomCarl05}
Zomorodian A.,  Carlsson G.,  2005, Discrete \& Computational Geometry, 33, 249

\bibitem[\protect\citeauthoryear{Zomorodian, Ablowitz, Davis, Hinch, Iserles,
  Ockendon \& Olver}{Zomorodian et~al.}{2005}]{Zom05}
Zomorodian A.~J.,  Ablowitz M.~J.,  Davis S.~H.,  Hinch E.~J.,  Iserles A.,
  Ockendon J.,    Olver P.~J.,  2005, Topology for Computing (Cambridge
  Monographs on Applied and Computational Mathematics).
Cambridge University Press, New York, NY, USA

\end{thebibliography}
 

\end{document}